\documentclass[a4paper,12pt]{book}
\usepackage[dvips]{graphicx}
\usepackage{amssymb}
\usepackage{amsmath}
\usepackage{latexsym}
\usepackage{amsthm}
\usepackage{eucal}
\usepackage{eufrak}

\def\cstok#1{\leavevmode\thinspace\hbox{\vrule\vtop{\vbox{\hrule\kern1pt
\hbox{\vphantom{\tt/}\thinspace{\tt#1}\thinspace}}
\kern1pt\hrule}\vrule}\thinspace}
\begin{document}
{\thispagestyle{empty}

\begin{center}
\begin{large}
{\Large Universit\`{a} degli Studi di Napoli ``Federico II''}

\begin{center}
\includegraphics[scale=0.16]{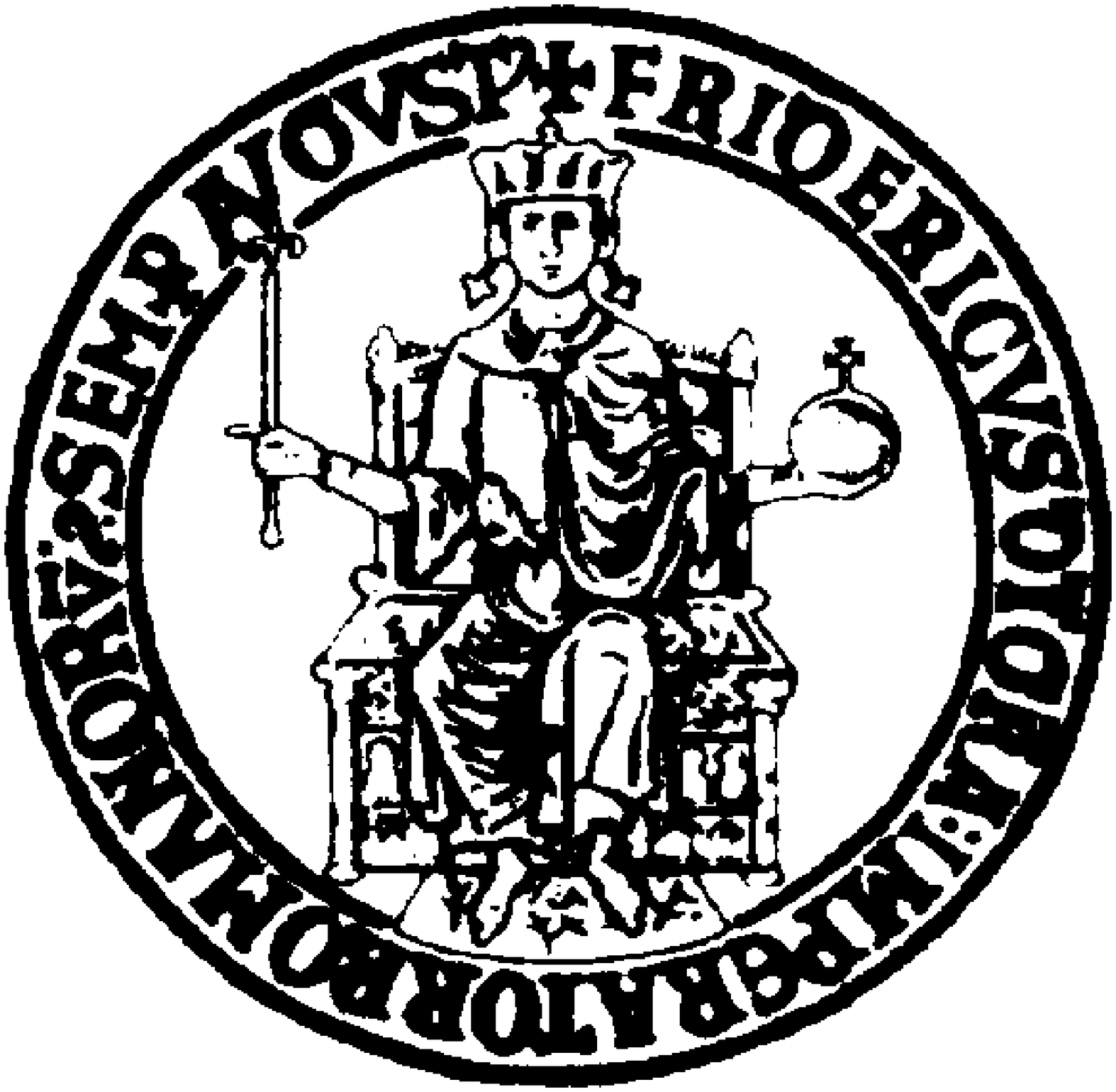}
\end{center}

Dottorato di Ricerca in Fisica Fondamentale ed Applicata\\ XXI ciclo

\vfill

{\Large\textbf{Electromagnetic and Gravitational Waves\\ on a de Sitter background}}

\vfill

Dissertation submitted for the degree of Philosophiae Doctor

Academic Year 2008/2009

\vfill

{\Large\textbf{Roberto Valentino Montaquila}}

\vfill

Under the supervision of

\textbf{Dr D. Bini - Dr G. Esposito}

\vfill\eject
\end{large}
\end{center}

\tableofcontents
\chapter*{Introduction}
\addcontentsline{toc}{chapter}{Introduction}

Within a few years another new window on the universe is expected to open up, with the first direct
detection of gravitational waves. There is keen interest in observing gravitational waves directly,
in order to test Einstein's theory of general relativity and to observe some of the most exotic
objects in nature, like black holes. In addition, the power of gravitational wave
observations to produce more surprises is very high.

The gravitational wave spectrum is completely distinct and, at the same time, complementary to the electromagnetic spectrum. The primary emitters of electromagnetic radiation are charged elementary
particles, mainly electrons; because of overall charge neutrality, electromagnetic radiation is typically
emitted in small regions, with short wavelengths, and conveys direct information about the
physical conditions of small portions of the astronomical sources. By contrast, gravitational waves
are emitted by the cumulative mass and momentum of entire systems, so they have long wavelengths
and convey direct information about large-scale regions. Moreover, electromagnetic waves couple
strongly to charges and so it is easy to detect them, but they are also easily scattered or absorbed by material
between us and the source; gravitational waves instead couple extremely weakly to matter, making them
very hard to detect, but also allowing them to travel to us substantially unaffected by intervening
matter, even from the earliest moments of the Big Bang.

These contrasts, and the history of serendipitous discovery in astronomy, suggest that electromagnetic
observations may be poor predictors of the phenomena that gravitational wave detectors
will eventually discover. Given that 96\% of the mass-energy of the universe carries no charge,
gravitational waves provide a first opportunity to observe directly a major part of the
universe. It might turn out to be as complex and interesting as the charged minor component, the
part that we call ``normal'' matter.

One of the longstanding problems of modern gravitational physics is the detection of gravitational waves, for which the standard theoretical analysis relies upon the split of the space-time metric $g_{ab}$ into a background metric plus perturbations. However, the background need not be Minkowskian in several cases of physical interest, nor it has to be always a solution of the vacuum Einstein equations. As a consequence, we are therefore aiming to investigate in more detail what happens if the background space-time has a non-vanishing Riemann curvature.

This issue has to be seriously considered from an experimental point of view since the gravitational wave detectors of new generation are also designed to investigate strong field regimes: this means that the physical situations, where only the standard Minkowski background is taken into account, could be misleading to achieve self-consistent results.

In particular, several ground-based laser interferometers have been built in the United States (LIGO) \cite{Abramovici, Barish}, Europe (VIRGO and GEO) \cite{Caron, Luck} and Japan (TAMA) \cite{Ando} and they are now in the data taking phase for frequency ranges about $10^{-1}$ kHz. However, new advanced optical configurations allow to reach sensitivities slightly above and below the standard quantum limit for free test-particles, hence we are now approaching the epoch of second \cite{Buonanno} and third \cite{Buonanno2} generation of gravitational wave detectors. This fact, in principle, allows to investigate wide ranges of frequencies where strong field regimes or alternative theories of gravity can be tested \cite{Babusci, Capozziello-Francaviglia, Capozziello-De Laurentis-Francaviglia}.

Besides, the laser interferometer space antenna (LISA) \cite{lisa-science} (which is mainly devoted to work in the range $10^{-4}\sim 10^{-2}$ Hz) should fly within the next decade to investigate the stochastic background of gravitational waves. At much lower frequencies ($10^{-17}$ Hz), cosmic microwave background (CMB) probes, like the forthcoming PLANCK satellite, are designed to detect also gravitational waves by measuring the CMB polarization \cite{rssd.esa.int/Planck} while millisecond pulsar timing can set interesting upper limits in the frequency range between $10^{-9}\sim 10^{-8}$ Hz \cite{Ferrari, Jenet}. At these frequencies, the large number of millisecond pulsars detectable by the square kilometer array would provide a natural ensemble of clocks which can be used as multiple arms of a gravitational wave detector \cite{skatelescope}.

This forthcoming experimental situation is intriguing, but deserves a serious theoretical analysis which cannot leave aside the rigorous investigation of strong field regimes and the possibility that further polarization states of gravitational waves could come out in such regimes. For example, if one takes into account scalar-tensor theories of gravity \cite{Babusci} or higher-order theories \cite{Capozziello-Francaviglia}, scalar-massive gravitons should be considered. This implies that the standard approach where gravitational waves are assumed as small perturbations (coming only from Einstein's general relativity) on a Minkowski background could be totally insufficient. On the other hand, the existence of these further polarization modes could be a straightforward solution of the dark matter problem since massive gravitons could be testable cold dark matter candidates as discussed in \cite{de Paula,Dubovsky}.

We want to face the issue of the rigorous formulation of gravitational wave problem in curved backgrounds. In particular, we want to perform a analysis of gravitational waves in the de Sitter space-time. Achieving solutions in this maximally symmetric background could constitute the paradigm to investigate any curved space-time by the same techniques and could have interesting cosmological applications if a conformal analysis is undertaken as, for example, in \cite{Capozziello-De Laurentis-Francaviglia}, where it is shown how the amplitude of cosmological gravitational waves strictly depends on the cosmological background.

Some important progress in the astronomical observations of the last ten years \cite{Riess,Perlmutter} have led in a progressively convincing way to the surprising conclusion that the recent universe is dominated by an almost spatially homogeneous exotic form of energy density to which there corresponds an effective negative pressure. Such negative pressure acts repulsively at large scales, opposing itself to the gravitational attraction. It has become customary to characterize such energy density by the term ``dark''.

The simplest and best known candidate for the ``dark energy'' is the cosmological constant. As of today, the $\Lambda$CDM (Cold Dark Matter) model, which is obtained by adding a cosmological constant to the standard model, is the one which is in better agreement with the cosmological observations, the latter being progressively more precise. Recent data show that dark energy behaves as a cosmological constant within a few percent error. In addition, if the description provided by the $\Lambda$CDM model is correct, Friedmann's equation shows that the remaining energy components must in the future progressively thin out and eventually vanish thus letting the cosmological constant term alone survive.

In the above scenario the de Sitter geometry, which is the homogeneous and isotropic solution of the vacuum Einstein equations with cosmological term, appears to take the double role of reference geometry of the universe, namely the geometry of space-time deprived of its matter and radiation content and of geometry that the universe approaches asymptotically.

It is now well-known that the problem of solving vector and tensor wave equations in curved space–time, motivated by physical problems such as those occurring in gravitational wave theory and relativistic astrophysics, is in general a challenge even for the modern computational resources. Within this framework, a striking problem is the coupled nature of the set of hyperbolic equations one arrives at.

The Maxwell equations for the electromagnetic potential, supplemented by the Lorenz gauge condition, are decoupled and solved exactly in de Sitter space–time studied in static spherical coordinates. There is no source besides the background. One component of the vector field is expressed, in its radial part, through the solution of a fourth-order ordinary differential equation obeying given initial conditions. The other components of the vector field are then found by acting with lower-order differential operators on the solution of the fourth-order equation (while the transverse part is decoupled and solved exactly from the beginning). The whole four-vector potential is eventually expressed through hypergeometric functions and spherical harmonics. Its radial part is plotted for given choices of initial conditions.

We have thus completely succeeded in solving the homogeneous vector wave equation for Maxwell theory in the Lorenz gauge when a de Sitter space–time is considered. The decoupling technique, analytic formulae and plots are completely original \cite{Montaquila}.

Thus, we have extended this method to the wave equation of metric perturbations on a de Sitter background. It is possible to show that, in a covariant formulation, the supplementary condition for gravitational waves can be described by a functional $\Phi_a$ acting on the space of symmetric rank-two tensors $h_{ab}$ (metric perturbations). For any choice of $\Phi_a$, one gets a different realization of the invertible operator $P_{ab}^{\phantom{ab}cd}$ (Lichnerowicz operator) on metric perturbations. The basic equations of the theory read therefore as
\begin{eqnarray*}
P_{ab}^{\phantom{ab}cd}h_{cd}&=&0,\\
\Phi_a(h)&=&0,
\end{eqnarray*}
where the Lichnerowicz operator $P_{ab}^{\phantom{ab}cd}$ results from the expansion of the Einstein-Hilbert action to quadratic order in the metric perturbations, subject to $\Phi_a(h)=0$. Eventually, a numerical analysis of solutions has been performed.

However, we want to solve explicitly the Einstein equations for metric perturbations on a de Sitter background. Thus, we consider the vacuum Einstein equations with cosmological constant $\Lambda$, i.e.
\begin{equation}
R_{ab}-\frac{1}{2}Rg_{ab}+\Lambda g_{ab}=0.\nonumber
\end{equation}

If we introduce $g_{ab}=\gamma_{ab}+\epsilon h_{ab}$, where $\epsilon$ is a dimensionless parameter which controls the perturbation, we get a coupled system of differential equations to first-order in the metric perturbation $h_{ab}$. At this stage, using the Regge-Wheeler gauge, we can solve this system exactly in terms of the Heun general functions \cite{Montaquila2}.

\chapter{Gravitational waves in \\ de Sitter space-time}

\def\cstok#1{\leavevmode\thinspace\hbox{\vrule\vtop{\vbox{\hrule\kern1pt
\hbox{\vphantom{\tt/}\thinspace{\tt#1}\thinspace}}
\kern1pt\hrule}\vrule}\thinspace}

\def\beq{\begin{equation}}
\def\eeq{\end{equation}}
\def\rmd{{\rm d}}

The non-linearity of the gravitational field in general relativity is one of its most characteristic properties and it is likely that at least some of the crucial properties of the field show themselves only through the non-linear terms. Moreover, it is never entirely clear whether solutions derived by the usual method of \emph{linear approximation} necessarily correspond in every case to exact solutions.

General relativity is a peculiarly complete theory and may not give sensible solutions for situations too far removed from what is physically reasonable. The simplest field due to a finite source is spherically symmetrical but Birkhoff's theorem shows that a spherically symmetrical empty-space field is necessarily static.

Therefore there cannot be truly spherically symmetrical waves and thus any description of radiation from a finite system must necessarily involve three coordinates significantly. This enormously complicates the mathematical difficulties and thus one has to make use of methods of approximation.

The standard theoretical analysis relies upon the split of the space-time metric $g_{ab}$ into "background plus perturbations", that is (unlike the contents section, here we don't write explicitily the $\epsilon$ parameter)
\begin{equation}
g_{ab}=\gamma_{ab}+h_{ab},\nonumber
\label{g}
\end{equation}
where $\gamma_{ab}$ is the background Lorentzian metric, often taken to be of the Minkowski form $\eta_{ab}$, while the symmetric tensor field $h_{ab}$ describes perturbations about $\gamma_{ab}$. The background $\gamma_{ab}$ needs not to be Minkowskian in several cases of physical interest, nor it has to be always a solution of the vacuum Einstein equations. As a consequence, we are therefore aiming to investigate in more detail what happens if the background space-time $(M,\gamma_{ab})$ has a non-vanishing Riemann curvature.

In this work, we perform a analysis of gravitational waves in de Sitter space-time.

\section{Einstein's equations and de Sitter space-time}
Any space-time metric satisfies Einstein's field equations
\begin{equation}
R_{ab}-\frac{1}{2}Rg_{ab}+\Lambda g_{ab}=8\pi T_{ab},
\end{equation}
where $\Lambda$ is the cosmological constant. We shall use $c=1$ and units of mass in which $G=1$ (geometric units). Since both sides are symmetric, these form a set of ten coupled non-linear partial differential equations in the metric tensor components and its first and second-order derivatives. However, due to the so-called Bianchi identity, the covariant divergence of each side vanishes identically, that is,
\begin{equation}
\nabla_{b}\left(R^{ab}-\frac{1}{2}Rg^{ab}+\Lambda g^{ab}\right)=0
\end{equation}
and
\begin{equation}
\nabla_{b}T^{ab}=0,
\end{equation}
hold independent of the field equations. Thus the field equations really provide only six independent differential equations for the metric. This is in fact the correct number of equations needed to determine the space-time, since four of ten components of the metric can be given arbitrary values by use of the four degrees of freedom associated with a coordinate transformation.

The space-time metrics of constant curvature are locally characterized by the condition
\begin{equation}
R_{abcd}=\frac{1}{12}{R(g_{ac}g_{bd}-g_{ad}g_{bc})}.
\end{equation}
and this equation is equivalent to
\begin{equation}
R_{ab}-\frac{1}{4}Rg_{ab}=0,
\end{equation}

Thus, the Riemann tensor is determined by the Ricci scalar $R$ alone and the Einstein tensor becomes
\begin{equation}
R_{ab}-\frac{1}{2}Rg_{ab}=-\frac{1}{4}Rg_{ab}.
\end{equation}

One can therefore regard these spaces as solutions of the field equations for an empty space with $\Lambda=\frac{1}{4}R$. The space of constant curvature with $R=0$ is Minkowski space-time. The space for $R>0$ is \emph{de Sitter space-time}, which has the topology $R^{1}\times S^{3}$. It is easiest visualized as the hyperboloid in five-dimensional Minkowski space given by
\begin{equation}
-(x^0)^2+(x^1)^2+(x^2)^2+(x^3)^2+(x^5)^2=\frac{3}{\Lambda},
\end{equation}
where $\Lambda$ is related to Hubble's constant, $H_0$, by
\begin{equation}
H_0^2=\frac{\Lambda}{3}.
\end{equation}

In the standard spherical coordinates, $(t,r,\theta,\phi)$, one has
\begin{eqnarray}
x^1&=&r\sin\theta\cos\phi, \nonumber\\
x^2&=&r\sin\theta\sin\phi, \nonumber\\
x^3&=&r\cos\theta, \nonumber\\
x^5&=&\sqrt{\frac{1}{H_0^2}-r^2}\cosh(H t), \nonumber\\
x^0&=&\sqrt{\frac{1}{H_0^2}-r^2}\sinh(H t).
\end{eqnarray}

Thus, the metric becomes
\begin{equation}
ds^{2}=-fdt^{2}+\frac{1}{f}dr^{2}
+r^{2}(d\theta^{2}+\sin^{2}\theta d\phi^{2}),
\label{metrica}
\end{equation}
where
\begin{equation}
f\equiv 1-H_0^{2}r^{2}.
\end{equation}

Now, consider de Sitter metric in Gaussian confomally flat umbilical coordinates
\beq
\label{metricds}
\rmd s^2=-\rmd t^2+e^{2H_0t}[\rmd x^2+\rmd y^2+\rmd z^2]\ ,
\eeq

This metric satisfies matter-free Einstein's equations with a non-vanishing cosmological constant $\Lambda$ such that $H_0^2=\Lambda/3$. Moreover, the timelike unit normal vector field to the $t=$constant hypersurfaces
\beq
n=\partial_t
\eeq
form a geodesic and irrotational congruence;
the 3-metric induced on the $t=$constant hypersurfaces results conformally flat:
\beq
g_{ab}=e^{2H_0t}\delta_{ab},\quad (a,b=1,2,3);
\eeq
finally the extrinsic curvature of these hypersurfaces is
\beq
K(n)_{ab}=-H_0 g_{ab} \ .
\eeq

An orthonormal frame associated with $n$ is given by
\beq
\label{frame}
n=\partial_t, \quad e_{\hat a}=e^{-H_0t}\partial_a,\quad (a=1,2,3)
%e_{\hat x}=e^{-H_0t}\partial_x,
%\quad e_{\hat y}=e^{-H_0t}\partial_y,
%\quad e_{\hat z}=e^{-H_0t}\partial_z.
\eeq

For this metric the geodesic equations can be integrated exactly. In fact, they reduce to:
\beq
\label{geneqs}
\frac{\rmd x^i}{\rmd \lambda}=C^ie^{-2H_0t}\ , \quad
\left(\frac{\rmd t}{\rmd \lambda}\right)^2=-\epsilon+C^2e^{-2H_0t}\ ,
\eeq
where the parameter $\epsilon=0,1,-1$ for null, spacelike (with proper length parametrization, say $\lambda=s$) and timelike (with proper time parametrization, say $\lambda=\tau$) geodesics respectively, and $C^i$, $i=1,2,3$ are constants with $C^2=\delta_{ij}C^iC^j$.

The physical components of the tangent vector to the geodesics with respect to the frame (\ref{frame}) result then in
\beq
U_{(\epsilon)}=U_{(\epsilon)}^\alpha \partial_\alpha=
\sqrt{-\epsilon+C^2e^{-2H_0t}}\left[n +\frac{C^ie^{-H_0t}}{\sqrt{-\epsilon+C^2e^{-2H_0t}}}e_{\hat i}\right].
\eeq

It is convenient to discuss the three cases $\epsilon=-1,0,1$ separately, denoting the three different tangent vectors by $U_{(-1)}=U$, $U_{(0)}=P$ and $U_{(1)}=T$, respectively. For timelike geodesics we have
\begin{eqnarray}
U&=&\gamma(U,n) \left[n+\nu(U,n) \hat \nu (U,n)\right]\nonumber \\
&=& \cosh \alpha(t) n +\sinh \alpha(t) \frac{C^i}{C} e_{\hat i},\nonumber \\
\cosh \alpha(t)&=&\sqrt{1+C^2e^{-2H_0t}},
\end{eqnarray}
identifying the speed
\beq
\nu(U,n)=\tanh \alpha(t)=\frac{Ce^{-H_0t}}{\sqrt{1+C^2e^{-2H_0t}}}\ ,
\eeq
as well as its direction (unit spacelike vector)
\beq
\hat \nu (U,n)^{\hat i}=\frac{C^i}{C}.
\eeq

For null geodesics we have
\beq
P=E(P,n) \left[n+ \hat \nu (P,n)\right]=Ce^{-H_0t}\left[n +\frac{C^i}{C} e_{\hat i}\right], \qquad \hat \nu (P,n)^{\hat i}=\frac{C^i}{C},
\eeq
identifying the relative energy
\beq
E(P,n)=Ce^{-H_0t}\ .
\eeq

\subsection{Null geodesics}
Let us consider first the null case. The general solution of Eq. (\ref{geneqs}) is given by
\beq
\label{nullgeos}
e^{H_0t}=H_0C\lambda+c_1\ , \quad
x^i=-\frac{C^i}{H_0C}\frac{1}{H_0C\lambda+c_1}+c_2^i\ .
\eeq

The integration constants $c_1,c_2^i$ can be chosen in such a way that $x^\alpha(\lambda=0)=x^\alpha_0$, whence
\beq
c_1=e^{H_0t_0}\ , \quad c_2^i=x^i_0+\frac{C^i}{H_0C}e^{-H_0t_0}\ ,
\eeq
so that the solution (\ref{nullgeos}) becomes
\beq
\label{nullgeos2}
t=\frac1{H_0}\ln[H_0C\lambda+e^{H_0t_0}]\ , \quad
x^i=-\frac{C^i}{H_0C}\left[-e^{-H_0t_0}+\frac{1}{H_0C\lambda+e^{H_0t_0}}\right]+x^i_0\ .
\eeq

The latter equation can also be cast in the form
\beq
\label{nullgeos3}
%t=\frac1{H_0}\ln[H_0C\lambda+e^{H_0t_0}]\ , \quad
x^i=x^i_0-\frac{C^i}{C}R(t,t_0)\ ,
\eeq
being
\beq
\label{Rdef}
R(t,t_0)=\frac{1}{H_0}(e^{-H_0t}-e^{-H_0t_0})
\eeq

\subsection{Timelike geodesics}
Let us consider now the timelike case. The general solution of Eq. (\ref{geneqs}) is given by
\beq
\label{timegeos}
e^{H_0t}=C\sinh(H_0\tau-c_1)\ , \quad
x^i=-\frac{C^i}{H_0C^2}\coth(H_0\tau-c_1)+c_2^i\ .
\eeq

The integration constants $c_1,c_2^i$ can be chosen in such a way that $x^\alpha(\lambda=0)=x^\alpha_0$, whence
\beq
c_1=-{\rm arcsinh}\left(\frac{e^{H_0t_0}}{C}\right)\ , \quad c_2^i=x^i_0+\frac{C^i}{H_0C^2}\sqrt{1+C^2e^{-2H_0t_0}}\ ,
\eeq
so that the solution (\ref{timegeos}) becomes
\begin{eqnarray}
\label{timegeos2}
t&=&\frac1{H_0}\ln\Bigg[C\sinh\left(H_0\tau+{\rm arcsinh}\left(\frac{e^{H_0t_0}}{C}\right)\right)\Bigg],\nonumber\\
x^i&=&-\frac{C^i}{H_0C^2}\Bigg[\coth\left(H_0\tau+{\rm arcsinh}\left(\frac{e^{H_0t_0}}{C}\right)\right)\nonumber\\
&-&\sqrt{1+C^2e^{-2H_0t_0}}\Bigg]+x^i_0.
\end{eqnarray}

The latter equation can also be written as
\begin{eqnarray}
\label{timegeos3}
x^i&=&x^i_0-\frac{C^i}{H_0C^2}\left( \sqrt{1+C^2e^{-2H_0t}}-\sqrt{1+C^2e^{-2H_0t_0}} \right).
\end{eqnarray}

\section{Conformal form of de Sitter metric}

It is well known that the de Sitter metric can be written as conformal to the Minkowski metric
\beq\qquad
\label{metricdsconf}
\rmd s^2=\left[1+\frac{H_0^2}{4}( x^2+y^2+z^2-t^2)\right]^{-2}(-\rmd t^2+\rmd x^2+\rmd y^2+\rmd z^2)\ .
\eeq

However, the explicit coordinate transformation allowing to cast the metric (\ref{metricds}) in the previous form is somehow hidden in the literature.
First of all introduce standard polar coordinates
\beq
\label{polarcoords}
x=\rho\sin\theta\cos\phi\ , \qquad
y=\rho\sin\theta\sin\phi\ , \qquad
z=\rho\cos\theta\ .
\eeq

The line element (\ref{metricds}) thus takes the form
\beq
\label{metricdspol}
\rmd s^2=-\rmd t^2+e^{2H_0t}[\rmd \rho^2+\rho^2(\rmd \theta^2+\sin^2\theta\rmd \phi^2)]\ .
\eeq

Applying then the following coordinate transformation
\beq\quad
\label{trasf1}
t=\frac{1}{H_0}\ln\left[e^{H_0\tau}\sqrt{1-H_0^2R^2}\right]\ , \quad
\rho=\frac{R e^{-H_0\tau}}{\sqrt{1-H_0^2R^2}}\ , \quad
\theta=\theta\ , \quad
\phi=\phi\ ,
\eeq
gives
\beq
\rmd s^2=-(1-H_0^2R^2)\rmd \tau^2+\frac{\rmd R^2}{1-H_0^2R^2}+R^2(\rmd \theta^2+\sin^2\theta\rmd \phi^2)\ .
\eeq

The further transformation
\beq
\label{trasf2}
\tau=\frac{1}{2H_0}\ln\left[\frac{H_0^2{\bar \rho}^2-(H_0{\bar t}-2)^2}{H_0^2{\bar \rho}^2-(H_0{\bar t}+2)^2}\right]\ , \quad
R=\frac{{\bar \rho}}{1+\frac{H_0^2}{4}({\bar \rho}^2-{\bar t}^2)}\ , \quad
\theta=\theta\ , \quad
\phi=\phi\ ,
\eeq
finally gets
\beq\qquad
\label{metricdsconfpol}
\rmd s^2=\left[1+\frac{H_0^2}{4}({\bar \rho}^2-{\bar t}^2)\right]^{-2}[-\rmd {\bar t}^2+\rmd {\bar \rho}^2+{\bar \rho}^2(\rmd \theta^2+\sin^2\theta\rmd \phi^2)]\ ,
\eeq
which reduces to the line element (\ref{metricdsconf}) once the Cartesian coordinates are restored by using standard relations as in Eq. (\ref{polarcoords}).

By combining the transformations (\ref{trasf1}) and (\ref{trasf2}) we get
\begin{eqnarray}\quad
t&=&\frac{1}{2H_0}\ln\left[\frac{H_0^2{\bar \rho}^2-(H_0{\bar t}-2)^2}{H_0^2{\bar \rho}^2-(H_0{\bar t}+2)^2}\left(1-\frac{H_0^2{\bar \rho}^2}{[1+\frac{H_0^2}{4}({\bar \rho}^2-{\bar t}^2)]^2}\right)\right], \nonumber\\
\rho&=&\frac{{\bar \rho}}{1+\frac{H_0^2}{4}({\bar \rho}^2-{\bar t}^2)}\left[\frac{H_0^2{\bar \rho}^2-(H_0{\bar t}-2)^2}{H_0^2{\bar \rho}^2-(H_0{\bar t}+2)^2}\left(1-\frac{H_0^2{\bar \rho}^2}{[1+\frac{H_0^2}{4}({\bar \rho}^2-{\bar t}^2)]^2}\right)\right]^{-1/2},\nonumber\\
\theta&=&\theta,\nonumber\\
\phi&=&\phi,
\end{eqnarray}
which allows to pass directly from the metric (\ref{metricdspol}) to the conformal one (\ref{metricdsconfpol}).

\section{Preservation of the de Donder supplementary condition}

In classical gauge theory with space-time metric of Lorentzian signature, the gauge-fixing (also called ``supplementary'' condition) leads to a convenient form of the field equation for the potential. For example, for classical electrodynamics in the Lorenz gauge\footnote{In \cite{Lore67}, the author L. Lorenz, who was studying the identity of the vibrations of light with electrical currents, built a set of retarded potentials for electrodynamics which, with hindsight, can be said to satisfy the gauge condition $\nabla^{b}A_{b}=0$, which therefore should not be ascribed to H. Lorentz.}, the wave equation reduces to equation (see Chapter \textbf{2})
$$\cstok{\ }A_{b}-R_{b}^{\; c}A_{c}=0.$$

However, while Maxwell Lagrangian
\begin{equation}
L_{EM}=-\frac{1}{4}F_{ab}F^{ab},
\end{equation}
is invariant under gauge transformations
\begin{equation}
A^f_b\equiv A_b+\nabla_b f,
\label{gt}
\end{equation}
where $f$ is a freely specifiable function of class $C^1$, the Lorenz gauge
$$\Phi(A)=\nabla^{b}A_{b}=0,$$
as well as any other admissible gauge, is not invariant under (\ref{gt}). Nevertheless, to achieve the desired wave equation on $A_b$, it is rather important to make sure that both $A_b$ and the gauge-transformed potential $A^f_b$ obey the same gauge-fixing condition, i.e. \cite{Raju}
\begin{equation}
\Phi(A)=0, \quad \Phi(A^f)=0.
\label{zz}
\end{equation}

A more general situation, here not considered, is instead the case when only the gauge-transformed potential obeys the gauge-fixing condition, i.e. \cite{Jackson}
\begin{equation}
\Phi(A)\neq 0, \quad \Phi(A^f)=0.
\end{equation}

The counterpart of (\ref{zz}) for pure gravity is the well known problem of imposing a gauge on metric perturbations and then requiring its invariance under infinitesimal diffeomorphisms. It is straightforward to show that, in a covariant formulation, the supplementary condition for gravitational waves can be described by a functional $\Phi_{a}$ acting on the space of symmetric rank-two tensors $h_{ab}$. For any choice of $\Phi_{a}$, one gets a different realization of the invertible operator $P_{ab}^{\; \; \; cd}$ on metric perturbations. The basic equations of the theory read therefore as
\begin{eqnarray*}
P_{ab}^{\phantom{ab}cd}h_{cd}&=&0,\\
\Phi_a(h)&=&0,
\end{eqnarray*}
where $P_{ab}^{\phantom{ab}cd}$ results from the expansion of the action functional to quadratic order in the metric perturbations. In general relativity, if one wants to obtain the standard covariant wave operator on metric perturbations, this is taken to be of the de Donder type \cite{Bini-Capozziello-Esposito}
\begin{equation}
\Phi_{a}(h)=\nabla^{b}\left(h_{ab}-\frac{1}{2}\gamma_{ab}h
\right), \label{deDonder}
\end{equation}
where $h \equiv \gamma^{cd}h_{cd}$ and $\nabla^{b}$ denotes covariant derivative with respect to the background metric $\gamma_{ab}$. Under infinitesimal space-time diffeomorphisms, the metric perturbations suffer the variation (the round brackets denoting symmetrization)
\begin{equation}
\delta h_{ab}=\nabla_{(a} \; \varphi_{b)},
\label{variation}
\end{equation}
where $\varphi_{b}$ is a covector, with associated one-form $\varphi_{b}dx^{b}$ and vector field $\varphi^{a}\frac{\partial}{\partial x^{a}}$ (having set $\varphi^{a} \equiv \gamma^{ab}\varphi_{b}$, which results from the isomorphism between tangent and cotangent space to the background space-time, that turns covectors into vectors, or the other way around). The change suffered from the de Donder gauge in (\ref{deDonder}) when metric
perturbations are varied according to (\ref{variation}) is then found to be
\begin{equation}
\delta \Phi_{a}(h)=-\left(\delta_{a}^{\; b}\cstok{\ }+R_{a}^{\; b} \right)\varphi_{b},
\label{(6)}
\end{equation}
where $\cstok{\ }$ is the standard d'Alembert operator in curved space-time, i.e.
\begin{equation}
\cstok{\ } \equiv\gamma^{cd}\nabla_{c}\nabla_{d}.
\end{equation}

By virtue of Eqs. (\ref{deDonder}) and (\ref{(6)}), if the de Donder gauge was originally satisfied, it is preserved
under space-time diffeomorphisms if and only if $\varphi_{b}$ solves the equation $\delta \Phi_{a}(h)=0$, that is
\begin{equation}
-\cstok{\ }\varphi_{a}=R_{a}^{\; b} \varphi_{b}.
\label{(7)}
\end{equation}

At this stage, adding $R_{a}^{\; b}\; \varphi_{b}$ to both sides of (\ref{(7)}), one has
\begin{equation}
P_{a}^{\; b} \; \varphi_{b}=2R_{a}^{\; b} \; \varphi_{b},
\label{(8)}
\end{equation}
where
\begin{equation}
P_{a}^{\; b} \equiv -\delta_{a}^{\; b}\cstok{\ }+R_{a}^{\; b}
\label{(9)}
\end{equation}
is the standard gauge-field operator in the Lorenz gauge (see Chapter \textbf{2}). Thus, we can solve Eq. (\ref{(8)}) in the form
\begin{equation}
\varphi_{c}=\varphi_{c}^{(0)}+2{\widetilde P}_{c}^{\; a} \;
R_{a}^{\; b} \; \varphi_{b},
\label{(10)}
\end{equation}
where $\varphi_{c}^{(0)}$ is a solution of the homogeneous wave equation \cite{Frie75}
\begin{equation}
P_{a}^{\; b} \; \varphi_{b}^{(0)}=0,
\label{(11)}
\end{equation}
while ${\widetilde P}_{c}^{\; a}$ is the inverse operator of $P_{a}^{\; b}$, satisfying
\begin{equation}
{\widetilde P}_{c}^{\; a} \; P_{a}^{\; b}=\delta_{c}^{\; b}.
\label{(12)}
\end{equation}

The operator ${\widetilde P}_{c}^{\; a}$ is an integral operator with kernel given by the massless spin-1 Green function $G_{ab}(x,x') \equiv G_{ab'}$. The latter can be chosen, for example, to be of the Feynman type, i.e. that solution of the equation (see Appendix \textbf{A} for the notation)
\begin{equation}
P_{a}^{\; b}G_{bc'}=g_{ac'}{\delta(x,x') \over \sqrt{-\gamma}},
\label{(13)}
\end{equation}
having the asymptotic expansion as $\sigma \rightarrow 0$ \cite{APNYA-9-220,Bimo04}
\begin{equation}
G_{ab'} \sim {{\rm i}\over 8 \pi^{2}}\left[\sqrt{\bigtriangleup}
{g_{ab'}\over (\sigma+{\rm i}\varepsilon)}
+V_{ab'}\log (\sigma+{\rm i}\varepsilon)+W_{ab'} \right],
\label{(14)}
\end{equation}
where $\sigma(x,x')$ is the Ruse-Synge world function \cite{Ruse31, Syng31, Syng60}, equal to half the square of the geodesic distance $\mu$ between the points $x$ and $x'$.

\section{Massless Green functions in de Sitter space-time}

This general scheme can be completely implemented in the relevant case \cite{Hawk00} of de Sitter space where, relying upon the work in \cite{Alle86}, we know that the massless spin-1 Green function reads as
\begin{equation}
G_{ab'}=\alpha(\mu)g_{ab'}+\beta(\mu)n_{a}n_{b'},
\label{(15)}
\end{equation}
where $\mu(x,x') \equiv \sqrt{2\sigma(x,x')}$ is the geodesic distance between $x$ and $x'$, $n^{a}(x,x')$ and $n^{a'}(x,x')$ are the unit tangents to the geodesic at $x$ and $x'$, respectively, for which
\begin{eqnarray}
n_{a}(x,x')=\nabla_{a} \mu(x,x'),\nonumber\\
n_{a'}(x,x') = \nabla_{a'} \mu(x,x'),
\label{(16)}
\end{eqnarray}
while, in terms of the new variable
\begin{equation}
z \equiv {1\over 2}\left(1+\cos {\mu \over \rho} \right),
\label{(17)}
\end{equation}
the coefficient functions $\alpha$ and $\beta$ are given by \cite{Alle86}
\begin{equation}
\alpha(z)={1\over 48 \pi^{2} \rho^{2}}\left[{3\over (1-z)}
+{1\over z}+\left({2\over z}+{1\over z^{2}}\right)\log(1-z) \right],
\label{(18)}
\end{equation}
\begin{equation}
\beta(z)={1\over 24 \pi^{2}\rho^{2}}\left[1-{1\over z}
+\left({1\over z}-{1\over z^{2}}\right)\log(1-z) \right].
\label{(19)}
\end{equation}

Strictly speaking, the formulae (\ref{(18)})--(\ref{(19)}) are first derived in the Euclidean de Sitter space. In the Lorentzian de Sitter space-time $M$ which is what we are interested in, one can define the set \cite{Alle86}
\begin{equation}
J_{x} \equiv \left \{ x' \in M: \exists \; {\rm geodesic} \;
{\rm from} \; x \; {\rm to} \; x' \right \}.
\label{(20)}
\end{equation}

Moreover, it is well-known that $M$ can be viewed as an hyperboloid imbedded in flat space, i.e. as the set of points $Y^{a} \in {\bf R}^{n+1}$ such that
\begin{equation}
Y^{a}Y^{b}\eta_{ab}=\rho^{2},
\label{(21)}
\end{equation}
where $\eta_{ab}={\rm diag}(-1,1,1,1)$, so that its induced metric reads as
\begin{equation}
ds^{2}=\eta_{ab}dY^{a}dY^{b}.
\label{(22)}
\end{equation}

As is stressed in Ref. \cite{Alle86}, the relation
\begin{equation}
z(x,x')={1\over 2} \left[1+{\eta_{ab}Y^{a}(x)Y^{b}(x')\over \rho^{2}}
\right]
\label{(23)}
\end{equation}
is well defined both inside and outside $J_{x}$, and it is an analytic function of the coordinates $Y^{a}$. Thus, Eq. (\ref{(23)}) makes it possible to define $z(x,x')$ everywhere on de Sitter, and one can define the geodesic distance
\begin{equation}
\mu(x,x') \equiv 2\rho \cos^{-1}(\sqrt{z})
\label{(24)}
\end{equation}
as the limiting value \cite{Alle86} above the standard branch cut of $\cos^{-1}$. Along similar lines, the equations defining $n_{a},n_{a'}$ and $g_{ab'}$ have right-hand sides which are analytic functions of the coordinates $Y^{a}$, and are hence well defined everywhere on Lorentzian de Sitter space-time \cite{Alle86}.

\section{Evaluation of the kernel}

In a de Sitter background the Ricci tensor is proportional to the metric through the cosmological constant: $R_{ab}=\Lambda g_{ab}$, and hence the formulae (\ref{(10)}), (\ref{(15)}), (\ref{(18)}) and (\ref{(19)}) lead to the following explicit expression for the solution of the inhomogeneous wave equation (\ref{(8)}):
\begin{eqnarray}
\varphi_{c}(x)&=&\varphi_{c}^{(0)}(x)+2 \Lambda \int\Bigr[\alpha(z(\mu(x,x')))g_{c}^{\; a'}\nonumber \\
&+&\beta(z(\mu(x,x')))n_{c}n^{a'}\Bigr]\varphi_{a'}(x')
\sqrt{-\gamma(x')}d^{4}x',
\label{(25)}
\end{eqnarray}
where, from Eq. (\ref{(24)}),
\begin{equation}
\mu(x,x')=2\rho \cos^{-1} \sqrt{{1\over 2}\left(1+{\eta_{ab}Y^{a}(x)Y^{b}(x')\over \rho^{2}}\right)},
\label{(26)}
\end{equation}
while Eqs. (\ref{(18)}) and (\ref{(19)}) should be exploited to express $\alpha$ and $\beta$, bearing in mind Eq. (\ref{(26)}) jointly with
\begin{equation}
z(x,x')={1\over 2}\left[1+\cos \left({\mu(x,x')\over \rho}\right)\right].
\label{(27)}
\end{equation}

Moreover, the bivector $g_{c}^{\; a'}$ in the integrand (\ref{(25)}) is given by \cite{Alle86}
\begin{eqnarray}
g_{a}^{\; b'}&=&C^{-1}(\mu)\nabla_{a}n^{b'}-n_{a}n^{b'},\nonumber \\
C(\mu)&=&-{1\over \rho \sin (\mu / \rho)}.
\label{(28)}
\end{eqnarray}

The right-hand side of the formula expressing $g_{a}^{\; b'}$ is an analytic function of the coordinates $Y^{a}$ and is therefore well defined everywhere on de Sitter \cite{Alle86}. The integral on the right-hand side of Eq. (\ref{(25)}) can be conveniently expressed
the form
\begin{eqnarray}
f_{c}(x)&=&\int \Bigr[\alpha(z)C^{-1}(\mu)\nabla_{c}\nabla^{a'}\mu\nonumber \\
&+&(\beta(z)-\alpha(z))(\nabla_{c}\mu)(\nabla^{a'}\mu)\Bigr]
\varphi_{a'}(x')\sqrt{-\gamma(x')}d^{4}x',
\label{(29)}
\end{eqnarray}
with $\alpha$ and $\beta-\alpha$ given by (cf. (\ref{(18)}) and (\ref{(19)}))
\begin{equation}
\alpha(z)={(1+2z)\over 48 \pi^{2}\rho^{2}}\left[{1\over z(1-z)}+{1\over z^{2}}\log(1-z)\right],
\label{(30)}
\end{equation}
\begin{equation}
\beta(z)-\alpha(z)={1\over 48 \pi^{2}\rho^{2}}\left[{(-3+2z-2z^{2})\over z(1-z)}-{3\over z^{2}}\log(1-z)\right].
\label{(31)}
\end{equation}

Equation (\ref{(25)}) is therefore an integral equation reading as
\begin{equation}
\varphi_{c}(x)=\varphi_{c}^{(0)}(x)+\Lambda \int K_{c}^{\; a'}\varphi_{a'} \sqrt{-\gamma(x')}d^{4}x',
\label{(32)}
\end{equation}
with unbounded kernel given by
\begin{equation}
K_{c}^{\; a'} \equiv 2 \Bigr[\alpha(z)C^{-1}(\mu) \nabla_{c}\nabla^{a'}\mu
+(\beta(z)-\alpha(z))(\nabla_{c}\mu)(\nabla^{a'}\mu)\Bigr].
\label{(33)}
\end{equation}

This kernel is indeed unbounded by virtue of the limits
\begin{equation}
48 \pi^{2} \rho^{2} \lim_{z \to 0} z \alpha(z)={1\over 2},
\label{(34)}
\end{equation}
\begin{equation}
48 \pi^{2} \rho^{2} \lim_{z \to 1} (1-z) \alpha(z)=1,
\label{(35)}
\end{equation}
\begin{equation}
48 \pi^{2} \rho^{2} \lim_{z \to 0} z (\beta(z)-\alpha(z))=-{3\over 2},
\label{(36)}
\end{equation}
\begin{equation}
48 \pi^{2} \rho^{2} \lim_{z \to 1} (1-z)(\beta(z)-\alpha(z))=-3.
\label{(37)}
\end{equation}

At this stage, we can exploit (\ref{(23)}) and (\ref{(33)}) to re-express the kernel in the form
\begin{eqnarray}
K_{c}^{\; a'}&=& {(\nabla_{c}z)(\nabla^{a'}z)\over 24 \pi^{2}\rho^{4}(1-z)}\biggr[2+
\left(-3 +{\sqrt{z}\over 2}(1+2z)\right)\left({1\over z(1-z)}+{1\over z^{2}}\log(1-z) \right)\biggr]\nonumber \\
&+& {(\nabla_{c}\nabla^{a'}z)\over 6 \pi^{2}}\sqrt{z}(1+2z)\left[{1\over z(1-z)}+{1\over z^{2}}\log(1-z)\right].
\label{(38)}
\end{eqnarray}

Note now that $\varphi_{c}^{(0)}(x)$ in Eq. (\ref{(32)}), being a solution of the homogeneous vector wave equation (\ref{(11)}), admits the Huygens principle representation \cite{APNYA-9-220}
\begin{equation}
\varphi_{c}^{(0)}(x)=\int_{\Sigma'}\sqrt{-\gamma(x')}\Bigr[G_{cb'}\varphi_{\; \; \; \; \; \; ;m'}^{(0)b'}
-G_{cb';m'}\varphi^{(0)b'}\Bigr]g^{m'l'} d\Sigma_{l'}^{'},
\label{(39)}
\end{equation}
where
\begin{equation}
G_{cb'}=\alpha g_{cb'}+\beta \mu_{;c} \mu_{;b'}={1\over 2}K_{cb'},
\label{(40)}
\end{equation}
\begin{equation}
G_{cb';m'}={1\over 2}K_{cb';m'}.
\label{(41)}
\end{equation}

Unlike the work in \cite{APNYA-9-220}, we here advocate the use of the Green function (\ref{(15)}) rather than the sum, over all distinct geodesics between $x$ and $x'$, of the Hadamard functions. To lowest order in the cosmological constant $\Lambda$, Eq. (\ref{(39)}) may be used to approximate the desired solution of Eq. (\ref{(32)}) in the form
\begin{equation}
\varphi_{c}(x)=\varphi_{c}^{(0)}(x)+\Lambda \int K_{c}^{\; \; a'} \varphi_{a'}^{(0)}
\sqrt{-\gamma(x')}d^{4}x'+{\rm O}(\Lambda^{2}).
\label{(42)}
\end{equation}

Omitting indices for simplicity, the general algorithm for solving Eq. (\ref{(32)}), here re-written in the form
\begin{equation}
\varphi=\varphi^{(0)}+\Lambda \int K \varphi,
\label{(43)}
\end{equation}
would be instead
\begin{equation}
\varphi_{1}=\varphi^{(0)}+\Lambda \int K \varphi^{(0)},
\label{(44)}
\end{equation}
\begin{equation}
\varphi_{2}=\varphi^{(0)}+\Lambda \int K \varphi_{1}=\varphi^{(0)}+\Lambda \int K \varphi^{(0)}
+\Lambda^{2} \int \int K K \varphi^{(0)},
\label{(45)}
\end{equation}
\begin{equation}
\varphi_{n}=\varphi^{(0)}+\sum_{j=1}^{n}\Lambda^{j}\int ... \int K^{j} \varphi^{(0)},
\label{(46)}
\end{equation}
\begin{equation}
\varphi= \lim_{n \to \infty} \varphi_{n}.
\label{(47)}
\end{equation}

In this Chapter, we have seen that when the de Donder gauge is imposed, its preservation under infinitesimal space-time diffeomorphisms is guaranteed if and only if the associated covector is ruled by a second-order hyperbolic operator which is the classical counterpart of the ghost operator in quantum gravity and the vector wave equation (\ref{(7)}) is been studied by using an integral representation.

However, a different approach is viable, that is, through a solution by factorization of a hyperbolic equation. In fact, in the equation (\ref{(7)}) the Ricci term has opposite sign with respect to the wave equation for Maxwell theory, in the Lorenz gauge. Thus, we are interested in the following generalized wave equation:
\begin{equation}
-\cstok{\ }X_{a}+\epsilon R_{a}^{\; b}X_{b}=0,
\end{equation}
where $\epsilon=\pm1$. In particular, $\epsilon=1$ corresponds to studying the Maxwell vector wave equation, whereas $\epsilon=-1$ provides our consistency equation (\ref{(7)}).

In the Chapter \textbf{2}, by virtue of the spherical symmetry of de Sitter space-time, these equations should be conveniently written by using the expansion of $X$ in vector harmonics. In this way, we solve the Maxwell equation in de Sitter space-time \cite{Montaquila} and, at this stage, it is possible to relate the solutions of the two problems \cite{Bini-Capozziello-Esposito}.
\chapter{The vector wave equation}
\def\beq{\begin{equation}}
\def\eeq{\end{equation}}
It is by now well known that the problem of solving vector and tensor wave equations in curved space-time is in general a challenge even for the modern computational resources. On using the Maxwell action functional
\begin{equation}
S=-{1\over 4}\int_{M}F_{ab}F^{ab}\sqrt{-g}\; d^{4}x
\end{equation}
where $F_{ab}$ is the electromagnetic field tensor, one gets the wave operator $P_{a}^{\; b}$, that is,
\begin{equation}
P_{a}^{\; b}=-\delta_{a}^{\; b}\cstok{\ }+R_{a}^{\; b}+\nabla_{a}\nabla^{b},
\end{equation}
but jointly with the Lorenz gauge condition
\begin{equation}
\nabla^{b}A_{b}=0,
\label{(gauge)}
\end{equation}
we have
\begin{equation}
P_{a}^{\; b}A_{b}=\left(-\delta_{a}^{\; b}\cstok{\ }+R_{a}^{\; b}\right)A_{b}.
\end{equation}

Thus, in vacuum, the coupled equations for the electromagnetic potential are
\begin{equation}
\left(-\delta_{a}^{\; b}\cstok{\ }+R_{a}^{\; b}\right)A_{b}=0
\end{equation}
and eventually
\begin{equation}
\cstok{\ }A_{b}-R_{b}^{\; c}A_{c}=0.
\label{(26)}
\end{equation}

We note that in the quantum case, one has, even on using the Lorenz gauge condition, the wave operator
\begin{equation}
\tilde{P_{a}^{\; b}}=-\delta_{a}^{\; b}\cstok{\ }+R_{a}^{\; b}+\left(1-\frac{1}{\alpha}\right)\nabla_{a}\nabla^{b},
\end{equation}
but, following Feynman, we put $\alpha=1$ and one has
\begin{equation}
\tilde{P_{a}^{\; b}}=-\delta_{a}^{\; b}\cstok{\ }+R_{a}^{\; b},
\end{equation}
that is, at least formally, the same wave operator of the classical case.

A deep link exists between classical and quantum theory, since in the latter, the one-loop analysis depends on the functional determinant of the operator $P_{a}^{\; b}$.

At this point, we want to study the vector wave equation in de Sitter space-time with static spherical coordinates, so that the line element is (\ref{metrica}). The vector field $X$ solving the vector wave equation can be expanded in spherical harmonics according to \cite{Cohen-Kegeles}
\begin{eqnarray}
X&=& {\widetilde Y}_{lm}(\theta)e^{-i(\omega t-m \phi)}\Bigr[f_{0}(r)dt+f_{1}(r)dr \Bigr] \nonumber \\
&+& e^{-i(\omega t-m \phi)}\left[-{mr \over \sin \theta}f_{2}(r){\widetilde Y}_{lm}(\theta)
+f_{3}(r){d{\widetilde Y}_{lm}\over d\theta}\right]d\theta \nonumber \\
&+& i e^{-i(\omega t-m \phi)}\left[-r \sin \theta f_{2}(r){d{\widetilde Y}_{lm}\over d\theta}
+m f_{3}(r){\widetilde Y}_{lm}(\theta)\right]d\phi,
\label{(210)}
\end{eqnarray}
where ${\widetilde Y}_{lm}(\theta)$ is the $\theta$-dependent part of the spherical harmonics $Y_{lm}(\theta,\phi)$, solution of the equation
\begin{equation}
\left[{d^{2}\over d\theta^{2}}+\cot \theta {d \over d\theta}
+\left(-{m^{2}\over \sin^{2}\theta}+L \right)\right]{\widetilde Y}_{lm}(\theta)=0,
\label{(210a)}
\end{equation}
with $L\equiv l(l+1)$. As one has shown in \cite{Bini-Capozziello-Esposito}, these equations lead to a system of coupled ordinary differential equations for the functions $f_{0},f_{1},f_{3}$, besides a decoupled equation for $f_{2}$ ($f_{2}$ being related to the transverse part of $X$). The equation for $f_{2}(r)$ can be easily integrated in terms of hypergeometric functions. In fact, assuming
\begin{equation}
f_{2}(r)=f^{-i \omega/(2H)}\psi(r),
\end{equation}
the resulting equation for $\psi$ reads as
\begin{equation}
{d^{2}\psi \over dr^{2}}=-{2i \over rf}\Bigr(2iH^{2}r^{2}-i+\omega Hr^{2}\Bigr){d\psi \over dr}
-{1\over r^{2}f^{2}}\Bigr[\omega^{2}r^{2}-L-2H^{2}r^{2}+3i \omega Hr^{2} \Bigr].
\end{equation}

Thus, the solution $f_{2}(r)$ is of the form
\begin{equation}
f_{2}(r)=f^{-i \Omega /2}\biggr[U_{1}r^{l}F\left(a_{-},a_{+};{3\over 2}+l;H^{2}r^{2}\right)
+ U_{2}r^{-l-1}F\left(a_{+},a_{-};{1\over 2}-l;H^{2}r^{2}\right)\biggr],
\label{(1.6)}
\end{equation}
where
\begin{equation}
\Omega \equiv {\omega \over H},\;a_{\pm}\equiv-{1\over 4}\left({2i \Omega}-3-2l \pm 1 \right).
\label{(1.7)}
\end{equation}

At this stage, however, the problem remained of solving explicitly also for $f_{0}(r),f_{1}(r),f_{3}(r)$ in the expansion (\ref{(210)}). For this purpose, we derive the decoupling procedure for such modes in de Sitter  and we write explicitly the decoupled equations. Eventually, we solve explicitly for $f_{0},f_{1},f_{3}$ in terms of hypergeometric functions and we plot such solutions for suitable initial conditions.

\section{Coupled modes}

Unlike $f_{2}$, the functions $f_{0},f_{1}$ and $f_{3}$ obey instead a coupled set, given by Eqs. (54), (55), (57) of \cite{Bini-Capozziello-Esposito}, which are here written, more conveniently, in matrix form as (we set $\epsilon=1$ in the Eqs. of \cite{Bini-Capozziello-Esposito}, which corresponds to studying the vector wave equation (\ref{(26)}))
\begin{equation}
\left(
\begin{array}{ccc}
P_{0} \quad & \alpha  \quad& 0 \\
f^{-2}\alpha  \quad & P_{1}   \quad &
r^{-2}f^{-1}L \beta  \\
0   \quad & \beta  \quad & P_{3}
\end{array}
\right)
\,
\left(
\begin{array}{l}
f_{0} \\
f_{1} \\
f_{3}
\end{array}
\right)
=0,
\label{(2.1)}
\end{equation}
having defined
\begin{eqnarray}
\alpha &\equiv& {2i \omega H^{2}r \over f},\\
\label{(2.2)}
\beta &\equiv& {2\over r},
\label{(2.3)}
\end{eqnarray}
\begin{eqnarray}
P_{0} &\equiv& {d^{2}\over dr^{2}}+Q_{1}{d\over dr}+Q_{2},\\
\label{(2.4)}
P_{1} &\equiv& {d^{2}\over dr^{2}}+Q_{3}{d\over dr}+Q_{4},\\
\label{(2.5)}
P_{3} &\equiv& {d^{2}\over dr^{2}}+Q_{5}{d\over dr}+Q_{6},
\label{(2.6)}
\end{eqnarray}
\begin{eqnarray}
Q_{1} &\equiv& \beta={2\over r},\\
\label{(2.7)}
Q_{2} &\equiv& {\omega^{2}\over f^{2}}-{L \over r^{2}f},\\
\label{(2.8)}
Q_{3} &\equiv& {6\over r}\left(1-{2\over 3}{1\over f}\right),\\
\label{(2.9)}
Q_{4} &\equiv& {\omega^{2}\over f^{2}}
-\left(4H^{2}+{(L+2) \over r^{2}}\right){1\over f},\\
\label{(2.10)}
Q_{5} &\equiv& {2\over r}\left(1-{1\over f}\right),\\
\label{(2.11)}
Q_{6} &\equiv& {\omega^{2}\over f^{2}}-{L\over r^{2}}{1\over f}.
\label{(2.12)}
\end{eqnarray}

With our notation, the three equations resulting from (\ref{(2.1)}) can be written as
\begin{eqnarray}
\label{(2.13)}
P_{0}f_{0}&=&-\alpha f_{1},\\
\label{(2.14)}
P_{1}f_{1}&=&-{\alpha\over f^{2}}f_{0}-{L \beta\over r^{2}f}f_{3},\\
\label{(2.15)}
P_{3}f_{3}&=&-\beta f_{1}.
\end{eqnarray}

\section{Decoupled equations}

We now express $f_{1}$ from Eq. (\ref{(2.13)}) and we insert it into Eq. (\ref{(2.14)}), i.e.
\begin{equation}
P_{1} \left(-{1\over \alpha}P_{0}f_{0}\right)=-{\alpha\over f^{2}}f_{0}-{L\beta\over r^{2}f}f_{3}.
\label{(3.1)}
\end{equation}

Next, we exploit the Lorenz gauge condition (\ref{(gauge)}), i.e. \cite{Bini-Capozziello-Esposito}
\begin{equation}
f_{3}={r^{2}f \over L}{d\over dr}\left(-{1\over \alpha}P_{0}f_{0}\right)-{2r(1-2f)\over L}
\left(-{1\over \alpha}P_{0}f_{0}\right)+i{\Omega H r^{2}\over L f}f_{0},
\label{(3.2)}
\end{equation}
and from Eqs. (\ref{(3.1)}) and (\ref{(3.2)}) we obtain, on defining the new independent and dimensionless variable $x=rH$, the following fourth-order equation for $f_{0}$:
\begin{equation}
\left[{d^{4}\over dx^{4}}+\kappa_{3}(x){d^{3}\over dx^{3}}+\kappa_{2}(x){d^{2}\over dx^{2}}+\kappa_{1}(x){d\over dx}
+\kappa_{0}(x) \right]f_{0}(x)=0,
\label{(3.3)}
\end{equation}
where
\begin{equation}
\kappa_{0}(x) \equiv {\kappa(x)\over x^{4}(x^{2}-1)^{4}},
\label{(3.4)}
\end{equation}
\begin{equation}
\kappa(x) \equiv L(L-2)+2L(2-L-\Omega^{2})x^{2}+\Bigr[\Omega^{4}+4 \Omega^{2}+L (L+2(\Omega^{2}-1))\Bigr]x^{4},
\label{(3.5)}
\end{equation}
\begin{equation}
\kappa_{1}(x) \equiv {4(\Omega^{2}+L-2+6x^{2})\over x(x^{2}-1)^{2}},
\label{(3.6)}
\end{equation}
\begin{equation}
\kappa_{2}(x) \equiv {2 \Bigr[-L+(\Omega^{2}+L-14)x^{2}+18 x^{4}\Bigr]\over x^{2}(x^{2}-1)^{2}},
\label{(3.7)}
\end{equation}
\begin{equation}
\kappa_{3}(x) \equiv {4(-1+3x^{2})\over x(x^{2}-1)}.
\label{(3.8)}
\end{equation}

Eventually, $f_{1}$ and $F_{3} \equiv H f_{3}$ can be obtained from Eqs. (\ref{(2.13)}) and (\ref{(3.2)}), i.e.
\begin{equation}
f_{1}(x)={i\over 2 \Omega}{(1-x^{2})\over x}\left({d^{2}\over dx^{2}}+{2\over x}{d\over dx}+{\Omega^{2}\over (1-x^{2})^{2}}-{L \over x^{2}(1-x^{2})}\right)f_{0}(x),
\label{(3.9)}
\end{equation}
\begin{equation}
F_{3}(x)=\left[{x^{2}(1-x^{2}) \over L}{d\over dx}-{2x(2x^{2}-1)\over L}\right]f_{1}(x)
+i{\Omega \over L}{x^{2}\over (1-x^{2})} f_{0}(x).
\label{(3.10)}
\end{equation}

Our $f_{1}$ and $f_{3}$ are purely imaginary, which means we are eventually going to take their imaginary part only.
Moreover, as a consistency check, Eqs. (\ref{(3.9)}) and (\ref{(3.10)}) have been found to agree with Eq. (\ref{(2.15)}), i.e. (\ref{(2.15)}) is then identically satisfied.

\section{Exact solutions}

Equation (\ref{(3.3)}) has four linearly independent integrals, so that its general solution involves four coefficients of linear combination $C_{1},C_{2},C_{3},C_{4}$, according to (hereafter, $F$ is the hypergeometric function already used in (\ref{(1.6)}))
\begin{eqnarray}
f_{0}(x)&=& C_{1}x^{-1-l}(1-x^{2})^{-{i\over 2}\Omega}F \left(-{i \over 2}\Omega-{l\over 2},
-{i \over 2}\Omega+{1\over 2}-{l\over 2};{1\over 2}-l;x^{2}\right)\nonumber \\
&+& C_{2}x^{-1-l}(1-x^{2})^{-{i\over 2}\Omega}F \left(-{i \over 2}\Omega+1-{l \over 2},
-{i \over 2}\Omega-{1\over 2}-{l\over 2};{1\over 2}-l;x^{2} \right)\nonumber \\
&+& C_{3}x^{l}(1-x^{2})^{-{i\over 2}\Omega}F \left(-{i \over 2}\Omega+{l \over 2},
-{i \over 2}\Omega+{3\over 2}+{l\over 2};{3\over 2}+l;x^{2} \right)\nonumber \\
&+& C_{4}x^{l}(1-x^{2})^{-{i\over 2}\Omega}F \left(-{i \over 2}\Omega+1+{l \over 2},
-{i \over 2}\Omega+{1\over 2}+{l\over 2};{3\over 2}+l;x^{2} \right).
\label{(4.1)}
\end{eqnarray}
Regularity at the origin (recall that $x=0$ should be included, since the event horizon for an observer situated at $x=0$ is given by $x=1$ \cite{BOUCHER}) implies that $C_{1}=C_{2}=0$, and hence, on defining
\begin{equation}
a_{1} \equiv -{i\over 2}\Omega+{l\over 2}, \;
b_{1} \equiv -{i\over 2}\Omega+{3\over 2}+{l \over 2}, \;
d_{1} \equiv {3\over 2}+l,
\label{(4.3)}
\end{equation}
we now re-express the regular solution in the form (the points $x=0,\pm 1$ being regular singular points of the equation (\ref{(3.3)}) satisfied by $f_{0}$)
\begin{equation}
f_{0}(x)=x^{l}(1-x^{2})^{-{i\over 2}\Omega}\Bigr[C_{3}F(a_{1},b_{1};d_{1};x^{2})
+C_{4}F(a_{1}+1,b_{1}-1;d_{1};x^{2})\Bigr],
\label{(4.4)}
\end{equation}
where the second term on the right-hand side of (\ref{(4.4)}) can be obtained from the first through the replacements
$$
C_{3} \rightarrow C_{4}, \;
a_{1} \rightarrow a_{1}+1, \;
b_{1} \rightarrow b_{1}-1
$$
and the series expressing the two hypergeometric functions are conditionally convergent, because they satisfy
${\it Re}(c-a-b)=i \Omega$, with
$$
a=a_{1},a_{1}+1; \;
b=b_{1},b_{1}-1; \;
c=d_{1}.
$$

Last, we exploit the identity
\begin{equation}
{d\over dz}F(a,b;c;z)={ab \over c}F(a+1,b+1;c+1;z)
\label{(4.5)}
\end{equation}
to find, in the formula (\ref{(3.9)}) for $f_{1}(x)$,
\begin{eqnarray}
\; & \; & {d\over dx}f_{0}(x)=C_{3} \biggr \{x^{l-1}(1-x^{2})^{-{i\over 2}\Omega -1}\Bigr[l(1-x^{2})
+i \Omega x^{2}\Bigr]F(a_{1},b_{1};d_{1};x^{2}) \nonumber \\
&+& {2a_{1}b_{1}\over d_{1}}x^{l+1}(1-x^{2})^{-{i\over 2}\Omega}F(a_{1}+1,b_{1}+1;d_{1}+1;x^{2})\biggr \} \nonumber\\
&+& \Bigr \{ C_{3} \rightarrow C_{4}, \; a_{1} \rightarrow a_{1}+1, \; b_{1} \rightarrow b_{1}-1 \Bigr \}.
\label{(4.6)}
\end{eqnarray}

It is then straightforward, although tedious, to obtain the second derivative of $f_{0}$ (see Eq. (\ref{(A1)}) of the Appendix \textbf{B}) in the equation for $f_{1}$ and the third derivative of $f_{0}$ in the formula (\ref{(3.10)}) for $H f_{3}$. The results are exploited to plot the solutions.

In general, for given initial conditions at $\bar{x} \in [0,1[$, one can evaluate $C_{3}$ and $C_{4}$ from
$$
f_{0}(\bar{x})=\bar{y}, \;
f_{0}'(\bar{x})=\bar{y}',
$$
i.e. $C_{3}=C_{3}(\bar{y},\bar{y}'),C_{4}=C_{4}(\bar{y},\bar{y}')$.

\section{Plot of the solutions}
To plot the solutions, we begin with $f_{0}$ as given by (\ref{(4.4)}), which is real-valued despite the many $i$ factors occurring therein. Figures (2.1) to (2.3) describe the solutions for the two choices $C_{3}=0,C_{4}=1$ or the other way around and various values of $l$ and $\Omega$.

\begin{figure}
\includegraphics[scale=0.25,angle=-90]{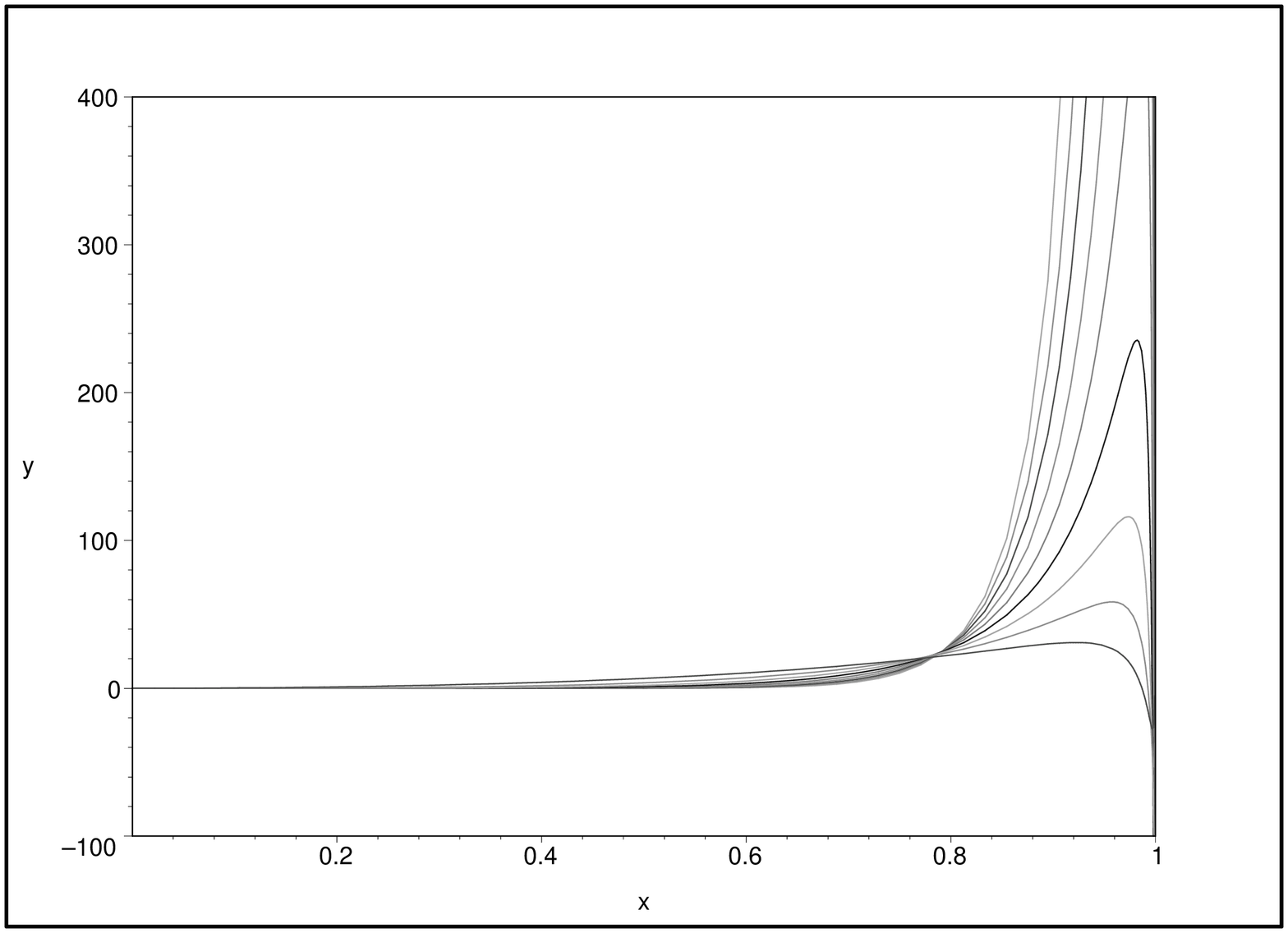}
\includegraphics[scale=0.25,angle=-90]{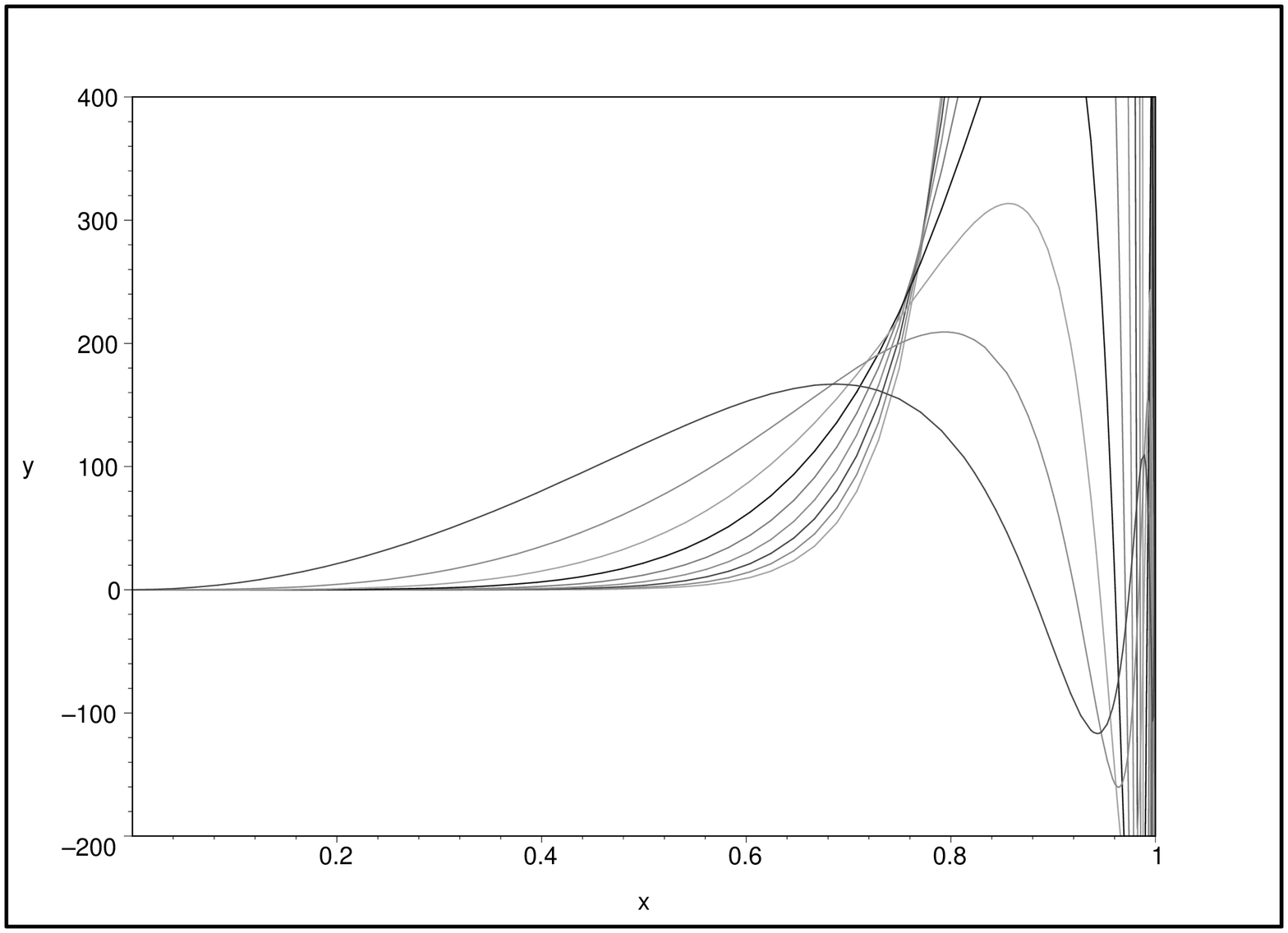}
\caption{Regular solution (\ref{(4.4)}) for $f_{0}$ with $C_{3}=0,C_{4}=1,
l=2,\ldots, 10$ with  $\Omega=2$ (left figure) and $\Omega=4$ (right figure).
Increasing values of $l$ correspond to more peaked curves on the right part of the plots.}
\end{figure}

\begin{figure}
\includegraphics[scale=0.25,angle=-90]{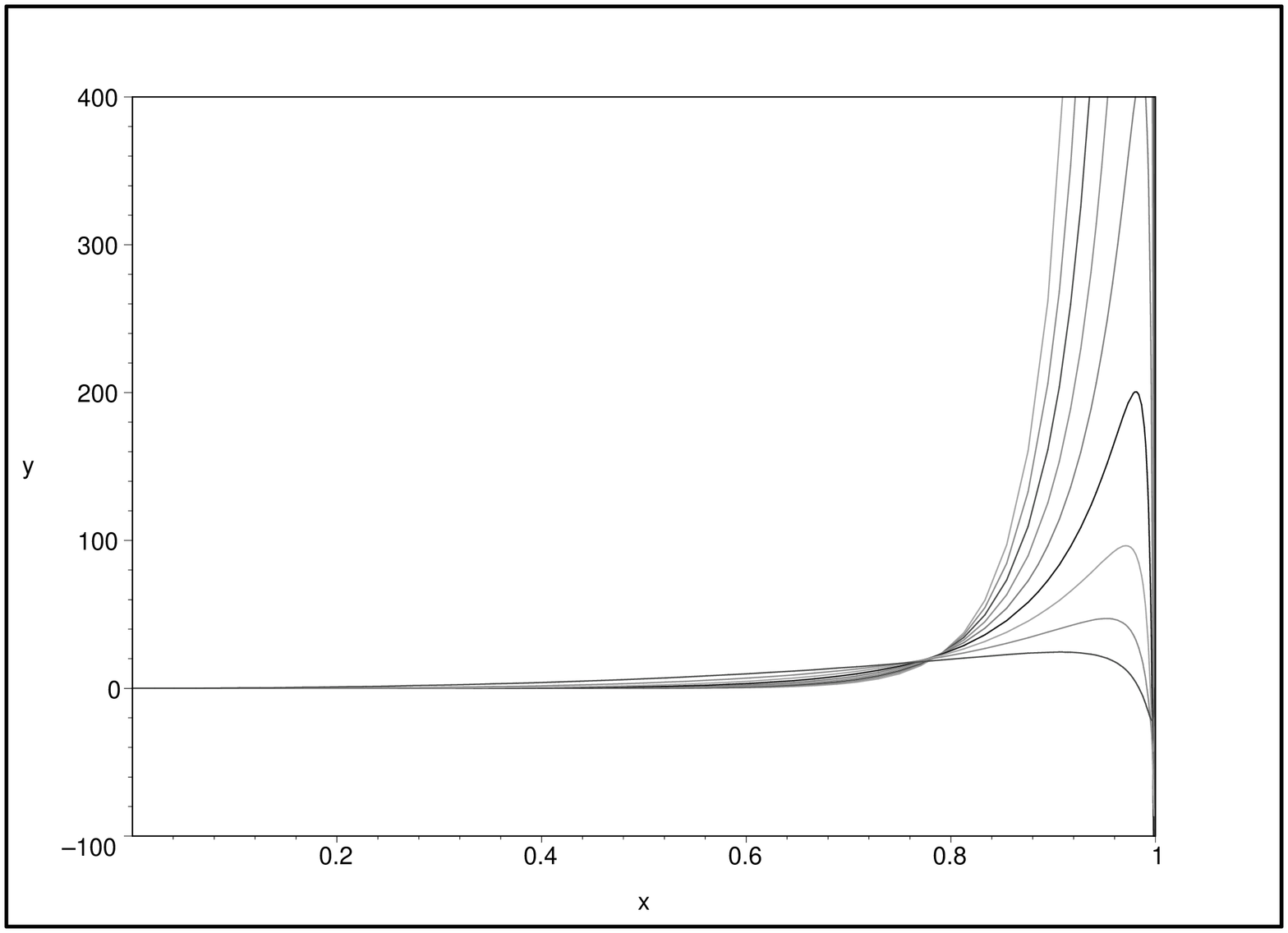}
\includegraphics[scale=0.25,angle=-90]{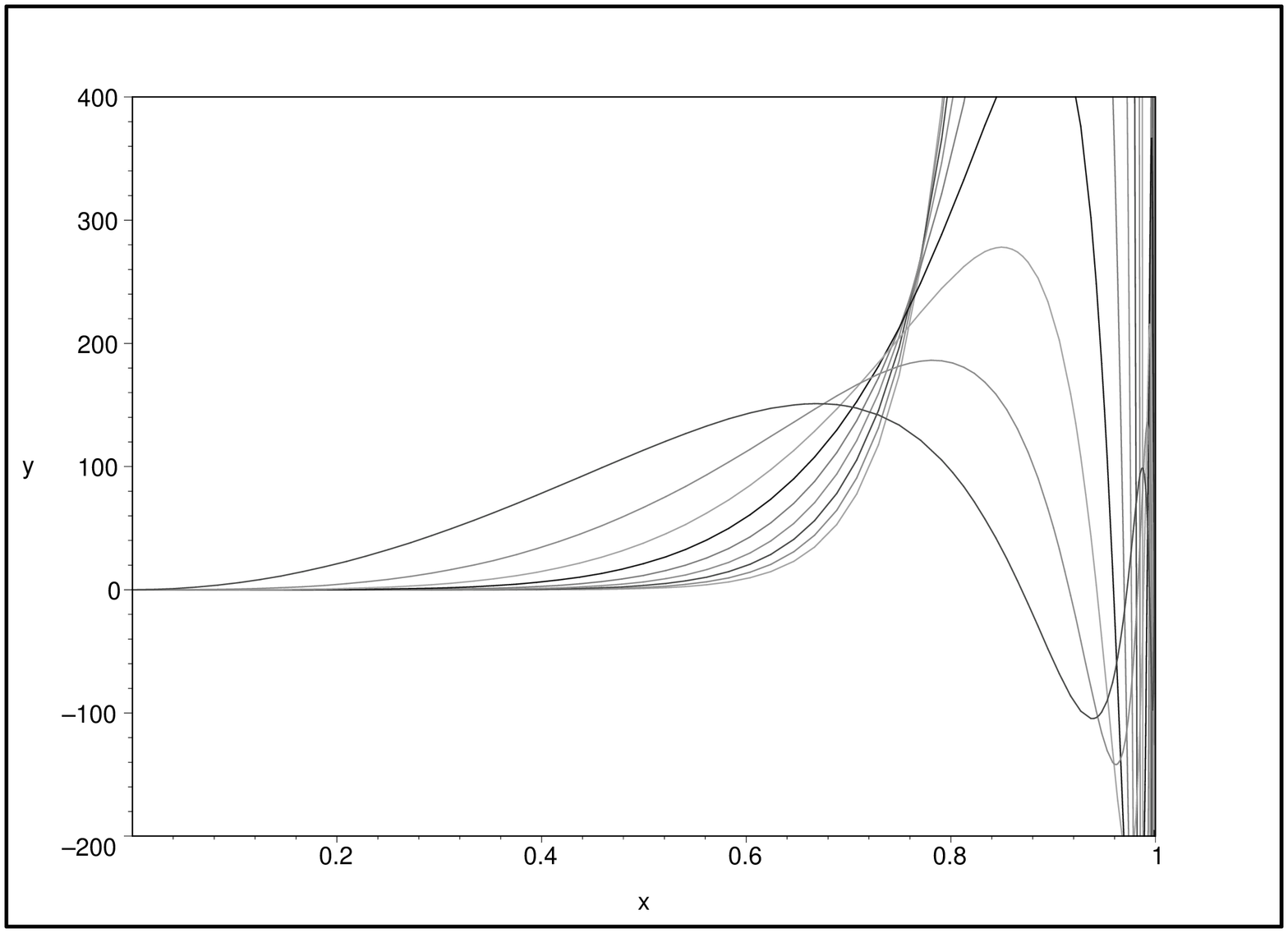}
\caption{Regular solution (\ref{(4.4)}) for $f_{0}$ with $C_{3}=1,C_{4}=0,
l=2,\ldots, 10$ with  $\Omega=2$ (left figure) and $\Omega=4$ (right figure).
Increasing values of $l$ correspond to more peaked curves on the right part of the plots.}
\end{figure}

\begin{figure}
\includegraphics[scale=0.30,angle=-90]{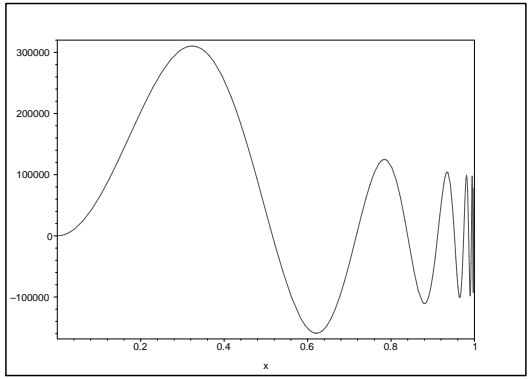}
\caption{Regular solution (\ref{(4.4)}) for $f_{0}$ with $C_{3}=0,C_{4}=1,
l=2$ and $\Omega=10$.}
\end{figure}

We next plot $f_{1}/i$ and $F_{3}/i \equiv H f_{3}/i$ by relying upon (\ref{(3.9)}) and (\ref{(3.10)}). As far as we can see, all solutions blow up at the event horizon, corresponding to $x=1$, since there are no static solutions
of the wave equation which are regular inside and on the event horizon other than the constant one \cite{BOUCHER}.

\begin{figure}
\includegraphics[scale=0.25,angle=-90]{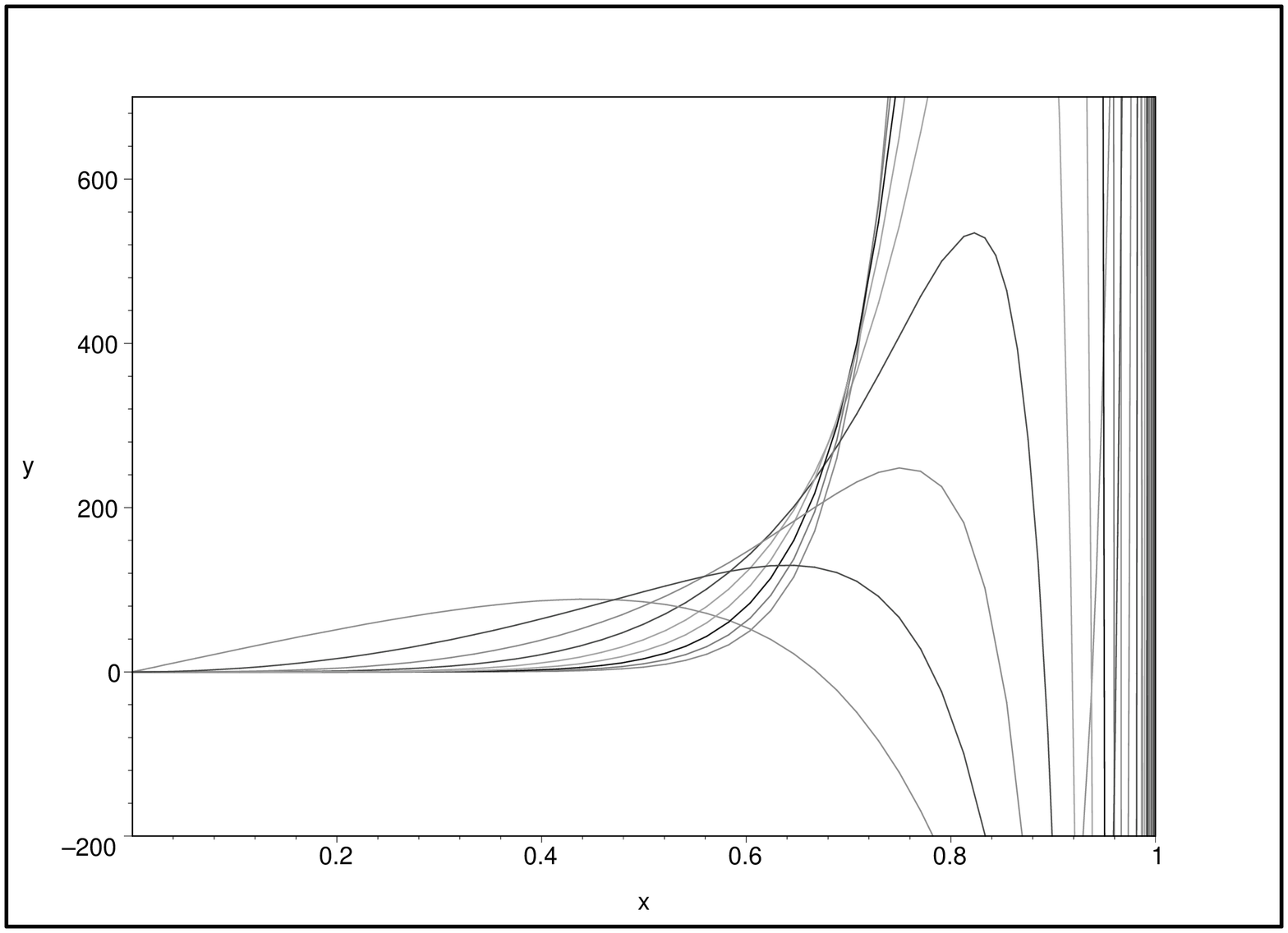}
\includegraphics[scale=0.25,angle=-90]{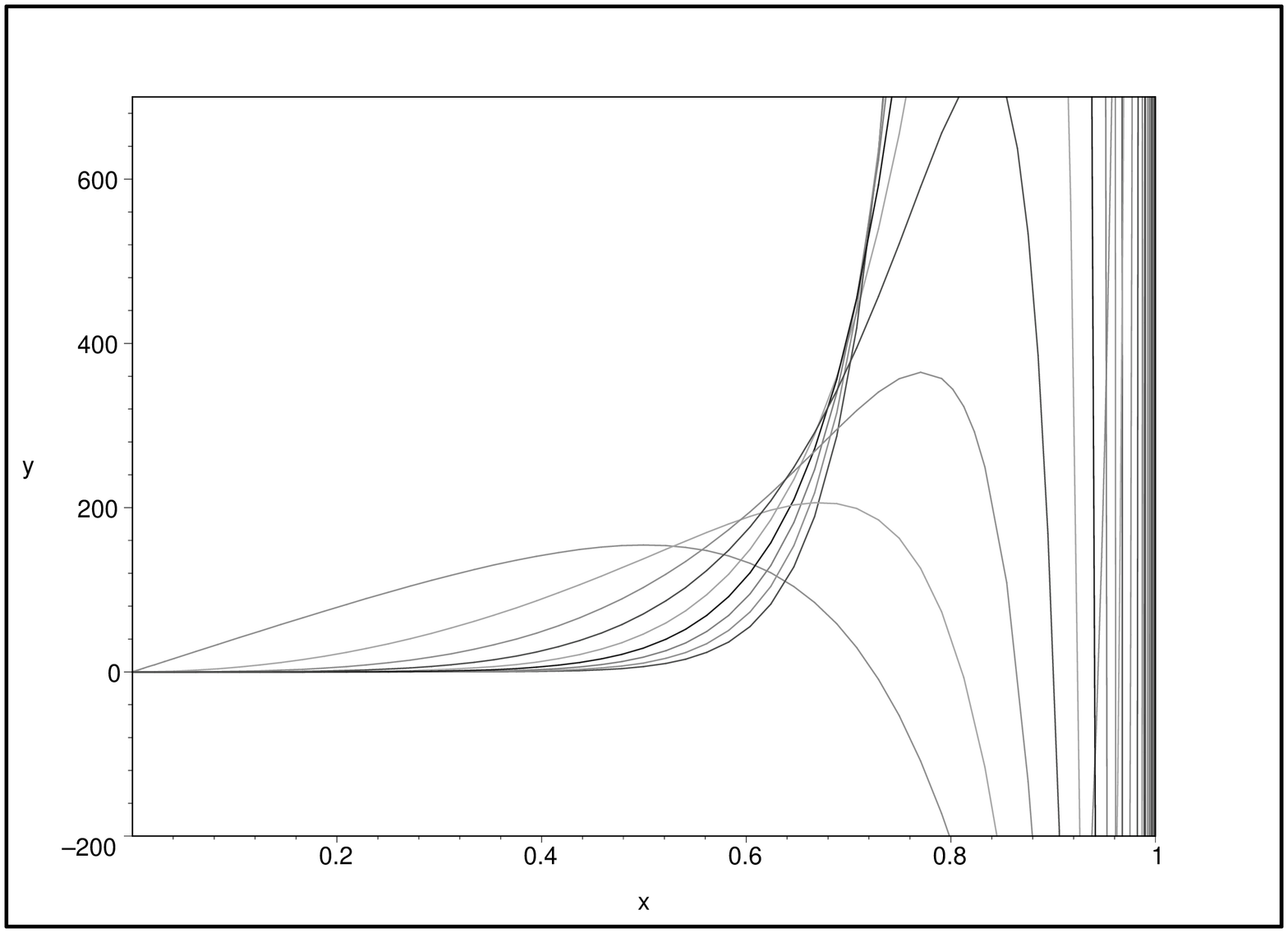}
\caption{Regular solution (\ref{(3.9)}) for $f_{1}/i$ with
$C_{3}=1,C_{4}=0$ (left figure) and $C_{3}=0,C_{4}=1$
(right figure) for $l=2\ldots, 10$ and $\Omega=4$.}
\end{figure}

\begin{figure}
\includegraphics[scale=0.25,angle=-90]{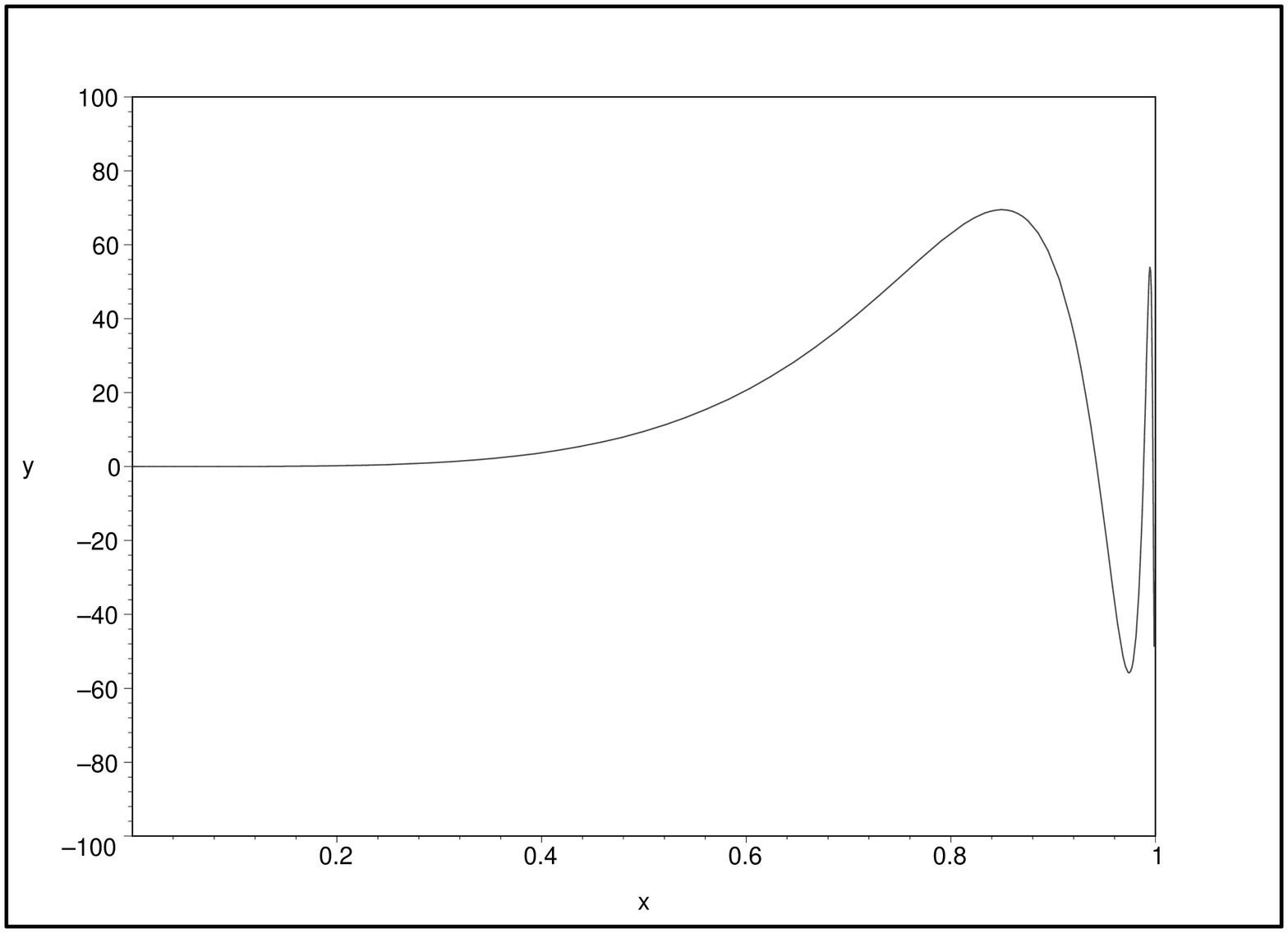}
\includegraphics[scale=0.25,angle=-90]{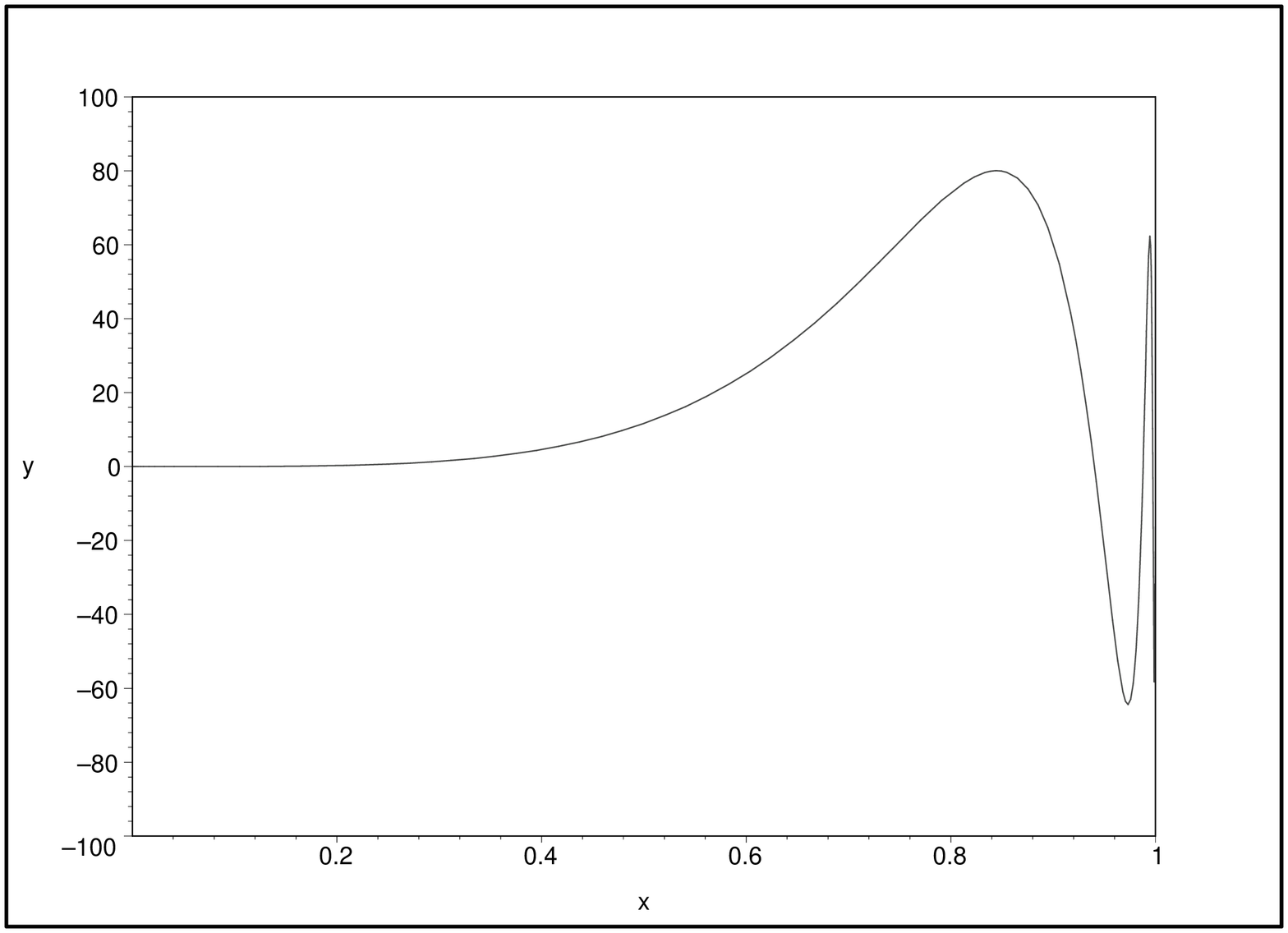}
\caption{Regular solution (\ref{(3.10)}) for $F_{3}/i$ with
$C_{3}=1,C_{4}=0$ (left figure) and $C_{3}=0,C_{4}=1$
(right figure) for $l=4$ and $\Omega=4$.}
\end{figure}

\section{Special cases: $l=0$ and $l=1$}
We list here the main equations for  $l=0,1$ for completeness.
\subsection{The case $l=0, m=0$}
In this case we have $Y(\theta)=(4\pi)^{-1/2}=$ constant and the only surviving functions are $f_0$ and $f_1$.
The main equations reduce then to
\beq
\frac{d^2f_0}{dx^2}=-\frac{2}{x}\frac{df_0}{dx}-\frac{\Omega^2-3\epsilon+3\epsilon x^2+3-3x^2}{(x^2-1)^2}f_0+\frac{2i\Omega x}{x^2-1}f_1
\eeq
and the Lorenz gauge condition which now becomes
\beq
\Omega f_0=i(x^2-1)^2\frac{df_1}{dx}+\frac{2i(x^2-1)(2x^2-1)}{x}f_1. \,
\eeq

These equations can be easily separated and explicitly solved in terms of hypergeometric functions.
\subsection{The case $l=1, m=0,1$}
In this case we have
\begin{eqnarray}
Y&=& \sqrt{\frac{3}{4\pi}}\cos \theta \,\quad l=1, m=0, \nonumber \\
Y&=& -\sqrt{\frac{3}{8\pi}}\sin \theta \,\quad l=1, m=1.
\end{eqnarray}

However, due to the spherical symmetry of the background the equations for  both cases $l=1, m=0$ and $l=1, m=1$  do coincide. We have
\begin{eqnarray}
\frac{d^2f_0}{dx^2}&=&-\frac{2}{x}\frac{df_0}{dx}-\frac{\Omega^2x^2+3\epsilon x^4+5x^2-3x^4-3\epsilon x^2-2}{x^2(x^2-1)^2}f_0+\frac{2i\Omega x}{x^2-1}f_1, \nonumber \\
\frac{d^2f_1}{dx^2}&=&-\frac{2(3x^2-1)}{x(x^2-1)}\frac{df_1}{dx}-\frac{3\epsilon x^4 +x^4 -3\epsilon x^2 +3x^2+\Omega^2x^2-4}{x^2(x^2-1)^2}f_1+\frac{2i\Omega x}{(x^2-1)^3}f_0\nonumber \\
&& +\frac{4F_3}{x^3(x^2-1)}, \nonumber \\
\frac{d^2f_2}{dx^2}&=& -\frac{2(2x^2-1)}{x(x^2-1)}\frac{df_2}{dx}-\frac{\Omega^2x^2 -x^4-3\epsilon x^2 -2+3\epsilon x^4+3x^2}{x^2(x^2-1)^2}f_2, \nonumber \\
\frac{d^2F_3}{dx^2}&=& -\frac{2x}{x^2-1}\frac{dF_3}{dx}-\frac{\Omega^2x^2+3\epsilon x^4+5x^2-3x^4-3\epsilon x^2-2}{x^2(x^2-1)^2}F_3-\frac{2}{x}f_1.\,
\end{eqnarray}

To this set one must add the Lorenz gauge condition, which now reads
\beq
\Omega f_0=i(x^2-1)^2\frac{df_1}{dx}+\frac{2i(x^2-1)}{x^2}[F_3+f_1x(2x^2-1)].\,
\eeq
Once more, the detailed discussion of this case can be performed repeating exactly the same steps done in the general case.
\chapter{The tensor wave equation}
\def\beq{\begin{equation}}
\def\eeq{\end{equation}}
\def\cstok#1{\leavevmode\thinspace\hbox{\vrule\vtop{\vbox{\hrule\kern1pt\hbox{\vphantom{\tt/}\thinspace{\tt#1}\thinspace}}
\kern1pt\hrule}\vrule}\thinspace}
\section{Metric perturbations in de Sitter space-time}
One of the longstanding problems of modern gravitational physics is the detection of gravitational waves, for which the standard theoretical analysis relies upon the split of the space-time metric $g_{ab}$ into ``background plus perturbations'', i.e.
\begin{equation}
g_{ab}=\gamma_{ab}+h_{ab}
\label{bpp}
\end{equation}
where $\gamma_{ab}$ is the background Lorentzian metric, often taken to be of the Minkowski form $\eta_{ab}$, while the symmetric tensor field $h_{ab}$ describes perturbations about $\gamma_{ab}$. However, the background $\gamma_{ab}$ need not be Minkowskian in several cases of physical interest. As a consequence, we are therefore aiming to investigate in more detail what happens if the background space-time $(M, \gamma_{ab})$ has a non-vanishing Riemann curvature. In particular, we perform an analysis of gravitational waves on a de Sitter space-time.

It is straightforward to show that, in a covariant formulation, the supplementary condition for gravitational waves can be described by a functional $\Phi_a$ acting on the space of symmetric rank-two tensors $h_{ab}$ occurring in Eq. (\ref{bpp}). For any choice of $\Phi_a$, one gets a different realization of the invertible operator $P_{ab}^{\phantom{ab}cd}$ (Lichnerowicz operator) on metric perturbations. The basic equations of the theory read therefore as (see also Chapter \textbf{1})
\begin{eqnarray}
P_{ab}^{\phantom{ab}cd}h_{cd}&=&0,\nonumber\\
\Phi_a(h)&=&0,
\end{eqnarray}
where the Lichnerowicz operator $P_{ab}^{\phantom{ab}cd}$ results from the expansion of the Einstein-Hilbert action to quadratic order in the metric perturbations, subject to $\Phi_a(h)=0$.

Consider de Sitter metric in standard spherical coordinates (\ref{metrica}). It satisfies the vacuum Einstein equations with non-vanishing cosmological constant $\Lambda$ such that $H^{2}={\Lambda \over 3}$. Moreover, the time-like unit normal vector fields $n$ to the $t={\rm constant}$ hypersurfaces,
\begin{equation}
n=\partial_t,
\label{(49)}
\end{equation}
form a geodesic and irrotational congruence. The 3-metric induced on the $t={\rm constant}$ hypersurfaces turns out to be conformally flat and the mixed form of the space-time Ricci tensor is simply given by
\begin{equation}
R_{a}^{\; b}=3H^{2}\delta_{a}{}^{b}.
\label{(50)}
\end{equation}

The aim of the present chapter consists in studying the invertible wave operator $P_{ab}^{\; \; \; cd}$ on metric perturbations. On considering the DeWitt supermetric
\begin{equation}
E^{abcd} \equiv \gamma^{a(c} \; \gamma^{d)b}-{1\over 2}\gamma^{ab}\gamma^{cd},
\label{(65)}
\end{equation}
the de Donder gauge in Eq. (\ref{(65)}) can be re-expressed in the form
\begin{equation}
\Phi_{a}(h)=E_{a}{}^{bcd}\nabla_{b}h_{cd},
\label{(66)}
\end{equation}
and the resulting Lichnerowicz operator on metric perturbations, obtained by expansion of the Einstein--Hilbert action to quadratic order in $h_{ab}$, subject to $\Phi_{a}(h)=0$, reads as (see \cite{BSMFA-92-11}, \cite{NUPHA-B146-90}, \cite{Moss1996} and Appendix \textbf{C})
\begin{equation}
P_{ab}^{\; \; \; cd} \equiv E_{ab}^{\; \; \; cd}(-\cstok{\ }+R)-2E_{ab}^{\; \; \; lf}R_{\; lhf}^{c}\gamma^{dh}
-E_{ab}^{\; \; \; ld}R_{l}^{\; c}-E_{ab}^{\; \; \; cl}R_{l}^{\; d}.
\label{(67)}
\end{equation}
A wave equation for metric perturbations is therefore given by
\begin{equation}
P_{ab}{}^{cd}h_{cd}=0.
\label{(68)}
\end{equation}

In de Sitter space-time the Lichnerowicz operator becomes \cite{Moss1996}
\begin{eqnarray}
P_{ab}{}^{cd}=E_{ab}{}^{cd}\left(-\cstok{\ }+\frac23 R\right)+\frac{R}{6}\gamma_{ab}\gamma^{cd},
\label{(69)}
\end{eqnarray}
so that the wave equation then reads as
\begin{equation}
0=P_{ab}{}^{cd}h_{cd}=\left(-\cstok{\ }+\frac23 R\right)\bar h_{ab}+\frac{R}{6}\gamma_{ab}h,
\label{(70)}
\end{equation}
or
\begin{equation}
\left(-\cstok{\ }+\frac23 R\right)\bar h_{ab}-\frac{R}{6}\gamma_{ab}{\bar h}=0 ,
\label{(71)}
\end{equation}
implying also
\begin{equation}
\left(-\cstok{\ }+\frac23 R\right)\bar h-\frac{2}{3}R{\bar h}=-\cstok{\ }{\bar h}=0,
\label{(72)}
\end{equation}
after contraction with $\gamma^{ab}$.
\subsection{Even metric perturbations}
Let us look for solutions of the tensorial wave equation (\ref{(68)}), metric perturbations expanded in tensor harmonics according to \cite{Zerilli}. Those of even parity can be written in the form
\\
\\
\begin{eqnarray}
h_{00}&=& f e^{-i(\omega t -m\phi)}H_0(r)Y(\theta),\nonumber \\
h_{01}&=& e^{-i(\omega t -m\phi)}H_1(r)Y(\theta),\nonumber \\
h_{02}&=& e^{-i(\omega t -m\phi)}h_0(r)\frac{dY}{d\theta},\nonumber \\
h_{03}&=& im e^{-i(\omega t -m\phi)}h_0(r)Y(\theta),\nonumber \\
h_{11}&=& \frac{1}{f} e^{-i(\omega t -m\phi)}H_2(r)Y(\theta),\nonumber \\
h_{12}&=& e^{-i(\omega t -m\phi)}h_1(r)\frac{dY}{d\theta},\nonumber \\
h_{13}&=& im e^{-i(\omega t -m\phi)}h_1(r)Y(\theta),\nonumber \\
h_{22}&=& r^2 e^{-i(\omega t -m\phi)}\left[K(r)Y(\theta)+G(r)\frac{d^2Y}{d\theta^2}\right],\nonumber \\
h_{23}&=& im r^2 G(r) e^{-i(\omega t -m\phi)}\left[\frac{dY}{d\theta}-\cot \theta Y(\theta)\right],\nonumber \\
h_{33}&=& r^2 e^{-i(\omega t -m\phi)}\left\{K(r)\sin^2\theta Y(\theta)\right. \nonumber \\
&& \left. +G(r)\left[-m^2 Y(\theta)+\sin\theta\cos\theta\frac{dY}{d\theta}\right]\right\}
\label{(74)}
\end{eqnarray}
and metric perturbations of odd parity are instead found to vanish identically. Here, $Y(\theta)$ is the $\theta$-dependent part of the spherical harmonics $Y_{lm}(\theta,\phi)$, solution of the equation (\ref{(210a)}). It is convenient to introduce the notation
\begin{eqnarray}
K(r)&=&W_1(r),\quad H_0(r)=W_2(r),\quad H_1(r)=W_3(r),\quad H_2(r)=W_4(r),\nonumber \\
G(r)&=&W_5(r),\quad h_0(r)=H^{-1}W_6(r),\quad h_1(r)=H^{-1}W_7(r),
\label{(74a)}
\end{eqnarray}
in place of standard Regge-Wheeler-Zerilli notation for metric perturbation quantities. The wave equation (\ref{(68)}), using the relations (\ref{(74)}), (\ref{(74a)}) and (\ref{(210a)}), leads to the following system of coupled differential equations:
\begin{eqnarray}
\frac{d^2W_1}{dr^2}&=&\frac{2(1-2f)}{rf}\frac{dW_1}{dr} -\frac{1}{r^2f^2}[\omega^2r^2-Lf-2f(2-f)]W_1-\frac{2L}{r^2}W_5\nonumber \\
&& -\frac{2}{r^2f}W_4+\frac{2H^2}{f}W_2,\nonumber \\
\frac{d^2W_2}{dr^2}&=&\frac{2}{rf}(2-3f)\frac{dW_2}{dr}-\frac{1}{r^2f^2}[\omega^2r^2 +2(1-f)(1-4f)-fL]W_2\nonumber \\
&& +\frac{2H^2r}{f}\frac{dW_4}{dr}-\frac{4H^2r}{f}\frac{dW_1}{dr}+\frac{2H^2rL}{f}\frac{dW_5}{dr}\nonumber \\
&& +\frac{2H^2L}{f}W_5-\frac{4HL}{rf}W_7-\frac{2H^{2}(1-6f)}{f^{2}}W_4-\frac{4H^2}{f}W_1,\nonumber \\
\frac{d^2W_3}{dr^2}&=& \frac{2}{rf}(1-2f)\frac{dW_3}{dr}-\frac{1}{r^2f^2}[\omega^2r^2-Lf-2(2-f^2)]W_3 \nonumber \\
&& -\frac{2L}{fr^3}\frac1H W_6 -\frac{2i\omega rH^2}{f^2}(W_2+W_4),\nonumber\\
\frac{d^2W_4}{dr^2}&=&\frac{2(2-3f)}{rf}\frac{dW_4}{dr}-\frac{1}{r^2f^2}[\omega^2r^2-Lf-10+6f+12(1-f)^2]W_4\nonumber \\
&& +\frac{2H^2r}{f}\frac{dW_2}{dr}-\frac{4H^2r}{f}\frac{dW_1}{dr}+\frac{2rLH^2}{f}\frac{dW_5}{dr}\nonumber \\
&& +\frac{2L(3-2f)}{r^2f}W_5-\frac{4L}{r^3f}\frac1H W_7-\frac{4(3-2f)}{r^2f}W_1-\frac{2H^2(1-2f)}{f^2}W_2,\nonumber \\
\frac{d^2W_5}{dr^2}&=& \frac{2}{rf}(1-2f)\frac{dW_5}{dr}-\frac{1}{r^2f^2}[\omega^2r^2-Lf+2f(3f-2)]W_5-\frac{4W_7}{Hr^3},\nonumber \\
\frac{d^2W_6}{dr^2}&=& -\frac{1}{r^2f^2}[\omega^2r^2-Lf-4f(1-f)]W_6-\frac{2i\omega rH^2}{f}W_7-\frac{2H}{r}W_3,\nonumber \\
\frac{d^2W_7}{dr^2}&=&\frac{6rH^2}{f}\frac{dW_7}{dr}-\frac{1}{r^2f^2}[\omega^2r^2-Lf+10(1-f)^{2}-6+2f]W_7\nonumber \\
&&+\frac{H}{rf^2}[2-(1+f)L]W_5-\frac{H(1+f)}{rf^2}W_4+\frac{2H}{rf}W_1+\frac{H^3r}{f^2}W_2.
\label{(75)}
\end{eqnarray}

To these equations one should add the de Donder gauge components
\begin{eqnarray}
\frac{dW_3}{dr}&=& \frac{i\omega L}{2f}W_5+\frac{L}{Hr^2f}W_6-\frac{i\omega}{2f}W_4-\frac{i\omega}{f}\left(W_1+{W_2\over 2}\right)
+\frac{2(1-2f)}{rf}W_3,\nonumber \\
\frac{dW_5}{dr}&=& \frac{2(1-3f)}{rf}W_4-\frac{dW_2}{dr}+2\frac{dW_1}{dr}-L\frac{dW_5}{dr}-\frac{2L}{r}W_5\nonumber \\
&& +\frac{2L}{Hr^2}W_7+\frac{4}{r}W_1+\frac{2H^2r}{f}W_2-\frac{2i\omega}{f}W_3,\nonumber \\
\frac{dW_7}{dr}&=&\frac{2(1-2f)}{rf}W_7+\frac{H(L-2)}{2f}W_5-\frac{i\omega}{f^2}W_6+\frac{H}{2f}(W_4-W_2).
\label{(76)}
\end{eqnarray}

Solving these two systems of coupled differential equations is clearly a hard task. A first progress can be done by introducing the same dimensionless radial variable seen in Chapter \textbf{2}, that is
\beq
x=Hr,
\label{x}
\eeq
as well as a dimensionless frequency parameter
\beq
\Omega=H^{-1}\omega.
\label{Omega}
\eeq

A second step consists in reducing the set of second order differential equations to the normal form, i.e. considering the following rescalings:
\begin{eqnarray}
\label{rescal}
\left(W_1,W_2,W_3,W_4,W_5\right)&=&\frac{1}{x\sqrt{1-x^2}}\left(f_1,f_2,if_3,f_4,f_5\right),\nonumber \\
\left(W_6,W_7\right)&=&  \left(if_6,\frac{f_7}{1-x^2} \right).
\end{eqnarray}

It turns out that the system (\ref{(75)}) can be cast in the form
\beq
\frac{d^2 f_i}{dx^2}=A_i{}^jf_j,
\label{second_x}
\eeq
where first order derivatives have been eliminated and with the only non-vanishing coefficients listed below
\begin{eqnarray}
&& A_1{}^1=-\frac{x^2\Omega^2 +3x^2+Lx^2-2-L}{x^2(1-x^2)^2},\,\quad A_1{}^2=\frac{2}{1-x^2},\,\quad A_1{}^4=-\frac{2}{x^2(1-x^2)},\nonumber \\
&& A_1{}^5=-\frac{2L}{x^2}.\,\quad \nonumber \\
&& A_2{}^1=\frac{4}{1-x^2},\,\quad A_2{}^2=-\frac{2x^4+x^2L-L+x^2\Omega^2-3x^2}{x^2(1-x^2)^2},\,\quad A_2{}^3=\frac{4\Omega x}{(1-x^2)^2},\nonumber \\
&& A_2{}^4=\frac{2}{(1-x^2)^2},\,\quad A_2{}^5=-\frac{2L}{1-x^2}.\,\quad \nonumber \\
&& A_3{}^2=-\frac{2\Omega x}{(1-x^2)^2},\,\quad A_3{}^3=-\frac{-x^2+\Omega^2x^2+x^2L-L-2}{x^2(1-x^2)^2},\,\quad A_3{}^4=-\frac{2\Omega x}{(1-x^2)^2},\nonumber \\
&& A_3{}^6=-\frac{2L}{x^2(1-x^2)^{1/2}}.\,\quad \nonumber \\
&& A_4{}^1=-\frac{4}{x^2(1-x^2)},\,\quad A_4{}^2=\frac{2}{(1-x^2)^2},\,\quad A_4{}^3=\frac{4\Omega x}{(1-x^2)^2},\nonumber \\
&& A_4{}^4=-\frac{-2x^4+x^2L+x^2\Omega^2+5x^2-4-L}{x^2(1-x^2)^2},\,\quad A_4{}^5=\frac{2L}{x^2(1-x^2)},\nonumber \\
&& A_4{}^7=-\frac{4L}{x^2(1-x^2)^{1/2}}.\,\quad \nonumber \\
&& A_5{}^5=-\frac{4x^4+x^2L-5x^2+x^2\Omega^2+2-L}{x^2(1-x^2)^2},\,\quad A_5{}^7=-\frac{4}{x^2(1-x^2)^{1/2}}.\nonumber\\
&& A_6{}^3=-\frac{2}{x^2(1-x^2)^{1/2}},\,\quad A_6{}^6=
-\frac{4x^4-L+x^2\Omega^2+x^2L-4x^2}{x^2(1-x^2)^2},\nonumber \\
&& A_6{}^7=-\frac{2x\Omega}{(1-x^2)^2}. \nonumber \\
&& A_7{}^1=\frac{2}{x^2(1-x^2)^{1/2}},\,\quad A_7{}^4=-\frac{2}{x^2(1-x^2)^{1/2}},\,\quad A_7{}^5=-\frac{2(L-1)}{x^2(1-x^2)^{1/2}},\nonumber \\
&& A_7{}^6=\frac{2x\Omega}{(1-x^2)^2},\,\quad  A_7{}^7=-\frac{-L+x^2L+4x^2+x^2\Omega^2-4}{x^2(1-x^2)^2}.\,
\end{eqnarray}

The rescaling (\ref{rescal}) also implies for the gauge equation the following form:
\begin{eqnarray}
\label{gauge_bis}
\frac{df_3}{dx}&=&B_3{}^jf_j, \nonumber \\
\frac{df_5}{dx}&=&B_5{}^jf_j+\frac{2}{L}\frac{df_1}{dx}
-\frac{1}{L}\frac{df_2}{dx}-\frac{1}{L}\frac{df_4}{dx},\nonumber \\
\frac{df_7}{dx}&=&B_7{}^jf_j,\,
\end{eqnarray}
with the only non-vanishing coefficients listed below
\begin{eqnarray}
&& B_3{}^1=-\frac{\Omega}{(1-x^2)},\,\quad B_3{}^2=-\frac{\Omega}{2(1-x^2)},\,\quad B_3{}^3=\frac{2x^2-1}{x(1-x^2)},\nonumber \\
&& B_3{}^4=-\frac{\Omega}{2(1-x^2)},\,\quad B_3{}^5=\frac{\Omega L}{2(1-x^2)}, \,\quad B_3{}^6= \frac{L}{x(1-x^2)^{1/2}}.\nonumber \\
&& B_5{}^1=\frac{2}{Lx(1-x^2)},\,\quad B_5{}^2=
\frac{1}{L x(1-x^2)},\,\quad B_5{}^3=\frac{2\Omega}{L (1-x^2)},\nonumber \\
&& B_5{}^4=\frac{4x^2-3}{xL(1-x^2)},\,\quad B_5{}^5=
-\frac{1}{x(1-x^2)},\,\quad B_5{}^7=\frac{2}{x(1-x^2)^{1/2}}. \nonumber \\
&& B_7{}^2=-\frac{1}{2x(1-x^2)^{1/2}},\,\quad B_7{}^4=\frac{1}{2x(1-x^2)^{1/2}},\,\quad B_7{}^5=\frac{L-2}{2x(1-x^2)^{1/2}},\nonumber \\
&& B_7{}^6=\frac{\Omega}{(1-x^2)}, \,\quad B_7{}^7=-\frac{2}{x}.\,
\end{eqnarray}

By differentiating Eqs. (\ref{gauge_bis}) and using into the new system Eqs. (\ref{second_x}) as well as Eqs. (\ref{gauge_bis}), gives a new system of first order equations
\begin{eqnarray}
\label{assoc_1}
\frac{df_2}{dx}&=&B_2{}^jf_j+C_2\frac{df_1}{dx}+D_2 f_1, \nonumber \\
\frac{df_4}{dx}&=&B_4{}^jf_j+C_4\frac{df_1}{dx}+D_4 f_1, \nonumber \\
\frac{df_6}{dx}&=&B_6{}^jf_j+C_6\frac{df_1}{dx}+D_6 f_1\,.
\end{eqnarray}

We list here the non-vanishing coefficients of this  system. It is convenient to introduce the quantity
\beq
A=2x^2L -2x^2+2x^2\Omega^2+2-L\,.
\eeq

We have then
\begin{eqnarray}
B_2{}^2&=&-\frac{1}{x(1-x^2)A}(4x^4L-2+x^4\Omega^2+3x^2-2x^2\Omega^2-x^4-\Omega^2x^2L+L-5x^2L), \nonumber \\
B_2{}^3&=& \frac{2\Omega}{(1-x^2)A}(2+2x^2\Omega^2+2x^2L-L+6x^4-5x^2), \nonumber \\
B_2{}^4&=& \frac{1}{x(1-x^2)A}(-4x^2L+5x^4\Omega^2+4x^4L-2-9x^4+11x^2+L), \nonumber \\
B_2{}^5&=& -\frac{xL}{(1-x^2)^2A}(-2x^2\Omega^2+\Omega^4x^2 -3-3x^4+6x^2+3x^4\Omega^2), \nonumber \\
B_2{}^6&=& -\frac{2\Omega L x^2 (1+\Omega^2)}{(1-x^2)^{3/2}A}, \nonumber \\
B_2{}^7&=& \frac{6Lx(1-x^2)(1-2x^2)}{(1-x^2)^{3/2}A}.\,
\end{eqnarray}
\begin{eqnarray}
B_4{}^2&=&\frac{1}{x(1-x^2)A}(2x^2L +x^2+L^2 -2L-x^2L^2-x^4-\Omega^2 x^2 L+x^4L +x^4 \Omega^2), \nonumber \\
B_4{}^3&=&-\frac{6x^2\Omega}{(1-x^2)A},\nonumber \\
B_4{}^4&=&-\frac{1}{x(1-x^2)A}(-3x^2-2L-x^4+x^4\Omega^2+x^4L+x^2L+2x^2\Omega^2+4), \nonumber \\
B_4{}^5&=&\frac{L}{x(1-x^2)^2A}(-4+9x^2-2x^2\Omega^2-6x^4-5x^2L+2L+4x^4\Omega^2+4x^4L \nonumber \\
&& +\,x^4\Omega^2L+\Omega^4x^4-x^6\Omega^2-x^6L-\Omega^2x^2L+x^6), \nonumber \\
B_4{}^6&=&\frac{2\Omega L}{(1-x^2)^{3/2}A}(-x^2+x^2L+x^2\Omega^2+2-L), \nonumber \\
B_4{}^7&=&\frac{6xL}{(1-x^2)^{1/2}A}.\,
\end{eqnarray}
\begin{eqnarray}
B_6{}^2&=& \frac{\Omega}{(1-x^2)^{1/2}A}(3x^2-2+L), \nonumber \\
B_6{}^3&=& \frac{1}{Lx(1-x^2)^{1/2}A}(-L^2+2L+2x^2L^2-8x^2L+2\Omega^2x^2L+12x^4L-12x^4+12x^2), \nonumber \\
B_6{}^4&=&\frac{\Omega}{L(1-x^2)^{1/2}A}(2L+x^2L-4+4x^2), \nonumber \\
B_6{}^5&=&-\frac{\Omega}{(1-x^2)^{3/2}A}(-x^2L-2x^2+3x^4L-L+2+\Omega^2x^2L), \nonumber \\
B_6{}^6&=&-\frac{2x}{(1-x^2)A}(-L+2+2x^2L+2x^2\Omega^2-2x^2-2\Omega^2+\Omega^2L), \nonumber \\
B_6{}^7&=&-\frac{\Omega}{(1-x^2)A}(-L+12x^4+2-14x^2+2x^2\Omega^2+2x^2L). \,
\end{eqnarray}
\begin{eqnarray}
&& C_2=\frac{(2x^2-1)(L-2)}{A}, \,\quad C_4=-\frac{(L-2)}{A}, \nonumber \\
&& C_6=-\frac{2(L-2)\Omega x (1-x^2)^{1/2}}{LA}.
\end{eqnarray}
\begin{eqnarray}
D_2&=&\frac{x}{(1-x^2)^2A}(2x^4-2x^2+2\Omega^2 -4x^4L+7x^2L-6x^2\Omega^2 +\Omega^2x^2 L \nonumber \\
&&-\Omega^2 L+6x^4 \Omega^2 +2\Omega^4 x^2-3L), \,\nonumber \\
D_4&=&-\frac{1}{x(1-x^2)^2A}(-4+2x^6 +8x^2 -6x^4-x^2L -2x^2 \Omega^2 +3x^4 L -2x^6 \Omega^2 \nonumber \\
&& +\,6 x^4 \Omega^2+2\Omega^4 x^4 -3\Omega^2x^2L -2x^2L^2+L^2 +3x^4\Omega^2 L+x^4L^2 -2x^6 L), \nonumber \\
D_6&=&\frac{\Omega}{L(1-x^2)^{3/2} A} (2\Omega^2x^2L +10x^4L +2L -10x^2L +x^2L^2 \nonumber \\
&&-L^2-8x^4+8x^2).\,
\end{eqnarray}

The final set of first order equations is then the union of Eqs. (\ref{gauge_bis}) and (\ref{assoc_1}), implying for Eqs. (\ref{gauge_bis}) the final form
\begin{eqnarray}
\label{gauge_ter}
\frac{df_3}{dx}&=&B_3{}^jf_j, \nonumber \\
\frac{df_5}{dx}&=&\bar B_5{}^jf_j+C_5\frac{df_1}{dx}+D_5f_1, \nonumber \\
\frac{df_7}{dx}&=&B_7{}^jf_j,
\end{eqnarray}
where
\begin{eqnarray}
\bar B_5{}^2&=& -\frac{3x^2+L-2}{xA}, \nonumber \\
\bar B_5{}^3&=& \frac{12\Omega x^2}{LA}, \nonumber \\
\bar B_5{}^4&=& -\frac{x(5L+4\Omega^2)}{LA}, \nonumber \\
\bar B_5{}^5&=& -\frac{1}{x(1-x^2)A}(-x^2L+4\Omega^2x^4-2+x^4L+6x^2-\Omega^2x^2L-4x^4+L), \nonumber \\
\bar B_5{}^6&=& -\frac{1}{Lx(1-x^2)^{3/2}A}(-4\Omega x^3L+4L\Omega x-2L^2\Omega x+2L^2\Omega x^3), \nonumber \\
\bar B_5{}^7&=& \frac{2}{x(1-x^2)^{1/2}A}(6x^4+2x^2\Omega^2+2x^2L-8x^2-L+2), \,
\end{eqnarray}
\beq
C_5=\frac{2x^2(2\Omega^2+L)}{LA}
\eeq
and
\beq
D_5=-\frac{1}{xL(1-x^2)A}(2\,\Omega^2x^2L +x^2L^2-8x^4\Omega^2+2L-6x^2L+2x^4L-L^2).\,
\eeq

Such a first-order set of equations, (\ref{assoc_1}) and (\ref{gauge_ter}), with a second-order equation for $f_1$, that is
\begin{eqnarray}
\frac{d^2 f_1}{dx^2}&=&\frac{1}{x^2(x^2-1)^2}\left\{\left[L+2-(\Omega^2+L+3)x^2\right]f_1-2x^2(x^2-1)f_2\right.\nonumber\\
&+&\left.2(x^2-1)f_4-2L(x^2-1)^2f_5\right\},
\label{34}
\end{eqnarray}
once used to replace derivatives into the system (\ref{second_x}), implies that the latter system is identically satisfied.

At this stage, a numerical analysis of solutions can be performed easily, using the set of equations (\ref{assoc_1}),(\ref{gauge_ter}) and (\ref{34}).

\section{Plot of the solutions}
We plot the solutions $f_1(x)-f_7(x)$, beginning with $f_i(0.1)=1$, where $i=1,...,7$, and $f'_1(0.1)=10$. Figures from (3.1) to (3.7) describe the solutions for various values of $l$ and $\Omega$. Figures from (3.8) to (3.14) describe the solutions for various values of $l$ and $\Omega$, when $f_i(0.1)=1$ with $i=1,...,7$, and $f'_1(0.1)=100$.

As far as we can see, all solutions blow up (as the electromagnetic waves in Chapter \textbf{2}) at the event horizon, corresponding to $x=1$, since there are no static solutions of the wave equation which are regular inside and on the event horizon other than the constant one \cite{BOUCHER}.

\begin{figure}
\includegraphics[scale=0.25,angle=-90]{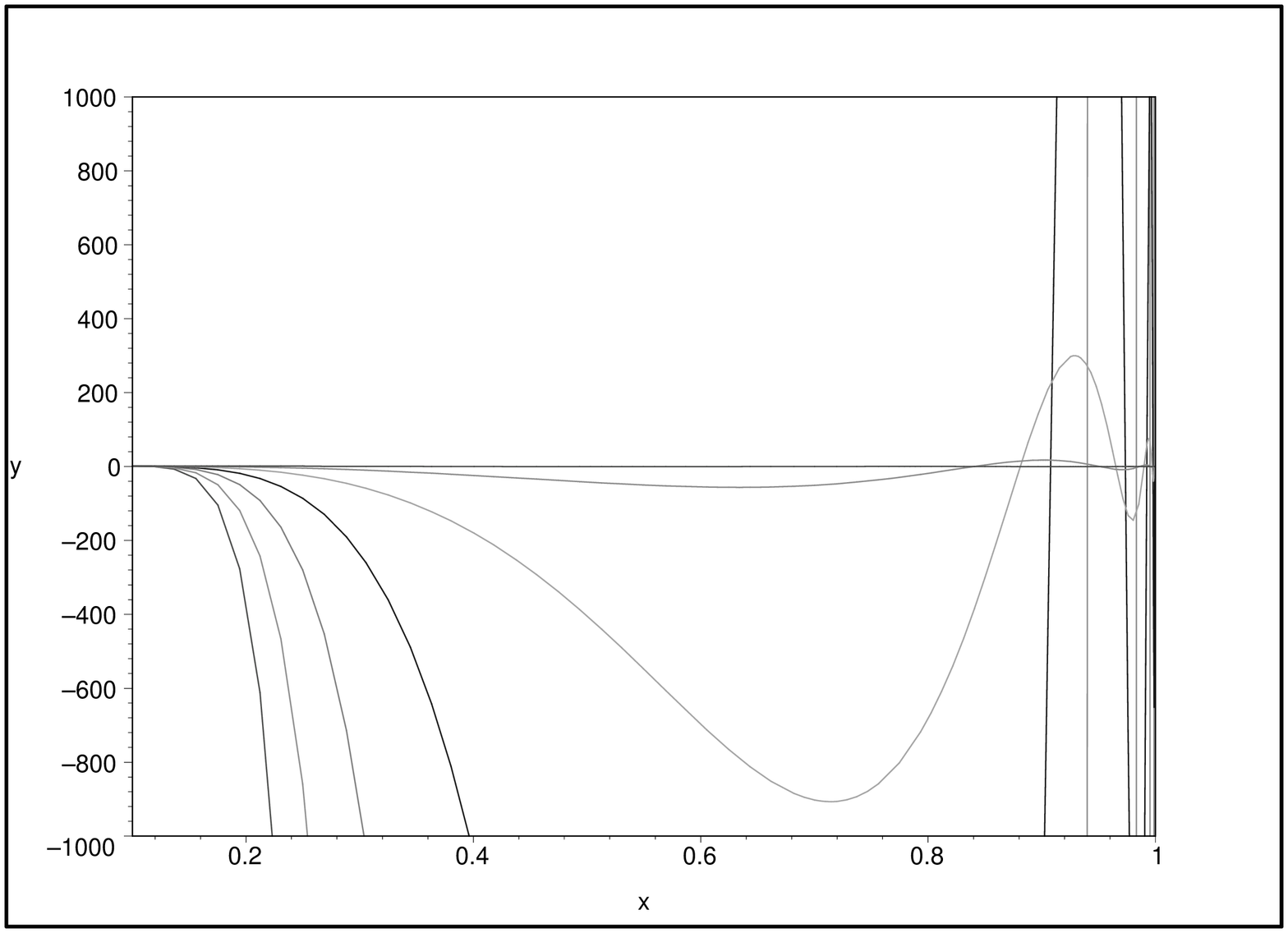}
\includegraphics[scale=0.25,angle=-90]{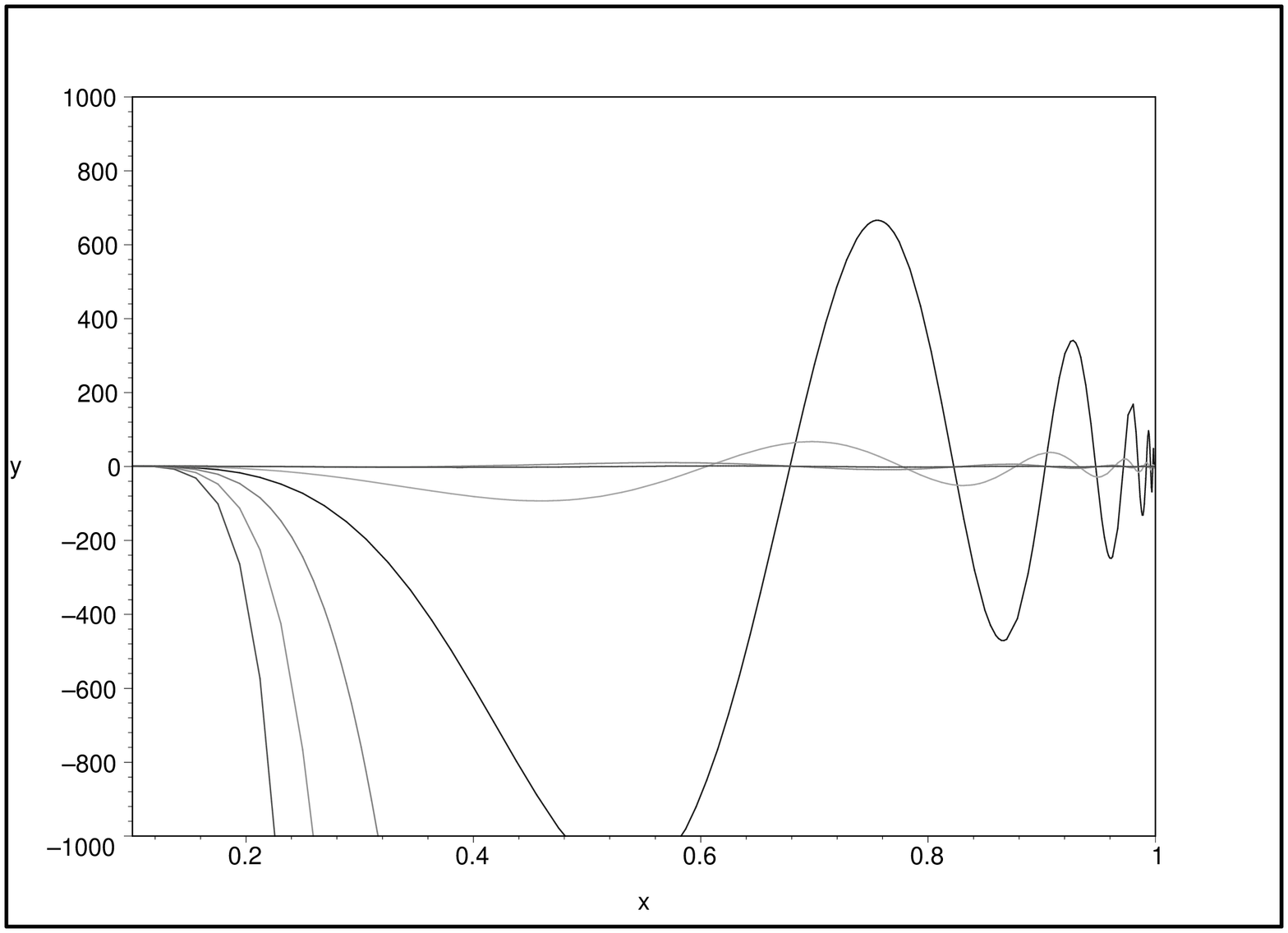}
\caption{Numerical solutions for $f_1(x)$ with $f_1(0.1)=1$, $f'_1(0.1)=10$, $l=1,\ldots, 7$
and with $\Omega=5$ (left figure) and $\Omega=10$ (right figure). Increasing values of $l$
correspond to more peaked curves on the right part of the plots.}
\end{figure}

\begin{figure}
\includegraphics[scale=0.25,angle=-90]{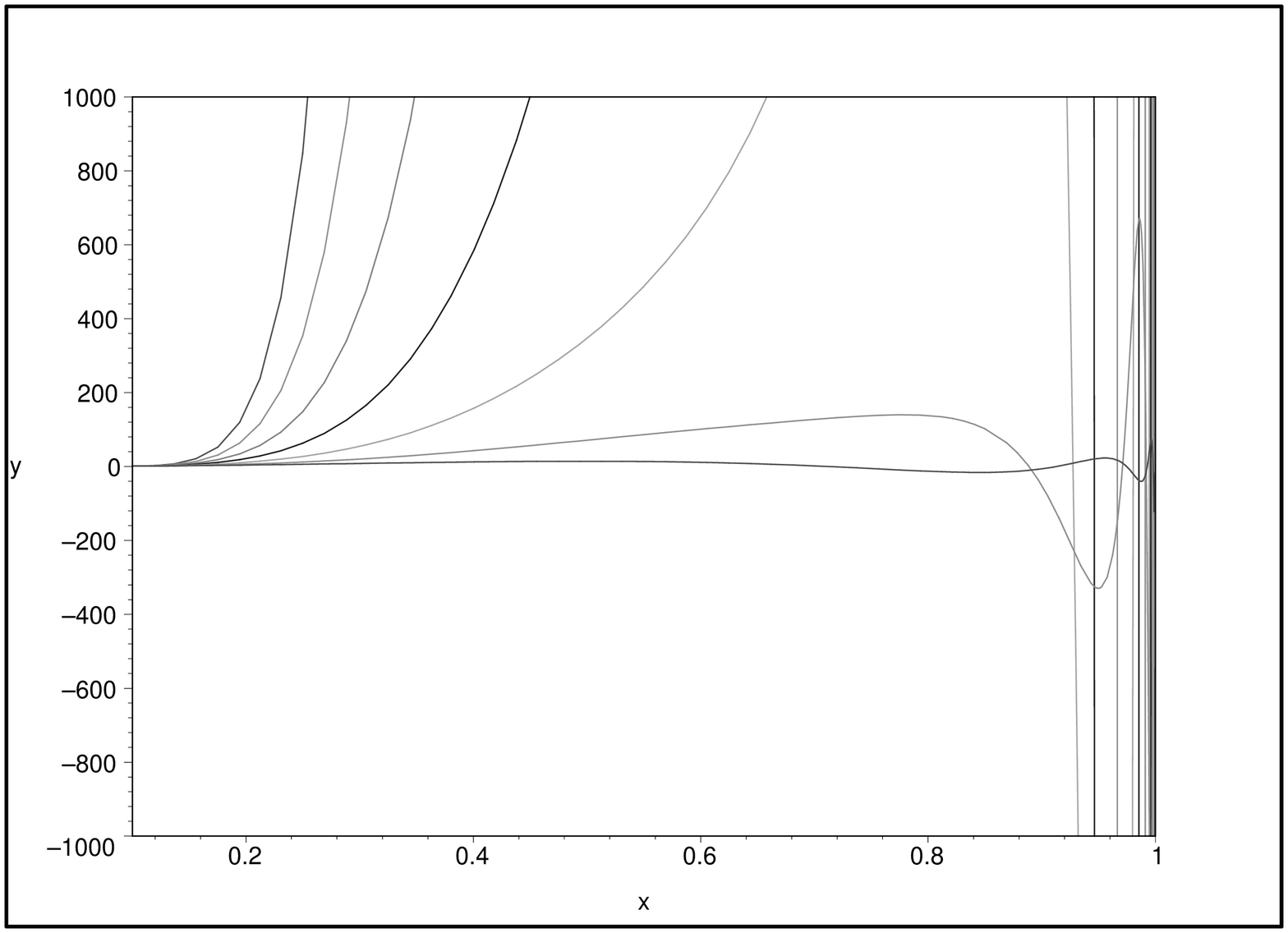}
\includegraphics[scale=0.25,angle=-90]{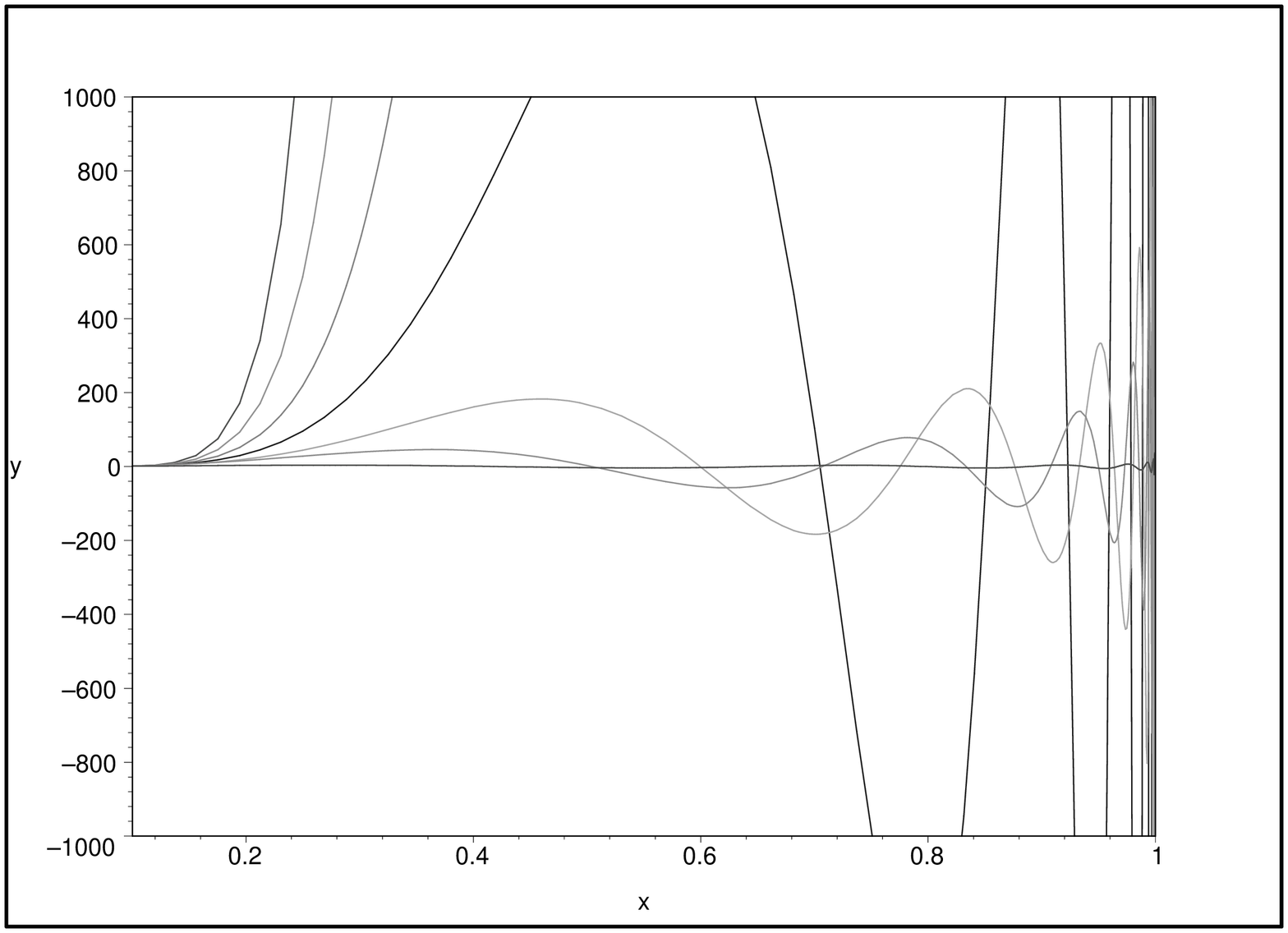}
\caption{Numerical solutions for $f_2(x)$ with $f_2(0.1)=1$, $f'_1(0.1)=10$, $l=1,\ldots, 7$
and with $\Omega=5$ (left figure) and $\Omega=10$ (right figure). Increasing values of $l$
correspond to more peaked curves on the right part of the plots.}
\end{figure}

\begin{figure}
\includegraphics[scale=0.25,angle=-90]{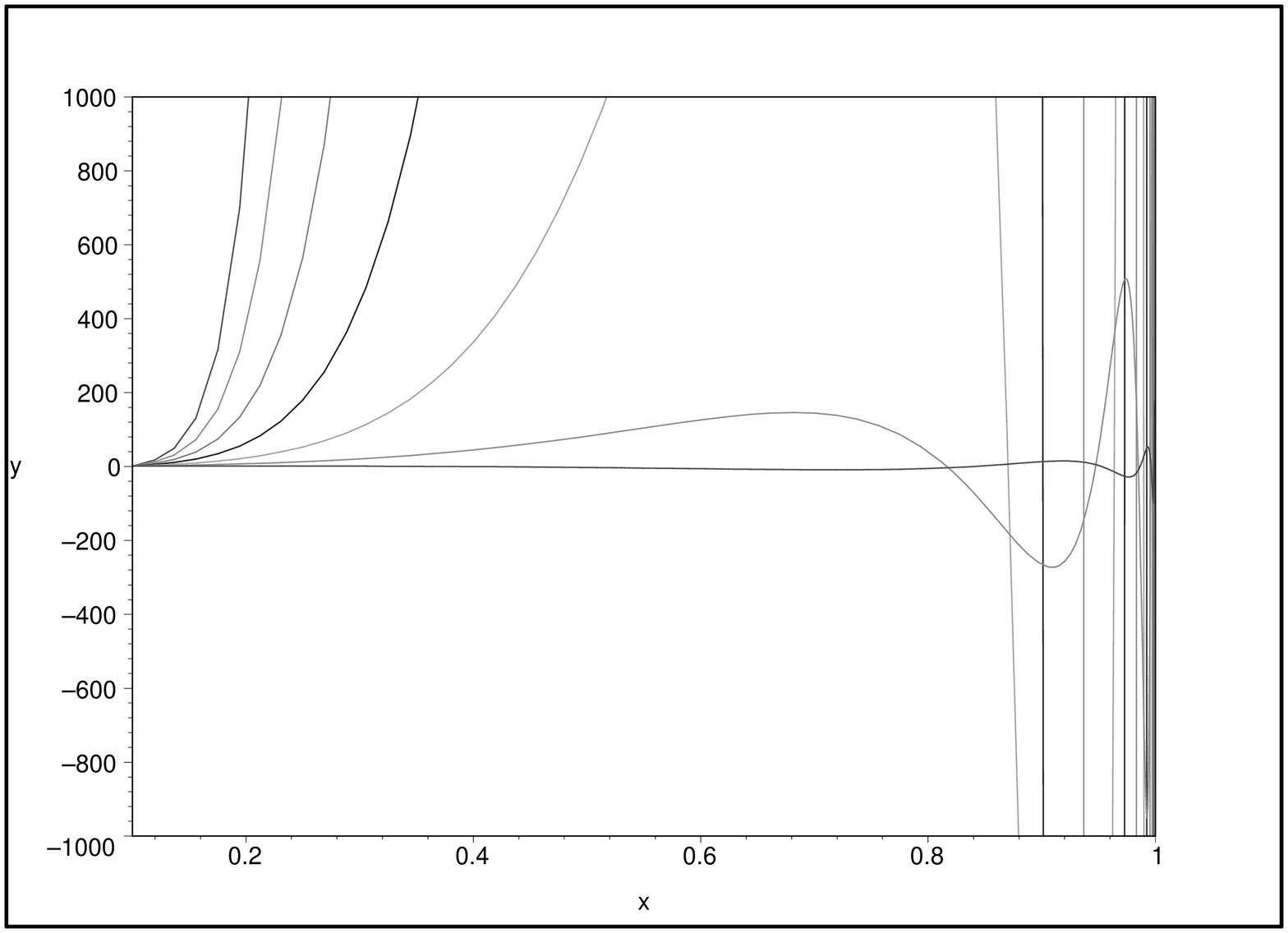}
\includegraphics[scale=0.25,angle=-90]{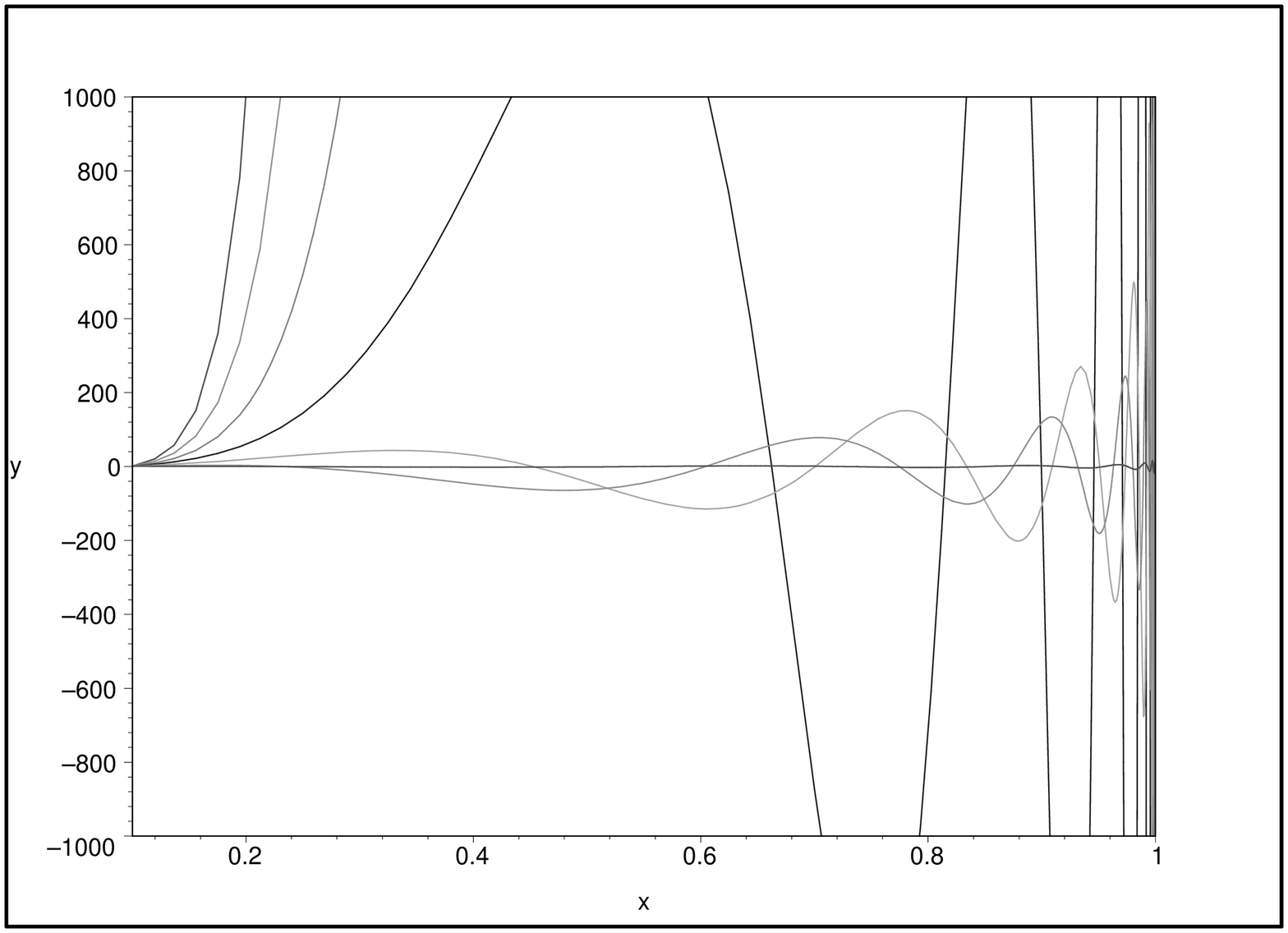}
\caption{Numerical solutions for $f_3(x)$ with $f_3(0.1)=1$, $f'_1(0.1)=10$, $l=1,\ldots, 7$
and with $\Omega=5$ (left figure) and $\Omega=10$ (right figure). Increasing values of $l$
correspond to more peaked curves on the right part of the plots.}
\end{figure}

\begin{figure}
\includegraphics[scale=0.25,angle=-90]{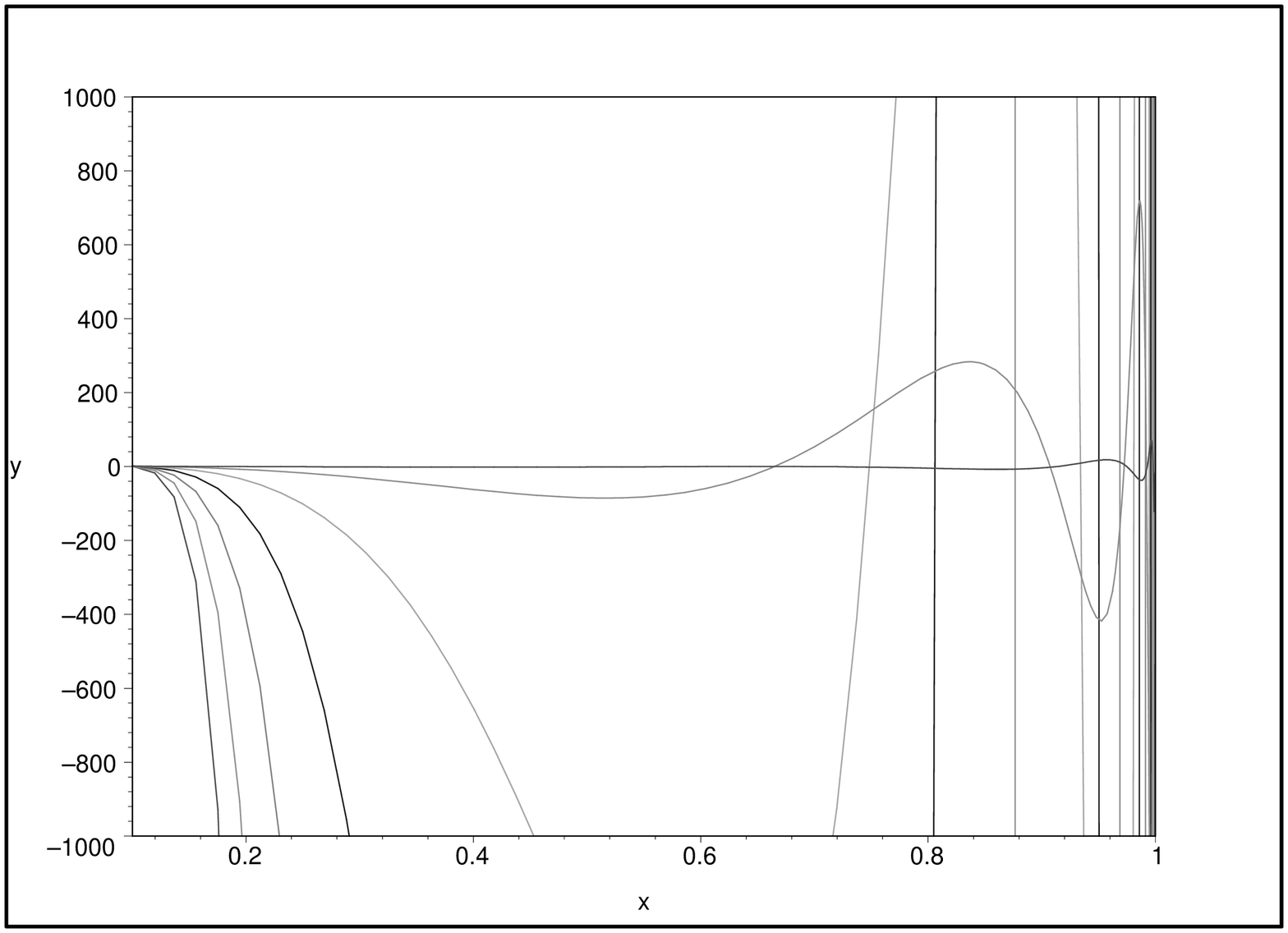}
\includegraphics[scale=0.25,angle=-90]{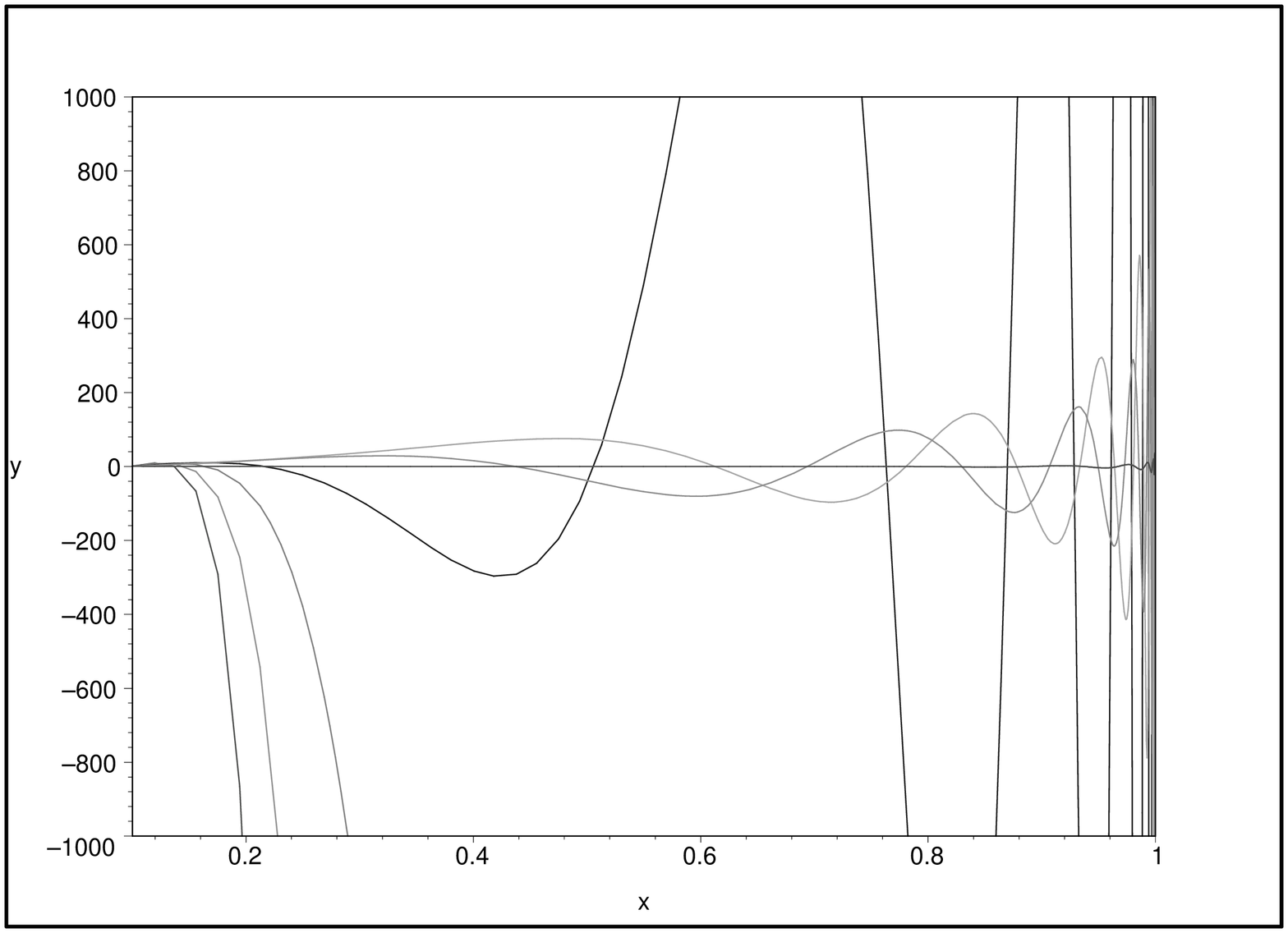}
\caption{Numerical solutions for $f_4(x)$ with $f_4(0.1)=1$, $f'_1(0.1)=10$, $l=1,\ldots, 7$
and with $\Omega=5$ (left figure) and $\Omega=10$ (right figure). Increasing values of $l$
correspond to more peaked curves on the right part of the plots.}
\end{figure}

\begin{figure}
\includegraphics[scale=0.25,angle=-90]{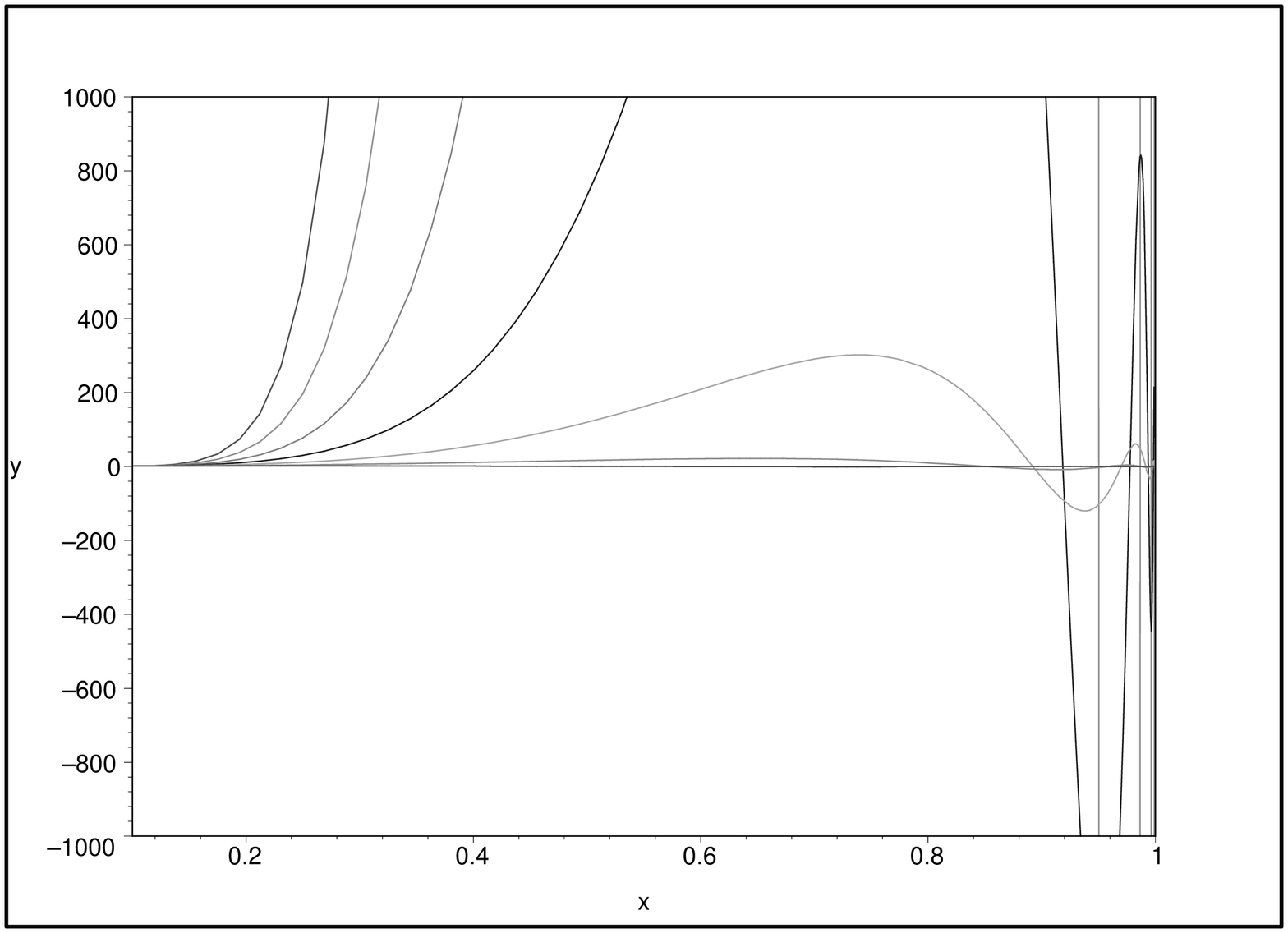}
\includegraphics[scale=0.25,angle=-90]{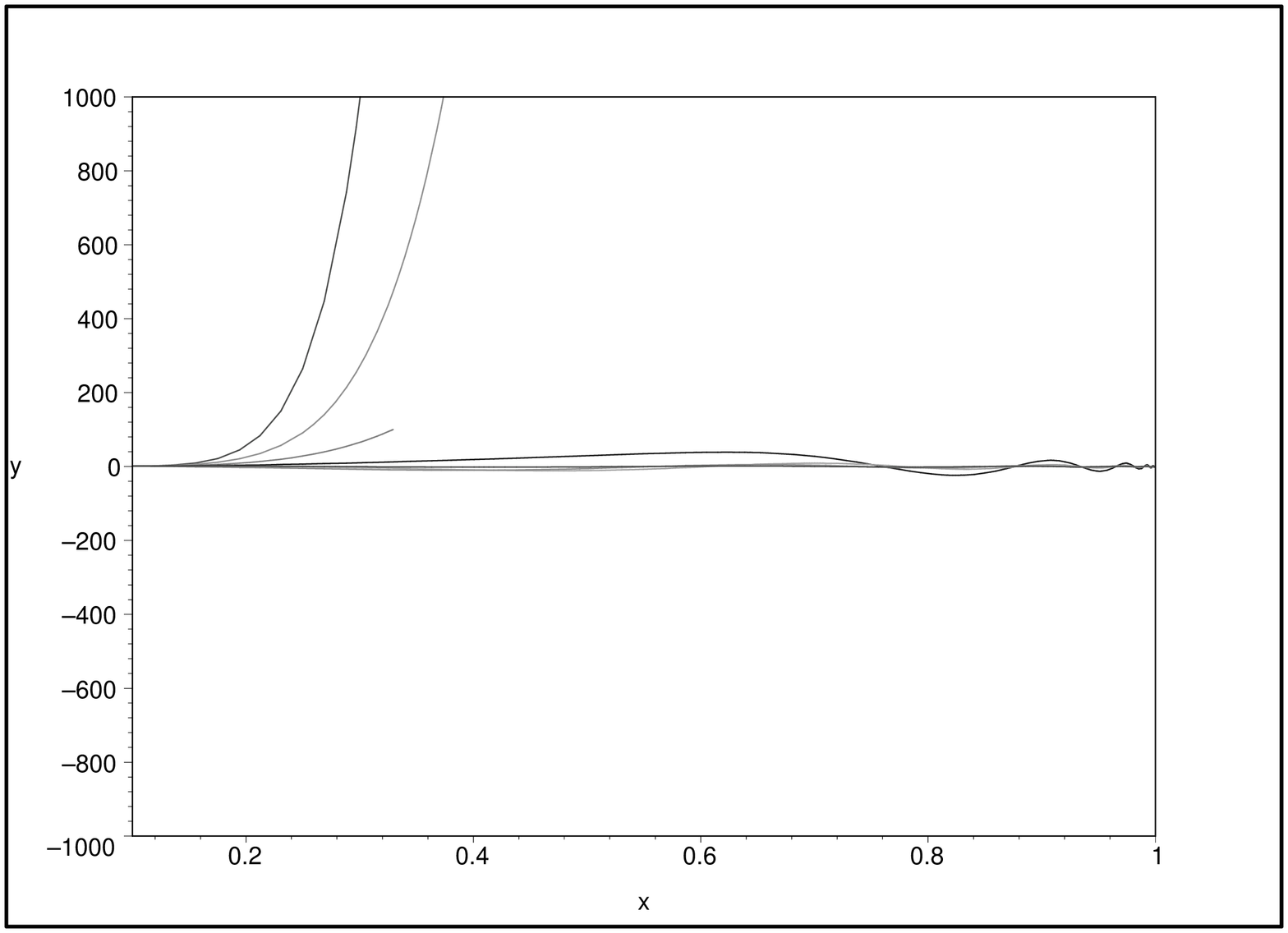}
\caption{Numerical solutions for $f_5(x)$ with $f_5(0.1)=1$, $f'_1(0.1)=10$, $l=1,\ldots, 7$
and with $\Omega=5$ (left figure) and $\Omega=10$ (right figure). Increasing values of $l$
correspond to more peaked curves on the right part of the plots.}
\end{figure}

\begin{figure}
\includegraphics[scale=0.25,angle=-90]{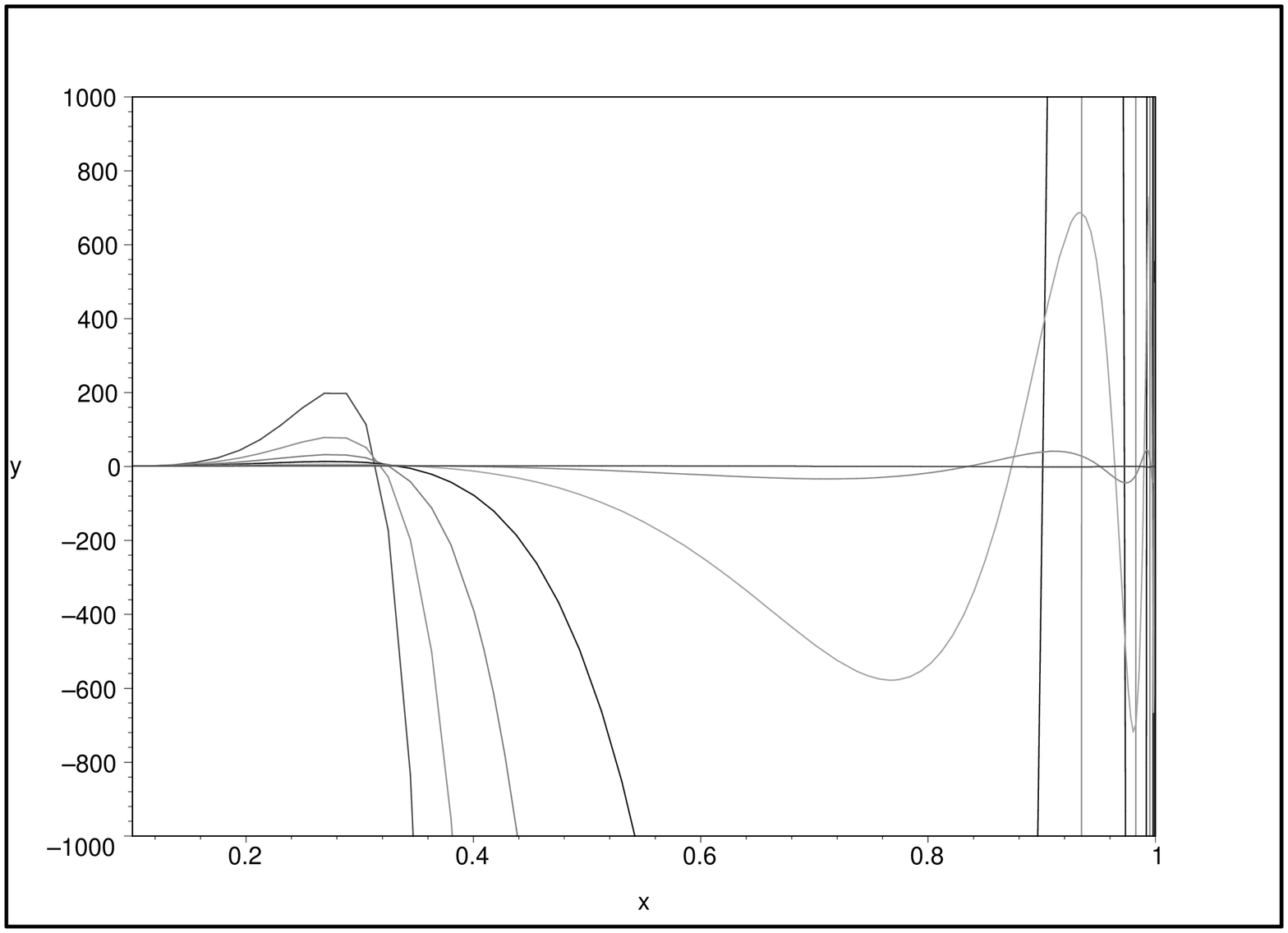}
\includegraphics[scale=0.25,angle=-90]{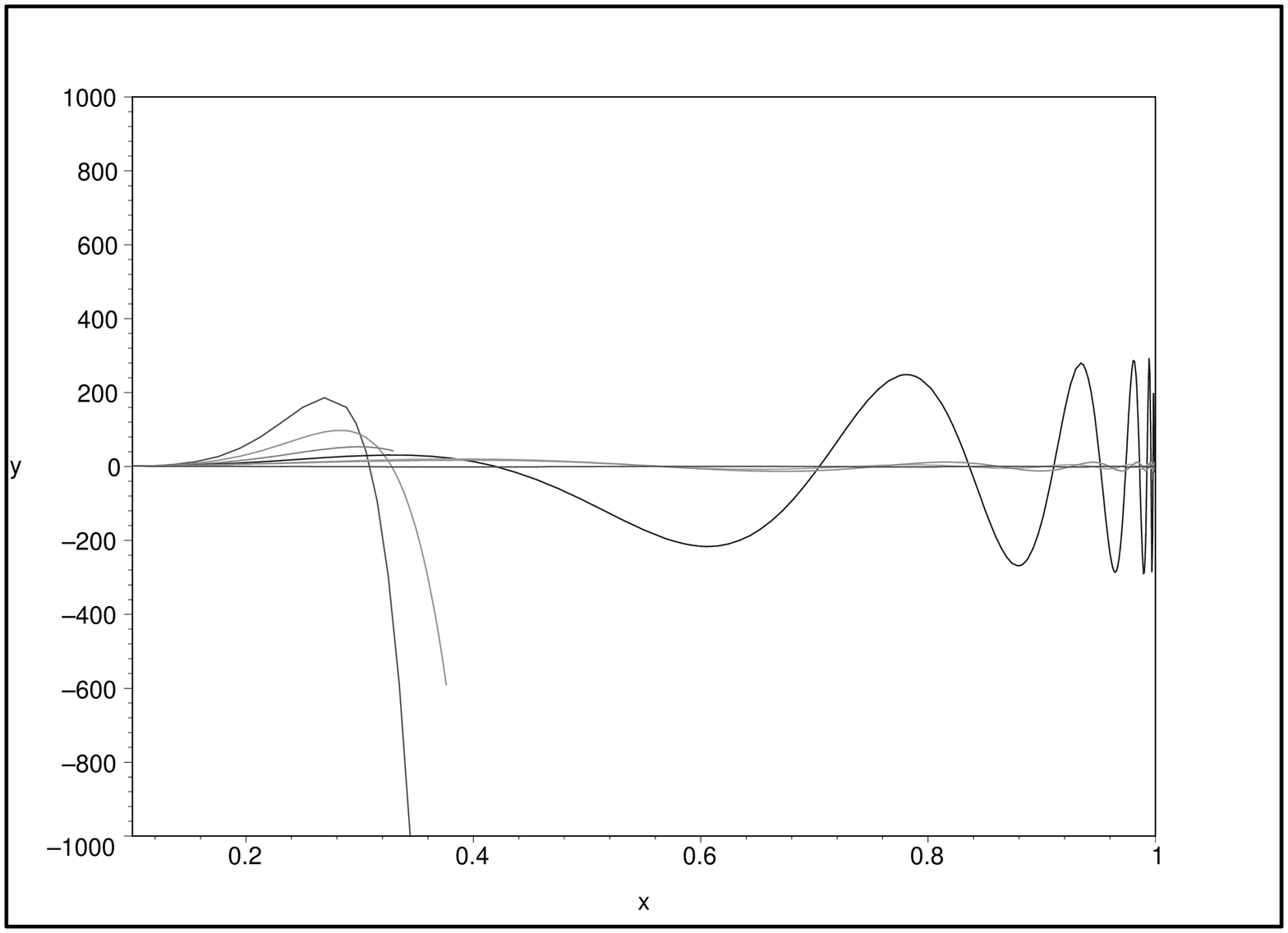}
\caption{Numerical solutions for $f_6(x)$ with $f_6(0.1)=1$, $f'_1(0.1)=10$, $l=1,\ldots, 7$
and with $\Omega=5$ (left figure) and $\Omega=10$ (right figure). Increasing values of $l$
correspond to more peaked curves on the right part of the plots.}
\end{figure}

\begin{figure}
\includegraphics[scale=0.25,angle=-90]{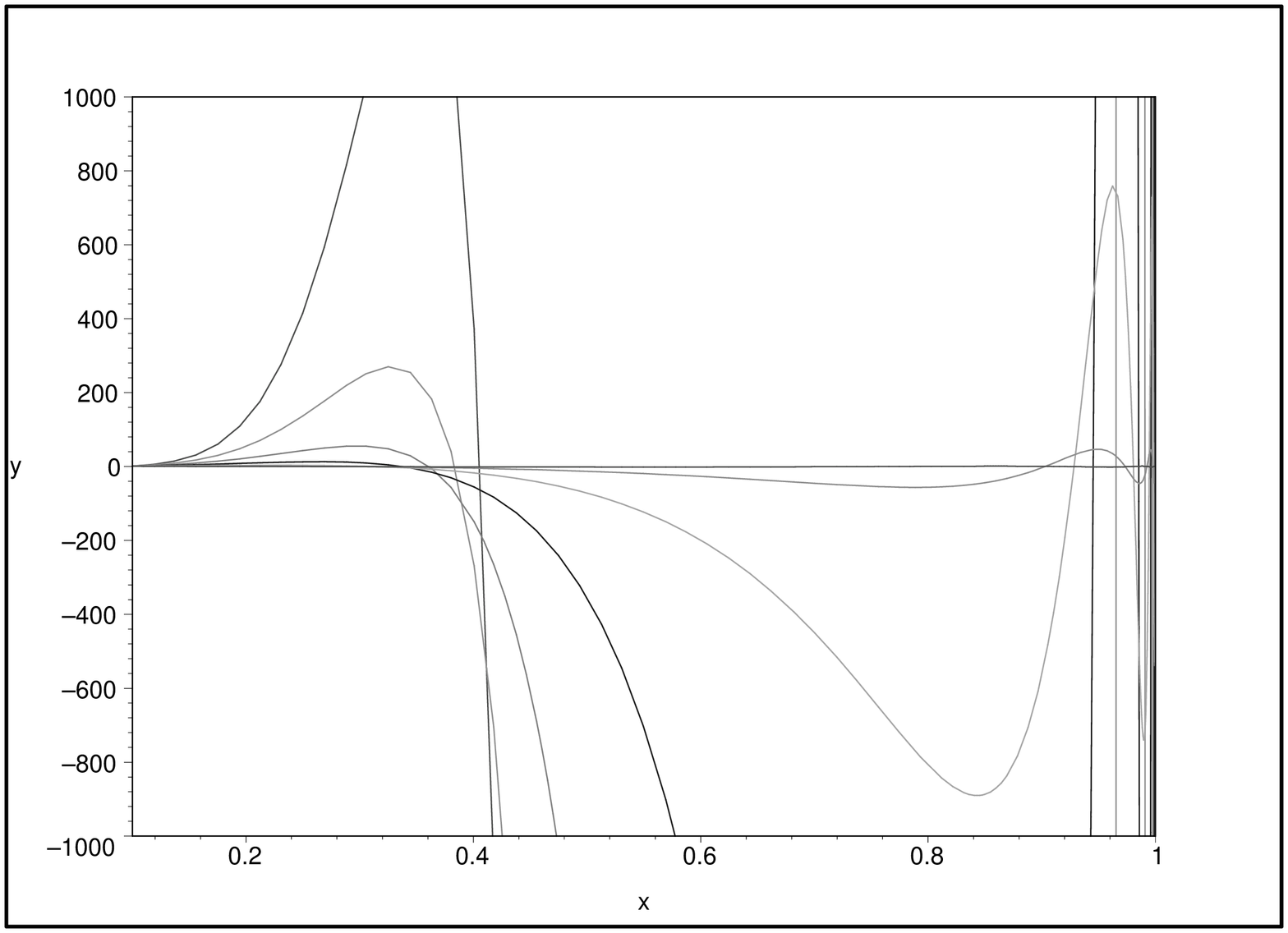}
\includegraphics[scale=0.25,angle=-90]{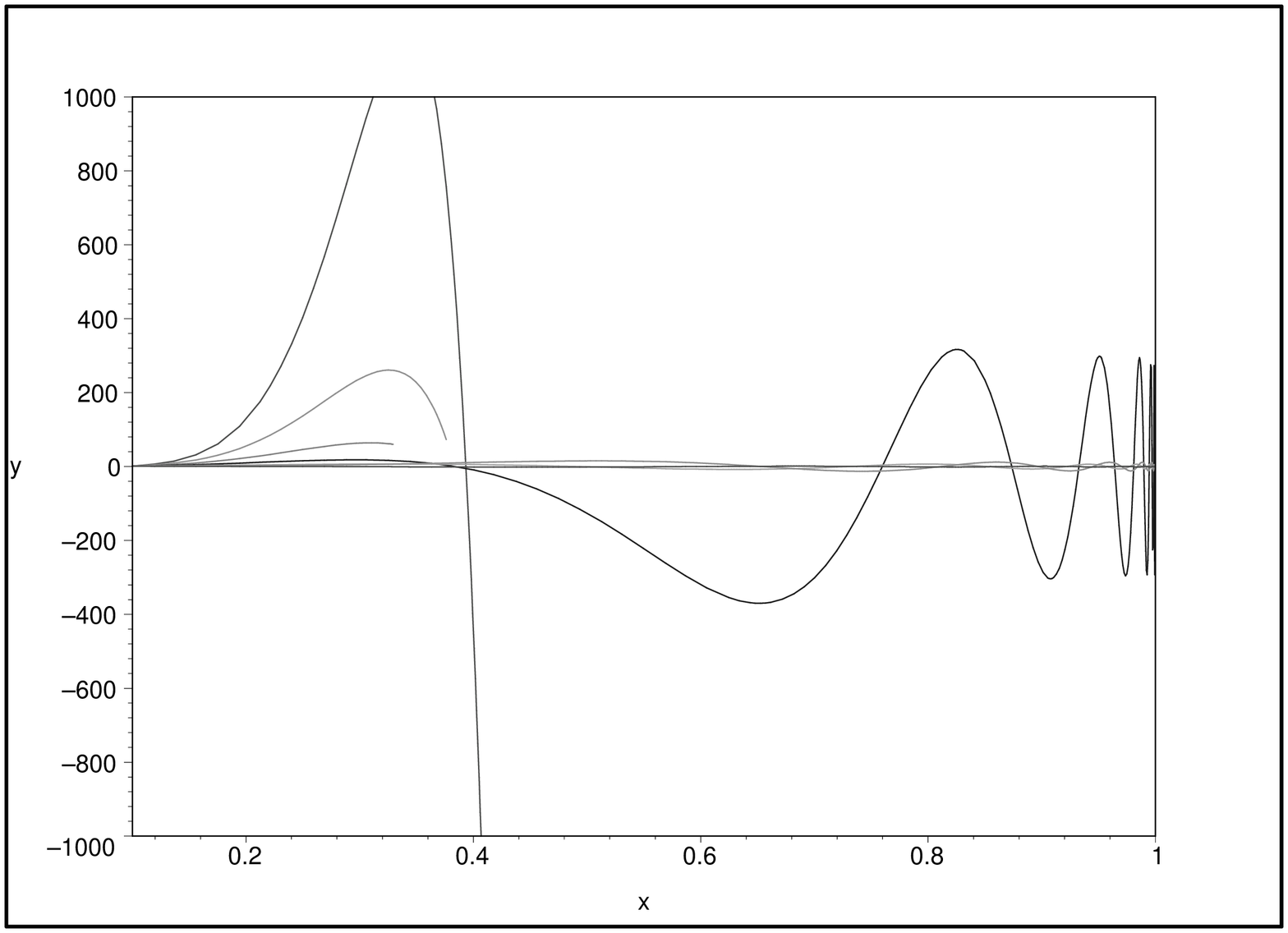}
\caption{Numerical solutions for $f_7(x)$ with $f_7(0.1)=1$, $f'_1(0.1)=10$, $l=1,\ldots, 7$
and with $\Omega=5$ (left figure) and $\Omega=10$ (right figure). Increasing values of $l$
correspond to more peaked curves on the right part of the plots.}
\end{figure}

\begin{figure}
\includegraphics[scale=0.25,angle=-90]{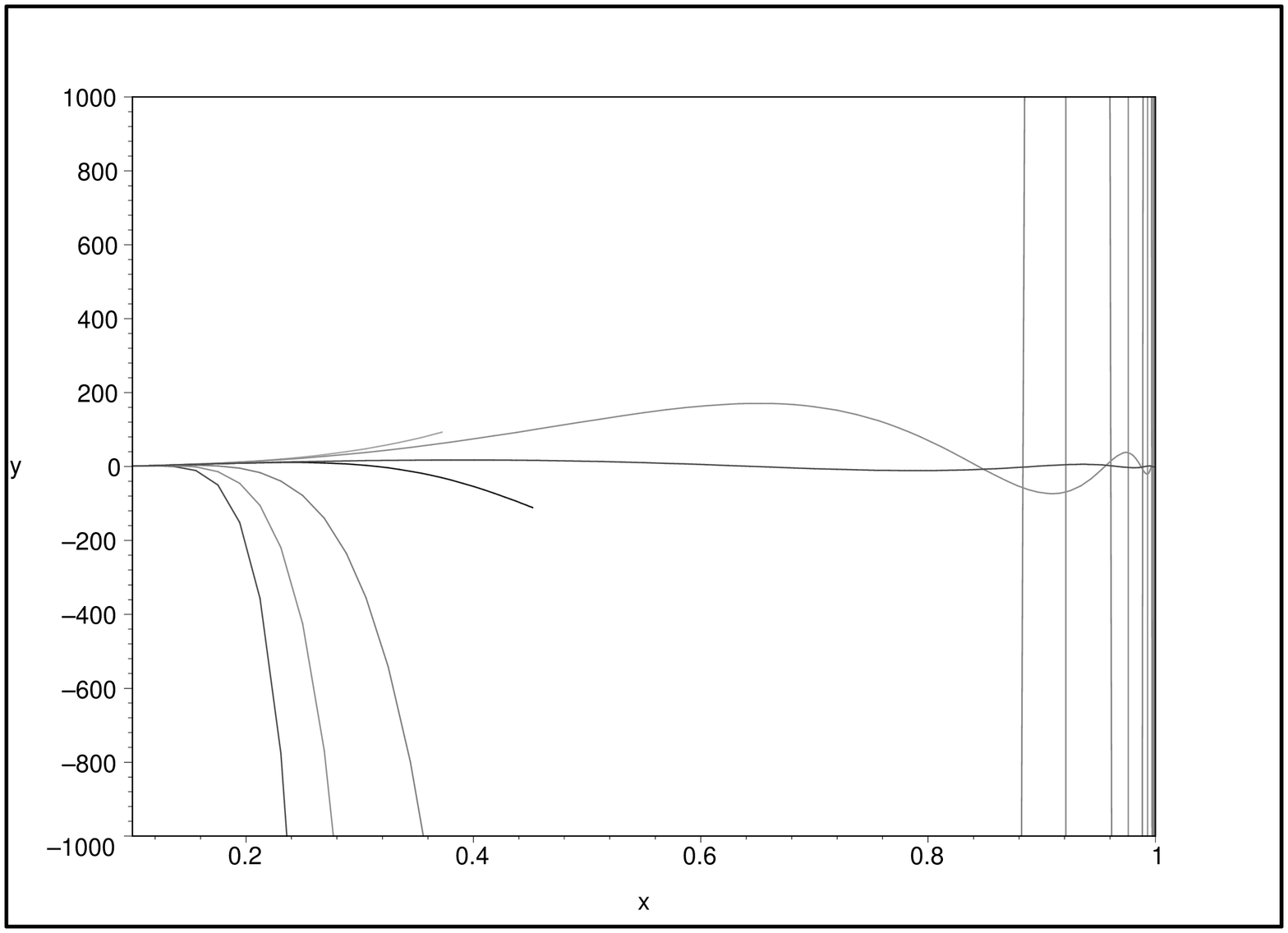}
\includegraphics[scale=0.25,angle=-90]{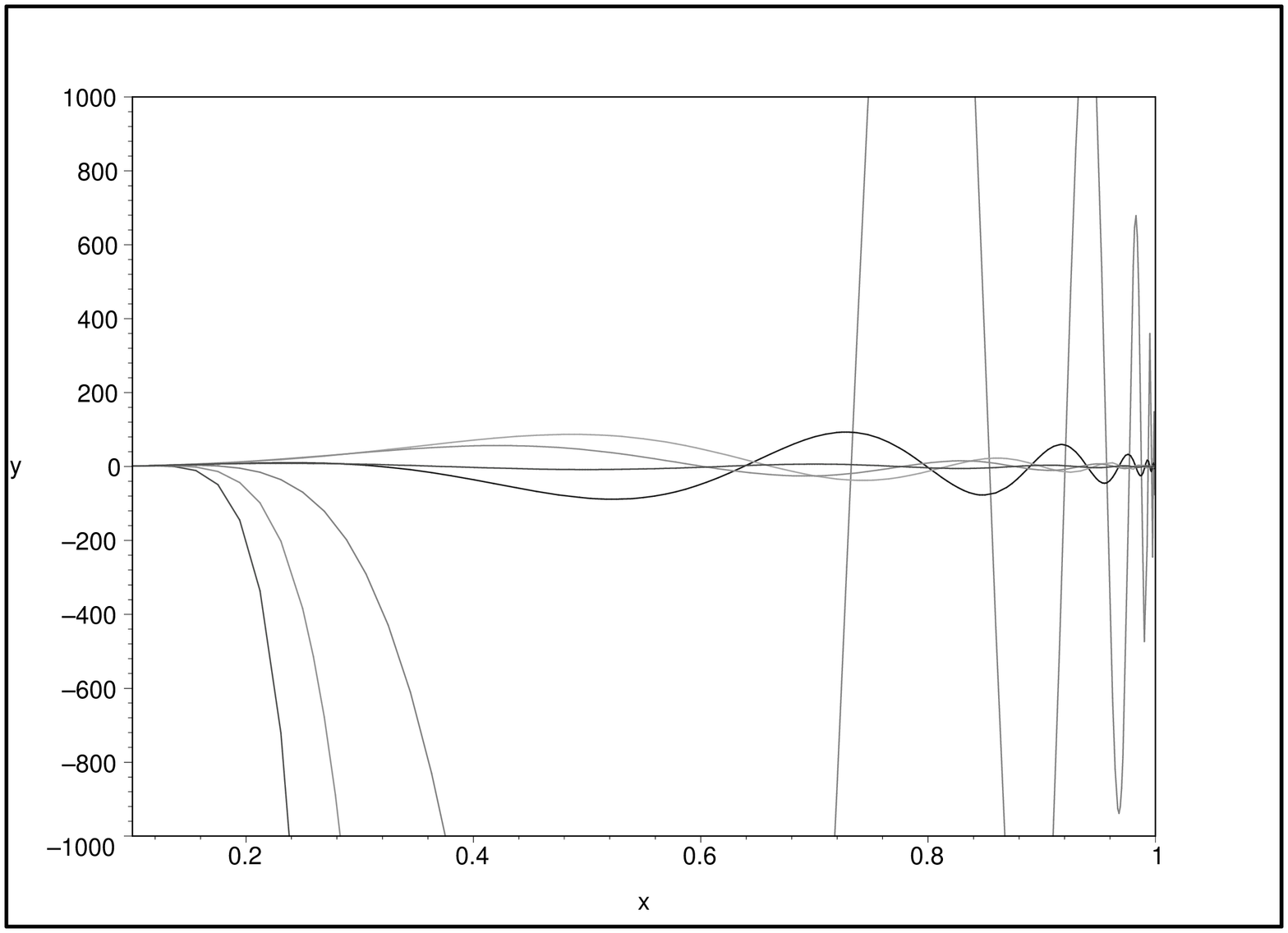}
\caption{Numerical solutions for $f_1(x)$ with $f_1(0.1)=1$, $f'_1(0.1)=100$, $l=1,\ldots, 7$
and with $\Omega=5$ (left figure) and $\Omega=10$ (right figure). Increasing values of $l$
correspond to more peaked curves on the right part of the plots.}
\end{figure}

\begin{figure}
\includegraphics[scale=0.25,angle=-90]{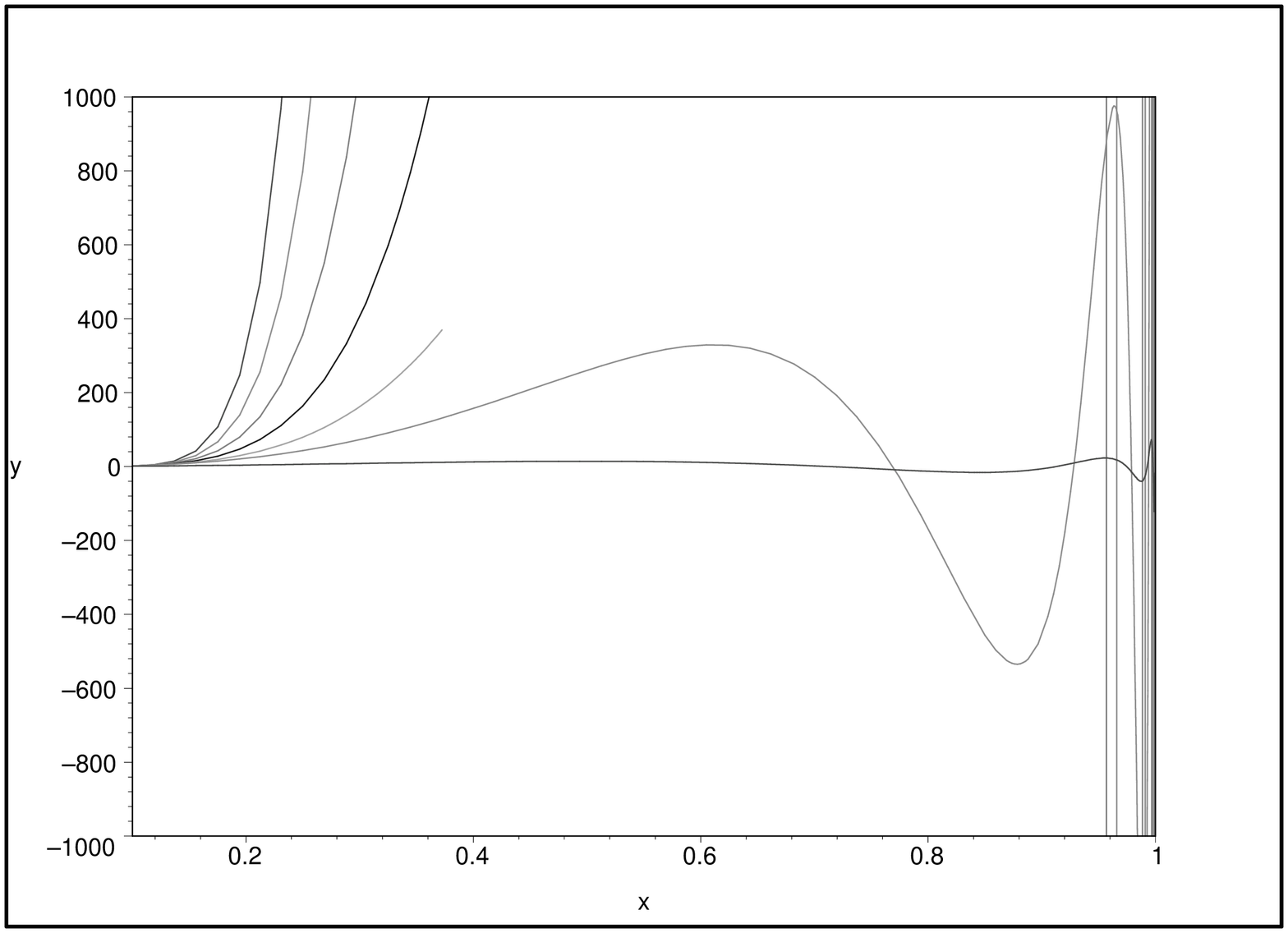}
\includegraphics[scale=0.25,angle=-90]{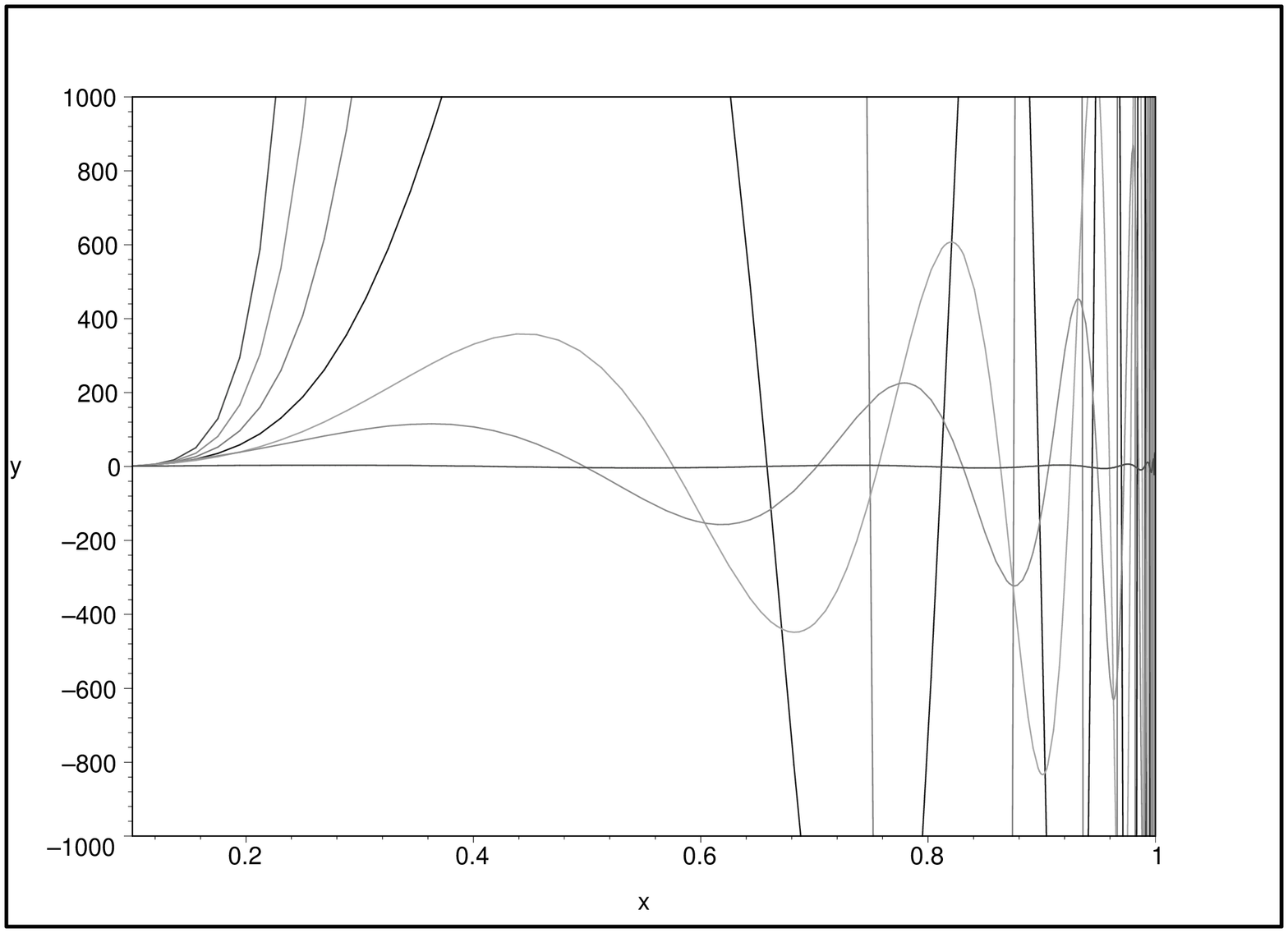}
\caption{Numerical solutions for $f_2(x)$ with $f_2(0.1)=1$, $f'_1(0.1)=100$, $l=1,\ldots, 7$
and with $\Omega=5$ (left figure) and $\Omega=10$ (right figure). Increasing values of $l$
correspond to more peaked curves on the right part of the plots.}
\end{figure}

\begin{figure}
\includegraphics[scale=0.25,angle=-90]{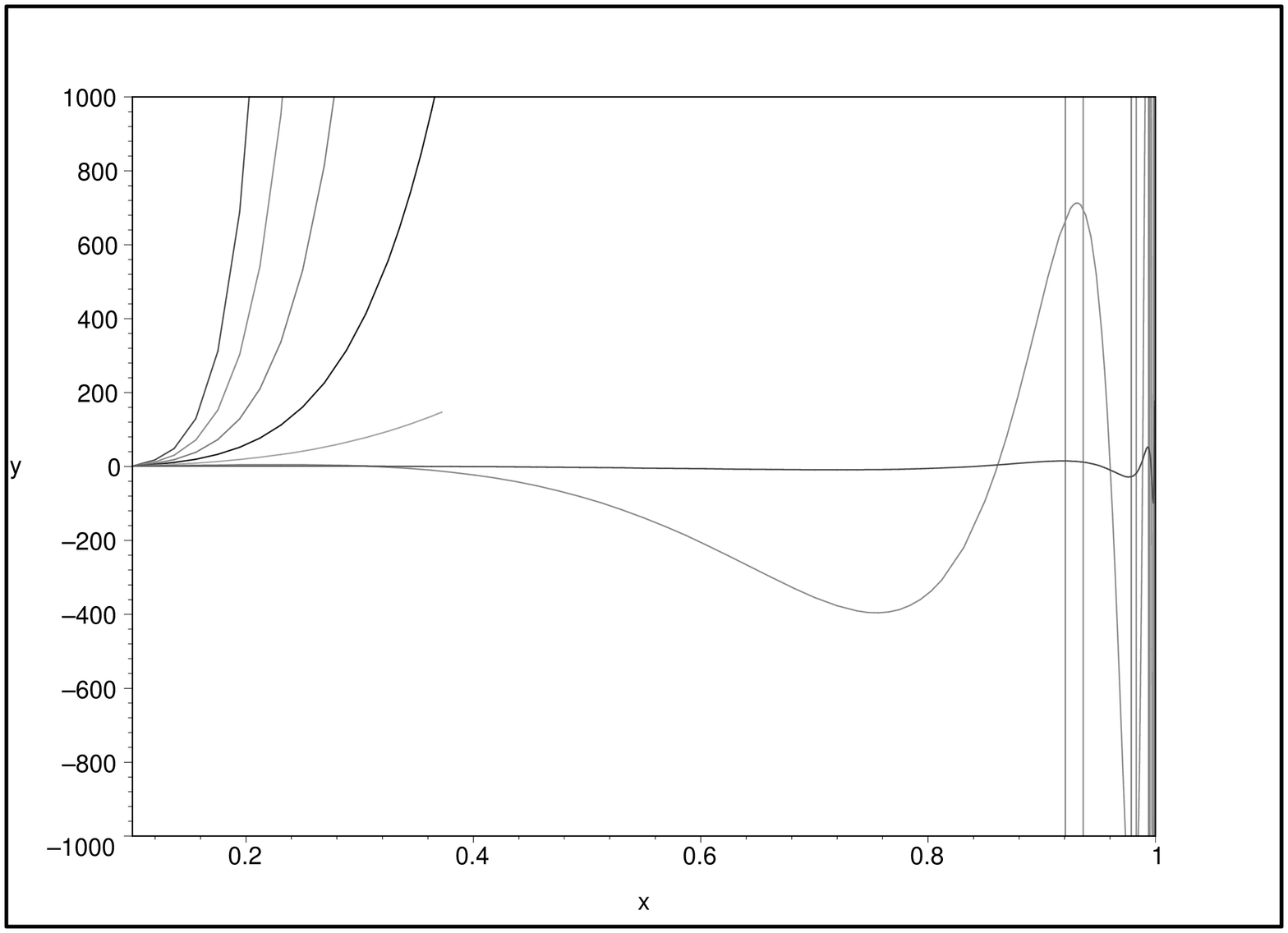}
\includegraphics[scale=0.25,angle=-90]{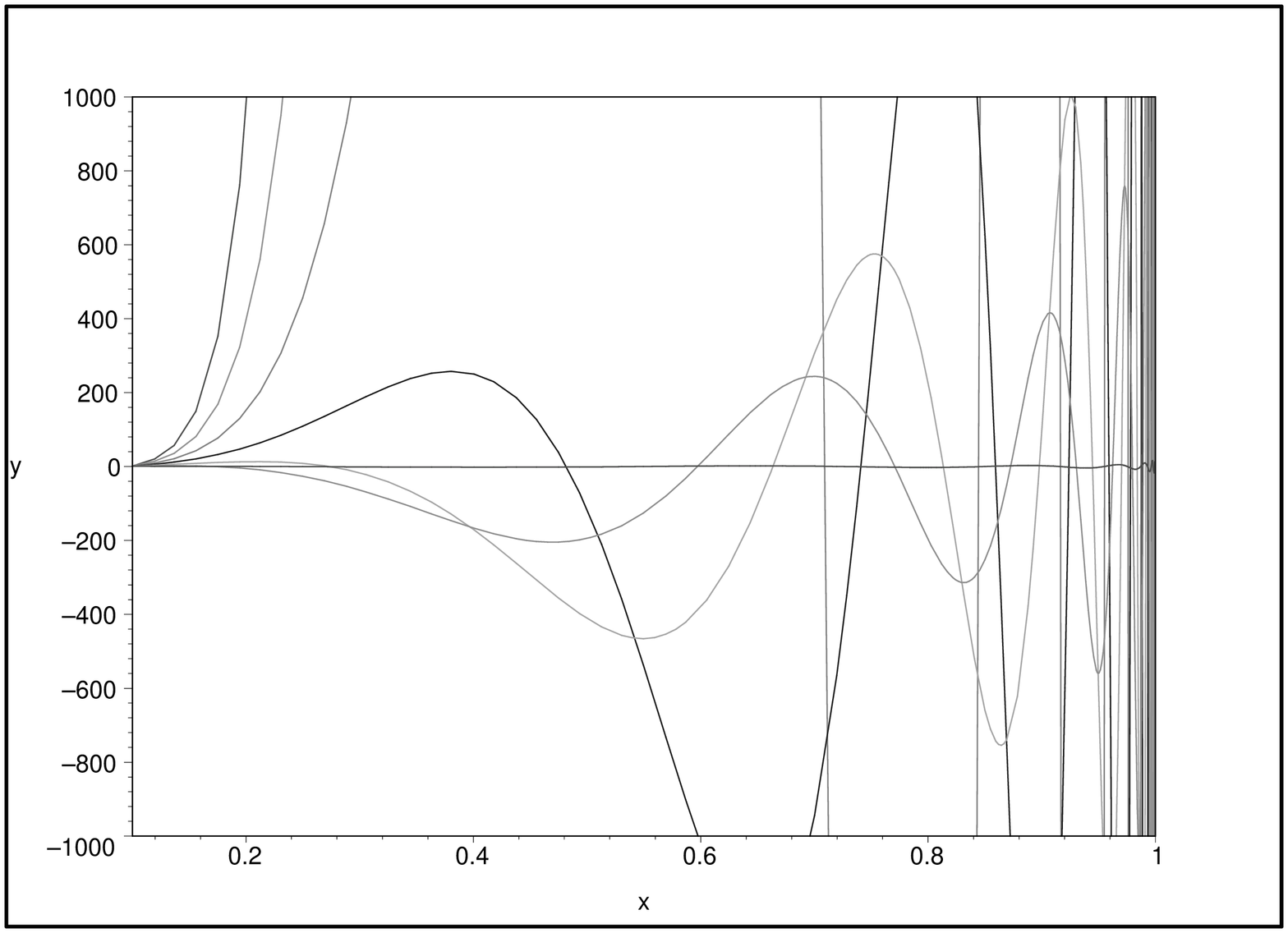}
\caption{Numerical solutions for $f_3(x)$ with $f_3(0.1)=1$, $f'_1(0.1)=100$, $l=1,\ldots, 7$
and with $\Omega=5$ (left figure) and $\Omega=10$ (right figure). Increasing values of $l$
correspond to more peaked curves on the right part of the plots.}
\end{figure}

\begin{figure}
\includegraphics[scale=0.25,angle=-90]{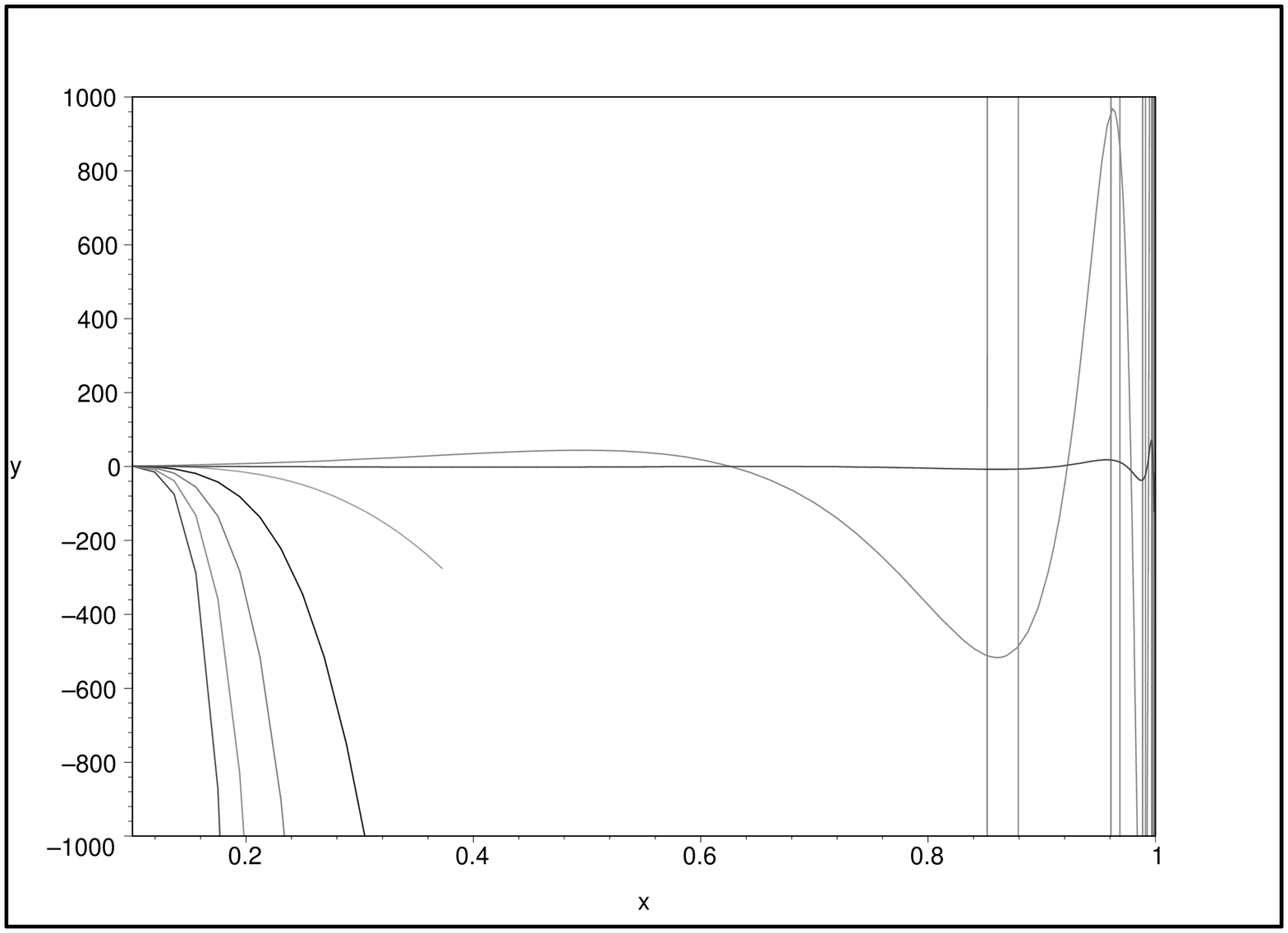}
\includegraphics[scale=0.25,angle=-90]{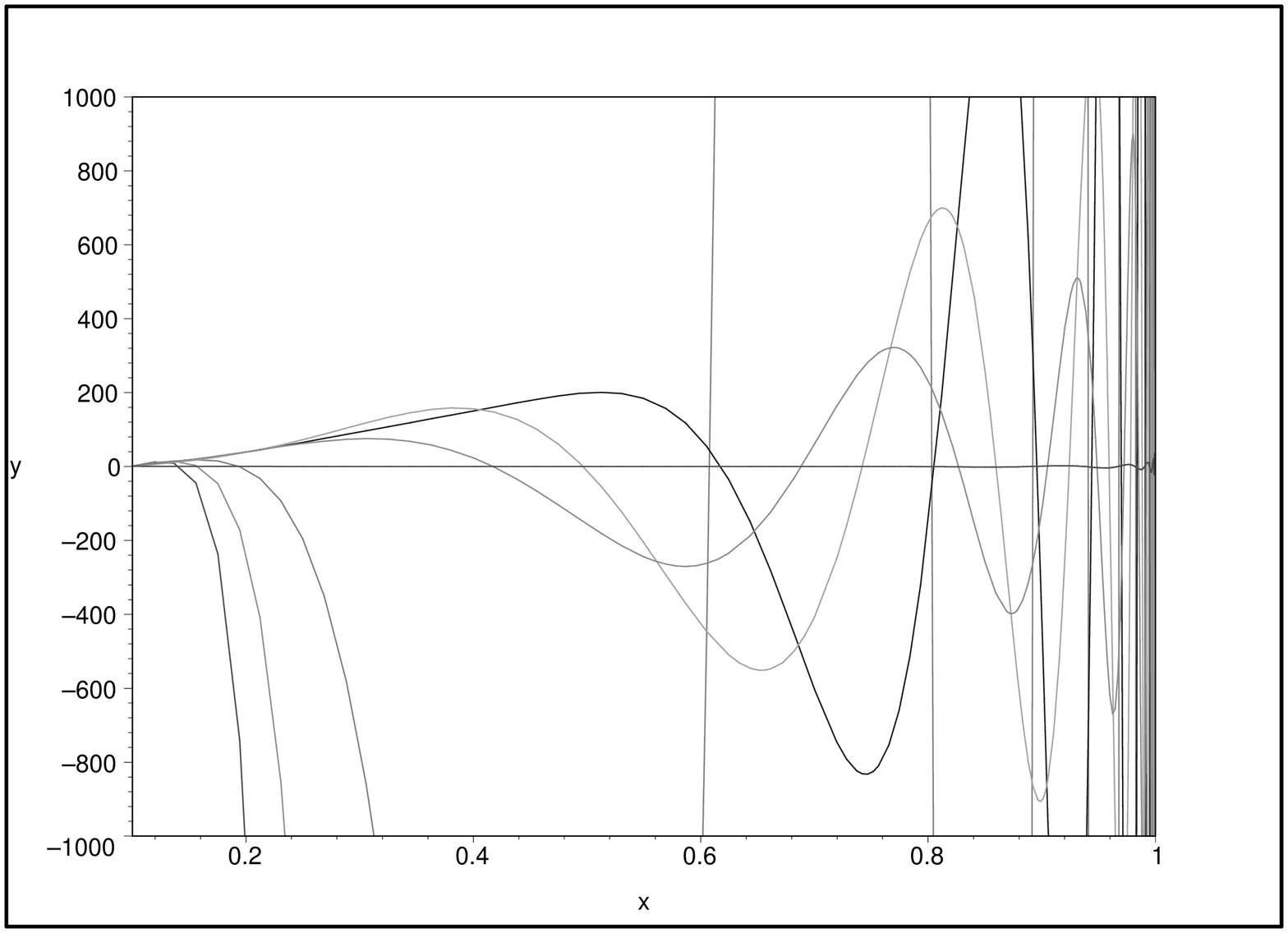}
\caption{Numerical solutions for $f_4(x)$ with $f_4(0.1)=1$, $f'_1(0.1)=100$, $l=1,\ldots, 7$
and with $\Omega=5$ (left figure) and $\Omega=10$ (right figure). Increasing values of $l$
correspond to more peaked curves on the right part of the plots.}
\end{figure}

\begin{figure}
\includegraphics[scale=0.25,angle=-90]{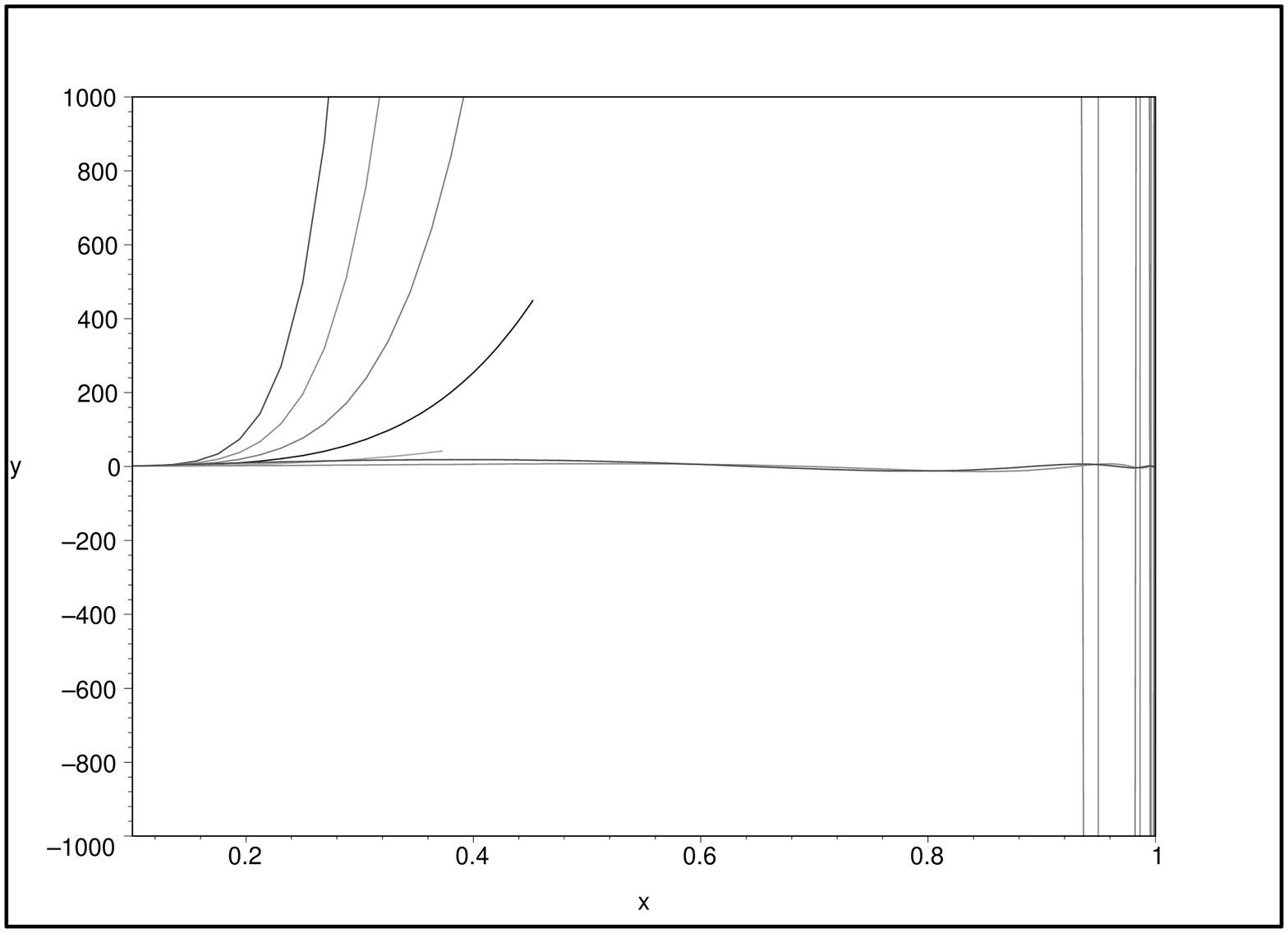}
\includegraphics[scale=0.25,angle=-90]{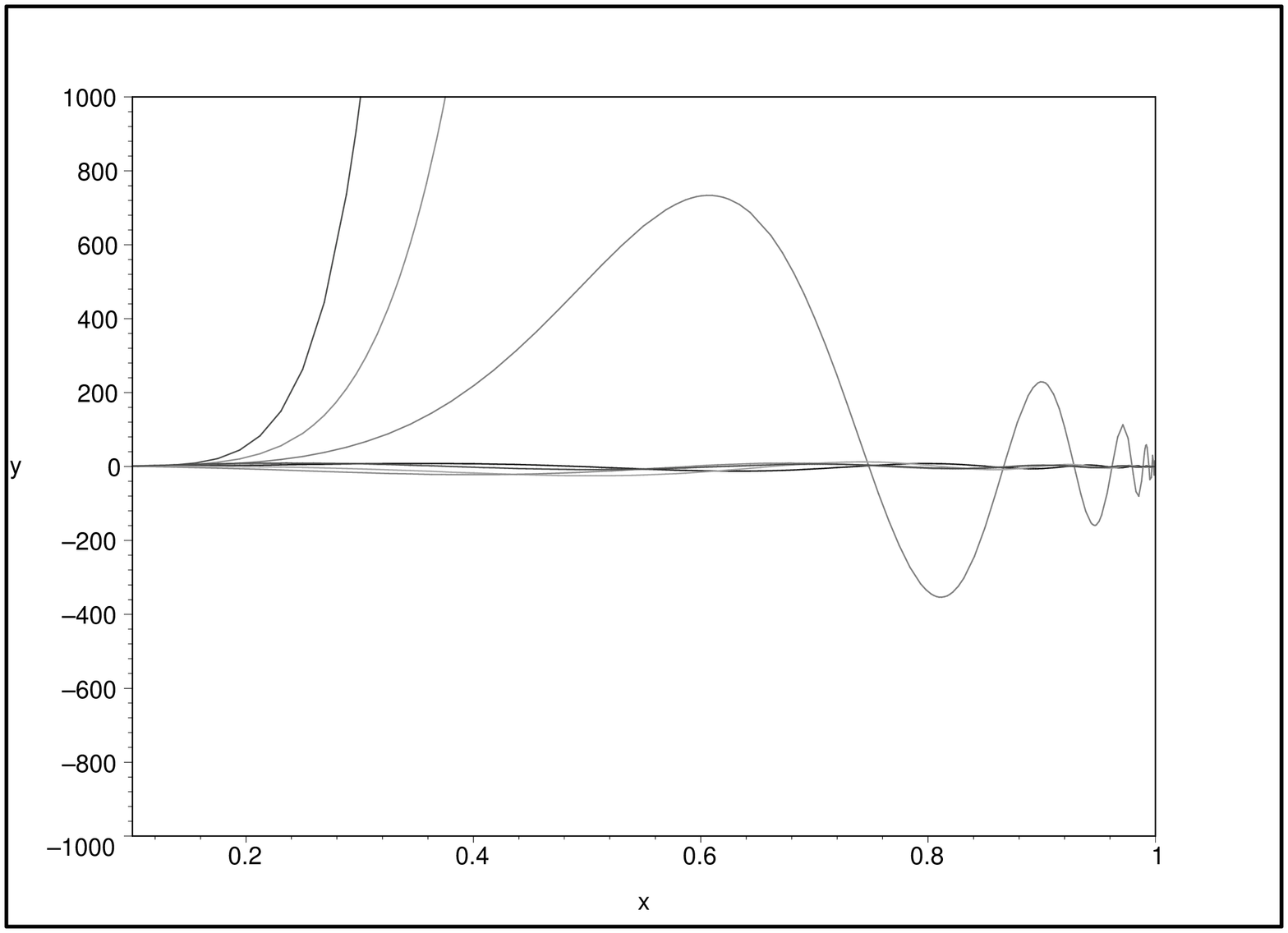}
\caption{Numerical solutions for $f_5(x)$ with $f_5(0.1)=1$, $f'_1(0.1)=100$, $l=1,\ldots, 7$
and with $\Omega=5$ (left figure) and $\Omega=10$ (right figure). Increasing values of $l$
correspond to more peaked curves on the right part of the plots.}
\end{figure}

\begin{figure}
\includegraphics[scale=0.25,angle=-90]{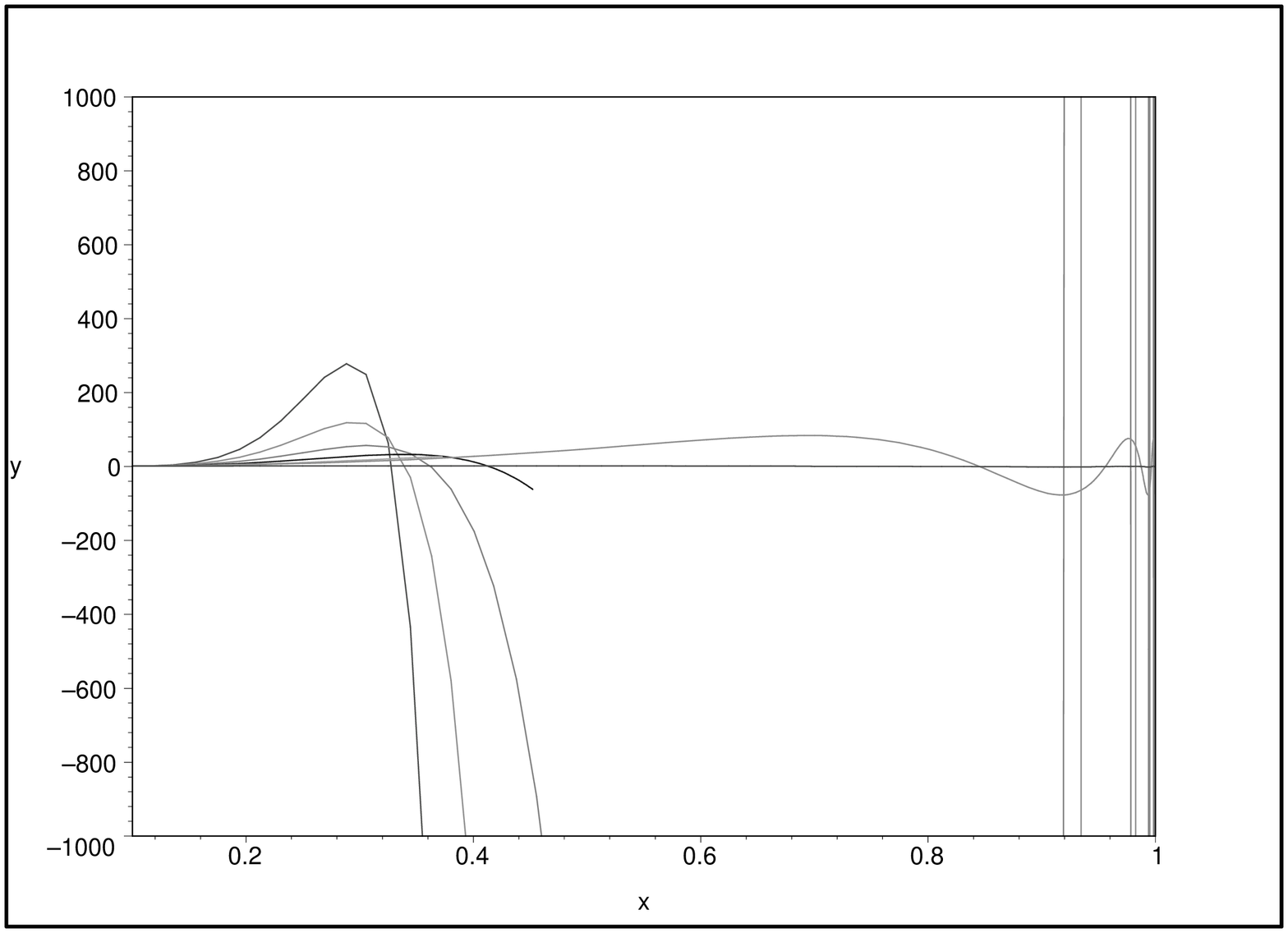}
\includegraphics[scale=0.25,angle=-90]{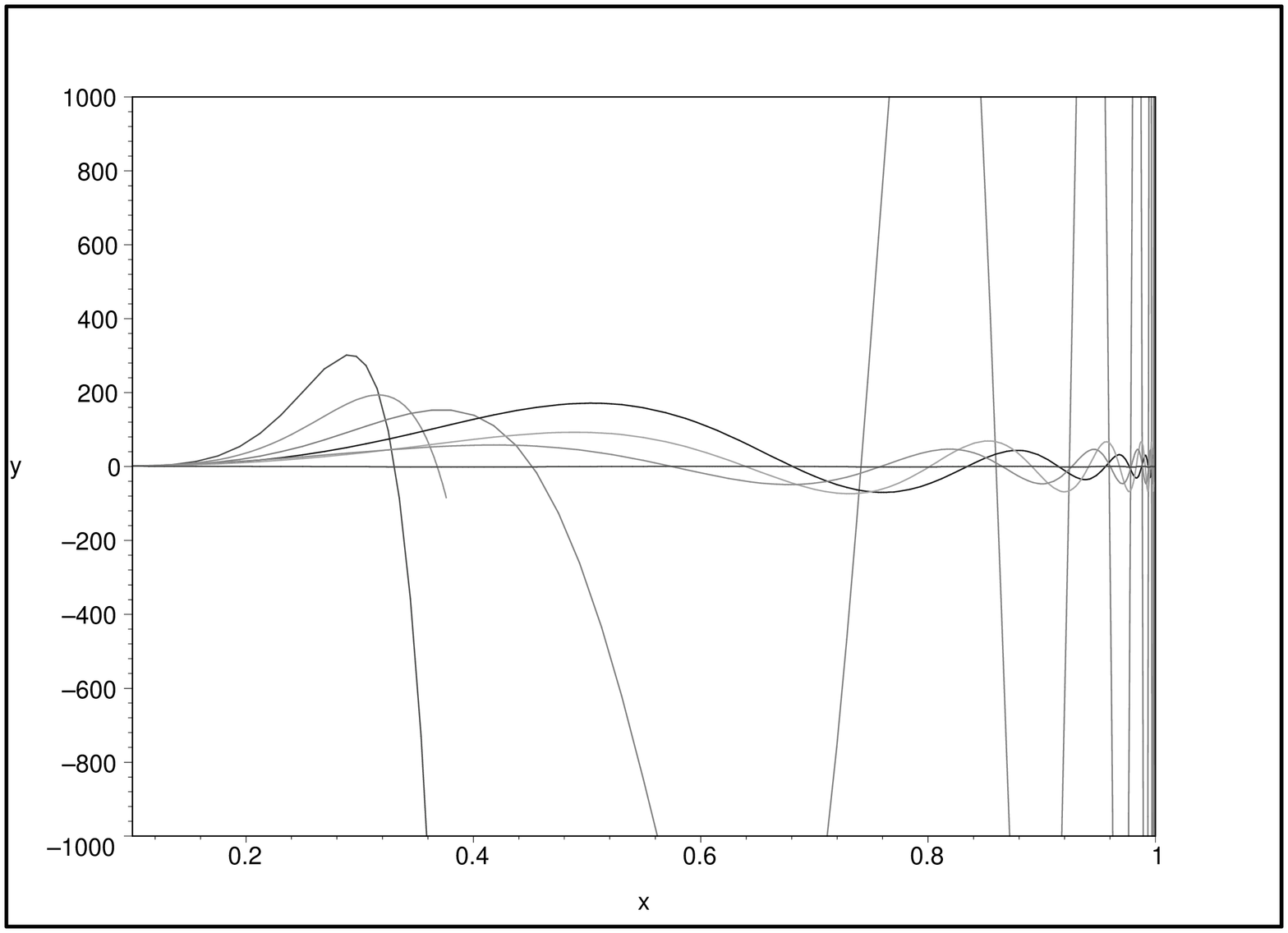}
\caption{Numerical solutions for $f_6(x)$ with $f_6(0.1)=1$, $f'_1(0.1)=100$, $l=1,\ldots, 7$
and with $\Omega=5$ (left figure) and $\Omega=10$ (right figure). Increasing values of $l$
correspond to more peaked curves on the right part of the plots.}
\end{figure}

\begin{figure}
\includegraphics[scale=0.25,angle=-90]{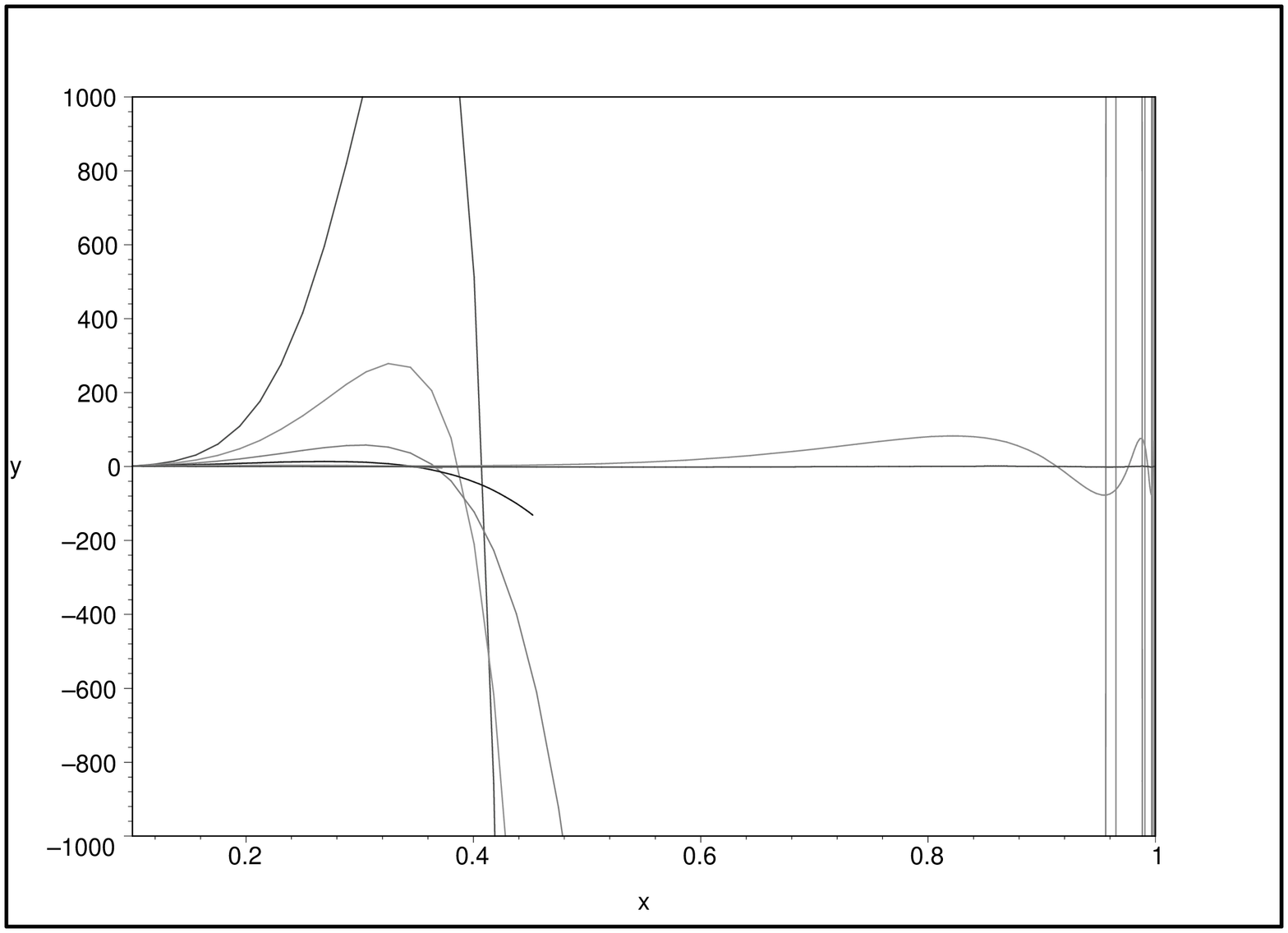}
\includegraphics[scale=0.25,angle=-90]{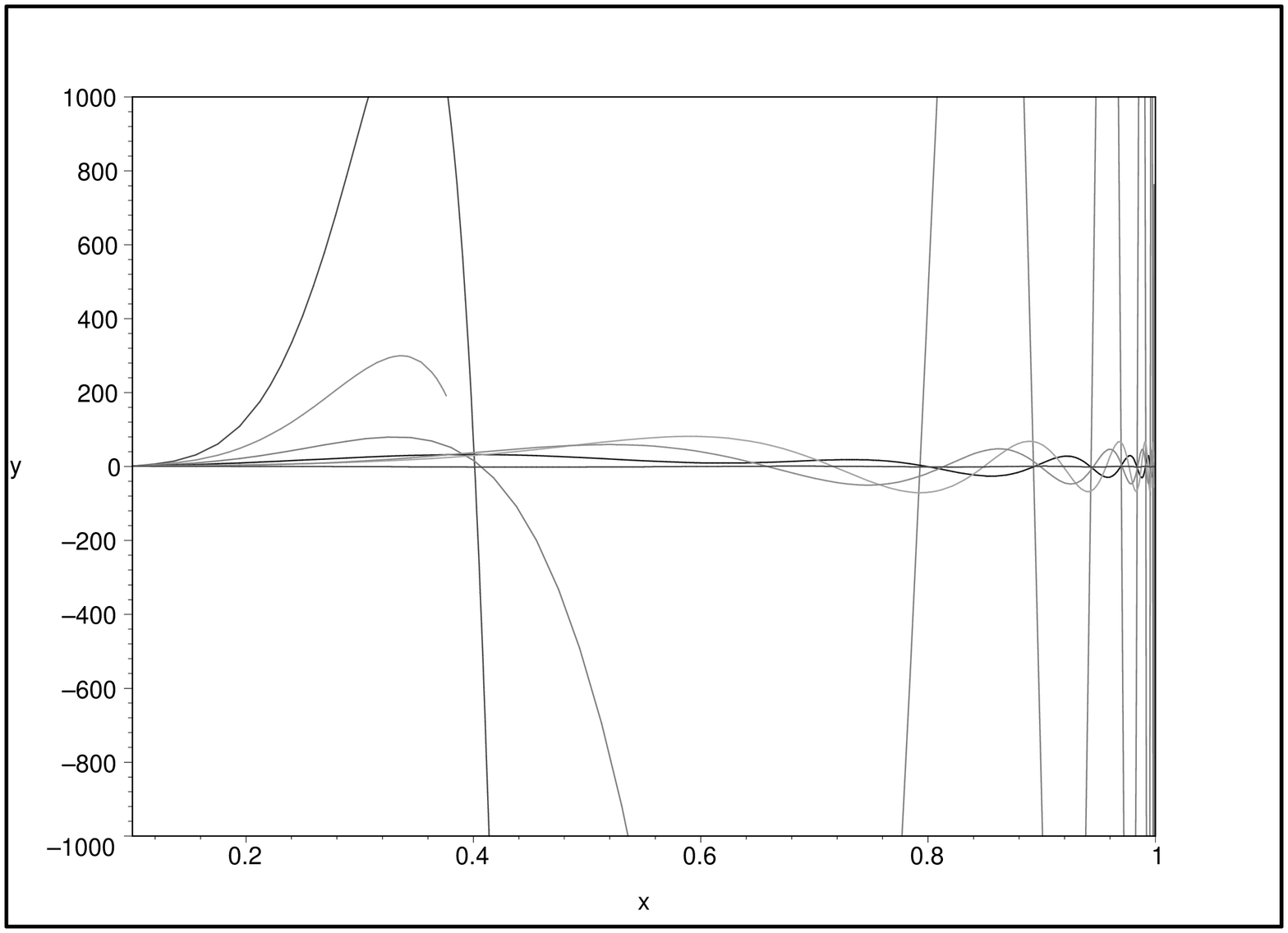}
\caption{Numerical solutions for $f_7(x)$ with $f_7(0.1)=1$, $f'_1(0.1)=100$, $l=1,\ldots, 7$
and with $\Omega=5$ (left figure) and $\Omega=10$ (right figure). Increasing values of $l$
correspond to more peaked curves on the right part of the plots.}
\end{figure}

\chapter{Exact solutions in a de Sitter background}

\section{Metric perturbations using the Regge-\\Wheeler gauge}

\def\b{\begin{equation}}
\def\e{\end{equation}}

In Chapter \textbf{3}, we have used the Lichnerowicz operator to derive the coupled system of differential equations for metric perturbations. However, it is possible to proceed in a different way. One considers the vacuum Einstein equations
\b
R_{ab}-\frac{1}{2}Rg_{ab}+\Lambda g_{ab}=0.
\e

If one introduces $g_{ab}=\gamma_{ab}+\epsilon h_{ab}$, where $\epsilon$ is a parameter which controls the perturbation, one has
\b
G_{ab}^0+\epsilon G_{ab}^1+\Lambda(\gamma_{ab}+\epsilon h_{ab})=0,
\e
where $G_{ab}^0$ is the Einstein tensor with respect to the background $\gamma_{ab}$ and $G_{ab}^1$ takes the form
\b
G_{ab}^1=R_{ab}^1-\frac{1}{2}(\gamma_{ab}R^1+h_{ab}R^0),
\e
where $R^0$ is the scalar curvature with respect to the metric $\gamma_{ab}$, whereas $R_{ab}^1$ and $R^1$ are the Ricci tensor and the scalar curvature valuated to first-order in the metric perturbation $h_{ab}$. Thus, one has
\b
G_{ab}^0+\Lambda\gamma_{ab}+\epsilon(G_{ab}^1+\Lambda h_{ab})=0
\e
and, since the background metric $\gamma_{ab}$ satisfies the vacuum Einstein equations, we must solve only the following equations:
\b
R_{ab}^1-\frac{1}{2}(\gamma_{ab}R^1+h_{ab}R^0)+\Lambda h_{ab}=0.
\label{Donato}
\e

\subsection{Regge-Wheeler gauge}

Solving this system of coupled differential equations is clearly a hard task. However, we can introduce the Regge-Wheeler gauge \cite{RW}. With the notation used in (\ref{(74)}), one has

\begin{eqnarray}
H_0(r)&=&W_1(r),\quad H_1(r)=W_2(r),\quad H_2(r)=W_3(r),\nonumber \\
K(r)&=&W_4(r),\quad G(r)=0,\quad h_0(r)=0,\quad h_1(r)=0.
\label{RW}
\end{eqnarray}

Thus, on using the relations (\ref{(210a)}), (\ref{(74)}), (\ref{x}), (\ref{Omega}) and (\ref{RW}), the wave equation (\ref{Donato}) leads to the following system of differential equations:

\begin{eqnarray}
\frac{dW_1(x)}{dx}&=&\frac{-2x^4+(L-4)\,x^2+2-L}{2x^3(x^2-1)}W_1(x)
+\frac{\imath\,\Omega}{x^2(x^2-1)}W_2(x)\nonumber\\
&+&\frac{(2-L-2\,\Omega^2)\,x^2-2+L}{2x^3(x^2-1)}W_4(x)\\
\label{I}
\frac{dW_2(x)}{dx}&=&\frac{\imath\,\Omega}{x^2-1}W_1(x)-\frac{2x}{x^2-1}W_2(x)
+\frac{\imath\,\Omega}{x^2-1}W_4(x)\\
\label{II}
\frac{dW_4(x)}{dx}&=&\frac{W_1(x)}{x}+\frac{\imath\,L}{2\,\Omega x^2}W_2(x)+\frac{W_4(x)}{x(x^2-1)}\\
\label{III}
W_1(x)&=&W_3(x)\\
\label{IV}
0&=&\frac{2-L}{2x^3}W_1(x)+\frac{\imath(L+2\,\Omega^2)}{2\,\Omega x^2}W_2(x)\nonumber\\
&+&\frac{(L+2\,\Omega^2)\,x^2+2-L}{2x^3(x^2-1)}W_4(x)
\label{V}
\end{eqnarray}

At this stage, we can solve this system exactly in terms of the Heun general functions (see \cite{Montaquila2} and Appendix \textbf{D}). In fact, if we use the equation (\ref{V}), we can isolate $W_1(x)$ and, by using the equations (\ref{I}) and (\ref{II}), we obtain a second-order differential equation for $W_4(x)$, that is

\b
\left[\frac{d^{\,2}}{dx^2}+A(x)\frac{d}{dx}+B(x)\right]W_4(x)=0
\e
where $A(x)$ and $B(x)$ take the form

\begin{eqnarray}
A(x)&=&\frac{4(L+2\,\Omega^2)x^4+2L(L-2)x^2-L(L-2)}{x(x^2-1)[2(L+2\,\Omega^2)x^2+L(L-2)]}\nonumber\\
B(x)&=&\frac{2\,\Omega^2(L+2\,\Omega^2)x^4+L(L-2)(L+\Omega^2)x^2-L^2(L-2)}{x^2(x^2-1)^2[2(L+2\,\Omega^2)x^2+L(L-2)]}
\end{eqnarray}

Eventually, we have

\begin{eqnarray}
W_4(x)&=&(x^2-1)^{\frac{i}{2}\Omega}\left[C_1x^{-1-l}\textit{HeunG}(a_1, q_1, \alpha_1, \beta_1, \gamma_1, \delta_1, x^2)\right.\nonumber\\
&+&\left.C_2x^l\textit{HeunG}(a_2, q_2, \alpha_2, \beta_2, \gamma_2, \delta_2, x^2)\right]
\end{eqnarray}
where
\begin{eqnarray}
a_1&=&\frac{l(2+l)(1-l^2)}{2(2\,\Omega^2+l^2+l)},\nonumber\\
q_1&=&\frac{(l-i\Omega)(l+2)(l+1)(i\Omega l^2-i\Omega l+4i\Omega+3l+2l^2-l^3)}{8(2\,\Omega^2+l^2+l)},\nonumber\\
\alpha_1&=&\frac{1}{2}(i\Omega-l),\nonumber\\
\beta_1&=&\frac{1}{2}(i\Omega-l-1),\nonumber\\
\gamma_1&=&\frac{1}{2}-l,\nonumber\\
\delta_1&=&i\Omega+1,
\end{eqnarray}
and
\begin{eqnarray}
a_2&=&\frac{l(2+l)(1-l^2)}{2(2\,\Omega^2+l^2+l)},\nonumber\\
q_2&=&\frac{l(1-l)(1+l+i\Omega)(i\Omega l^2+3i\Omega l+6i\Omega+4l+5l^2+l^3)}{8(2\,\Omega^2+l^2+l)},\nonumber\\
\alpha_2&=&\frac{1}{2}(i\Omega+l),\nonumber\\
\beta_2&=&\frac{1}{2}(i\Omega+l+1),\nonumber\\
\gamma_2&=&\frac{3}{2}+l,\nonumber\\
\delta_2&=&i\Omega+1.
\end{eqnarray}

At this stage, if we use the solution for $W_4(x)$ and the equations (\ref{I}) and (\ref{II}), we obtain the solution for $W_2(x)$. Moreover, by using the equations (\ref{III}) and (\ref{V}), we have the solutions for $W_1(x)$ and $W_3(x)$.

The decoupling technique and analytic formulae are original \cite{Montaquila2} but in this context, we want to solve explicitly the equation (\ref{Donato}) also in the de Donder gauge (see Chapter \textbf{3}) and, eventually, to compare this result with that obtained in this Chapter by using the Regge-Wheeler gauge.

Moreover, other powerful techniques are available for solving a tensor wave equation in curved space-time. The interested reader is referred, for example, to Appendix \textbf{E}.

\chapter*{Conclusions}
\addcontentsline{toc}{chapter}{Conclusions}

Gravitational waves are been here considered as metric perturbations in a curved background rather than in the flat Minkowski one, since several interesting physical situations can be discussed in this framework.

In Chapter \textbf{1}, we have seen that in the case in which the de Donder gauge is imposed, its preservation under infinitesimal space-time diffeomorphisms is guaranteed if and only if the associated covector is ruled by a second-order hyperbolic operator. Moreover, since in this case the Ricci term of the wave equation has opposite sign with respect to the wave equation of Maxwell theory, in the Lorenz gauge, it is possible to relate the solutions of the two problems \cite{Bini-Capozziello-Esposito}.

In Chapter \textbf{2}, we have completely succeeded in solving the homogeneous vector wave equation of Maxwell theory
in the Lorenz gauge when a de Sitter space-time is considered. One component of the vector field is expressed, in its radial part, through the solution of a fourth-order ordinary differential equation obeying given initial conditions. The other components of the vector field are then found by acting with lower-order differential operators on the solution of the fourth-order equation. The whole four-vector potential is eventually expressed through hypergeometric functions and spherical harmonics. The decoupling technique, analytic formulae and plots are completely original \cite{Montaquila}.

In Chapter \textbf{3}, we have extended this method to the wave equation of metric perturbations in de Sitter space-time and we have written the Lichnerowicz operator $P_{ab}^{\phantom{ab}cd}$ which results from the expansion of the action functional to quadratic order in the metric perturbations. The basic equations of the theory read therefore as
\begin{eqnarray*}
P_{ab}^{\phantom{ab}cd}h_{cd}&=&0\\
\Phi_a(h)&=&0
\end{eqnarray*}
where $\Phi_a$ is the supplementary condition for gravitational waves. Eventually, a numerical analysis of solutions has been performed.

In Chapter \textbf{4}, we have solved explicitly the Einstein equations for metric perturbations on a de Sitter background. In fact, by using the Regge-Wheeler gauge, the coupled system of differential equations to first-order in the metric perturbation $h_{ab}$ is been solved in terms of the Heun general functions.

In future work, we want to solve explicitly the equation (\ref{Donato}) also in the de Donder gauge (see Chapter \textbf{3}) and, eventually, to compare this result with that obtained in Chapter \textbf{4} by using the Regge-Wheeler gauge \cite{Montaquila2}.

\chapter*{Acknowledgments}
\addcontentsline{toc}{chapter}{Acknowledgments}

I would like to thank my supervisors, Dr Giampiero Esposito and Dr Donato Bini, for their support and patience throughout these years and for many discussions we had. Particular thanks go to my advisors, Professor Salvatore Capozziello and Dr Patrizia Vitale, whose questions and comments helped me to improve this work. I would like to thank CNR for partial support and Guido Celentano.

I want to thank my friends, Amedeo, Angelo, Enzo, Raffaele, Salvatore. I thank Patrizia who is ``simply'' my life.

\appendix
\chapter{Bivectors and biscalars}
In Eq. (\ref{(13)}), $g_{\; c'}^{a}$ is the geodesic parallel displacement
bivector (in general, bitensors behave as a tensor both at $x$ and
at $x'$) which effects parallel displacement of vectors along the
geodesic from $x'$ to $x$. In general, it is defined by the
differential equations
\begin{equation}
\sigma^{;b} \; g_{\; c';b}^{a}=\sigma^{;b'} \;
g_{\; c';b'}^{a}=0,
\label{(A1)}
\end{equation}
jointly with the coincidence limit
\begin{equation}
\lim_{x' \to x}g_{\; c'}^{a}=\Bigr[g_{\; c'}^{a} \Bigr]
=\delta_{\; c}^{a}.
\label{(A2)}
\end{equation}
The bivector $g_{\; c'}^{a}$, when acting on a vector $B^{c'}$ at
$x'$, gives therefore the vector ${\overline B}^{a}$ which is obtained by
parallel transport of $B^{c'}$ to $x$ along the geodesic connecting
$x$ and $x'$, i.e.
\begin{equation}
{\overline B}^{a}=g_{\; c'}^{a} \; B^{c'}.
\label{(A3)}
\end{equation}

In Eq. (\ref{(14)}), $\bigtriangleup(x,x')$ is a biscalar built from the
Van Vleck--Morette determinant
\begin{equation}
D(x,x') \equiv {\rm det} (\sigma_{; ab'})
\label{(A4)}
\end{equation}
according to
\begin{equation}
\bigtriangleup(x,x') \equiv {1\over \sqrt{-\gamma(x)}} D(x,x')
{1\over \sqrt{-\gamma(x')}}.
\label{(A5)}
\end{equation}
The biscalar $\bigtriangleup(x,x')$ has unit coincidence limit:
$[\bigtriangleup]=1$; as a function of $x$ (resp. $x'$), it becomes
infinite on any caustic formed by the geodesics emanating from $x'$
(resp. $x$). When $\bigtriangleup$ diverges in this way, $x$ and $x'$
are said to be conjugate points \cite{DeWi84}.
\chapter{Derivatives of $f_{0}$}
The higher-order derivatives of $f_{0}$ in Chapter \textbf{2} get increasingly cumbersome, but for completeness we write hereafter the result for $f_{0}''(x)$, i.e.
\begin{eqnarray}
{d^{2}\over dx^{2}}f_{0}(x)&=& C_{3}\biggr\{{4a_{1}(a_{1}+1)b_{1}(b_{1}+1)\over d_{1}(d_{1}+1)}x^{l+1}
(1-x^{2})^{-{i\over 2}\Omega}F(a_{1}+2,b_{1}+2;d_{1}+2;x^{2})\nonumber \\
&+& x^{l-1}(1-x^{2})^{-{i\over 2}\Omega-1}{2a_{1}b_{1}\over d_{1}}x\nonumber \\
& \times & [(2l+1)(1-x^{2})+2i \Omega x^{2}]F(a_{1}+1,b_{1}+1;d_{1}+1;x^{2})\nonumber \\
&+& x^{l-2}(1-x^{2})^{-{i\over 2}\Omega-2}\biggr[l(l-1)(x^{2}-1)^{2}-{i\Omega\over 2}
(x^{2}-1)(lx+2(l+1)x^{2})\nonumber \\
&+& (2i \Omega- \Omega^{2})x^{4}\biggr]F(a_{1},b_{1};d_{1};x^{2})\biggr\}\nonumber \\
&+&\biggr \{C_{3}\rightarrow C_{4},\;a_{1}\rightarrow a_{1}+1,\;b_{1}\rightarrow b_{1}-1\biggr\}.
\label{(A1)}
\end{eqnarray}

\chapter{Lichnerowicz operator}
The Lichnerowicz operator on metric perturbations is obtained by expansion of the Einstein-Hilbert action to quadratic order, i.e.
\begin{equation}
\frac{\delta^2S_{EH}}{\delta g^{\mu\nu}\delta g^{\rho\sigma}}\delta g^{\rho\sigma}=0,
\end{equation}
where
\begin{equation}
S_{EH}=\frac{1}{16\pi}\int_M d^4x\sqrt{g}\,R
\end{equation}
and $g=-det(g_{\mu\nu})$. The first functional derivative of the action is
\begin{equation}
\frac{\delta S}{\delta g^{\mu\nu}}=\int d^4x\frac{\delta\sqrt{g}}{\delta g^{\mu\nu}}R+\int d^4x\sqrt{g}\,\frac{\delta g_{\rho\sigma}}{\delta g^{\mu\nu}}R^{\rho\sigma}+\int d^4x\sqrt{g}\,g_{\rho\sigma}\frac{\delta R^{\rho\sigma}}{\delta g^{\mu\nu}},
\label{L1}
\end{equation}
where the first two integrals can be written as
\begin{equation}
\int d^4x\frac{\delta\sqrt{g}}{\delta g^{\mu\nu}}R+\int d^4x\sqrt{g}\,\frac{\delta g_{\rho\sigma}}{\delta g^{\mu\nu}}R^{\rho\sigma}=\int d^4x\sqrt{g}\left [R_{\mu\nu}-\frac{1}{2}g_{\mu\nu}R\right ].
\end{equation}

Thus, the first functional derivative provides the vacuum Einstein equations. In fact, the integrand of the third integral of (\ref{L1}) is a total derivative, that is
\begin{equation}
g_{\rho\sigma}\frac{\delta R^{\rho\sigma}}{\delta g^{\mu\nu}}=
\frac{1}{2}\nabla_\rho[\delta^\rho_\mu\nabla_\nu\delta(x,x')+\delta^\rho_\nu\nabla_\mu\delta(x,x')-
g_{\mu\nu}\nabla^\rho\delta(x,x')].
\end{equation}

At this stage, it's possible to write the second functional derivative of the Einstein-Hilbert action, i.e.
\begin{eqnarray}
\frac{\delta^2S}{\delta g^{\gamma\varepsilon}\delta g^{\mu\nu}}=\int d^4x\frac{\delta^2\sqrt{g}}{\delta g^{\gamma\varepsilon}\delta g^{\mu\nu}}R+\int d^4x \frac{\delta\sqrt{g}}{\delta g^{\mu\nu}}\frac{\delta g_{\rho\sigma}}{\delta g^{\gamma\varepsilon}}R^{\rho\sigma}+\int d^4x \frac{\delta\sqrt{g}}{\delta g^{\mu\nu}}g_{\rho\sigma}\frac{\delta R^{\rho\sigma}}{\delta g^{\gamma\varepsilon}}\nonumber \\+\int d^4x\frac{\delta\sqrt{g}}{\delta g^{\gamma\varepsilon}}\frac{\delta g_{\rho\sigma}}{\delta g^{\mu\nu}}R^{\rho\sigma}+\int d^4x\sqrt{g}\frac{\delta^2g_{\rho\sigma}}{\delta g^{\gamma\varepsilon}\delta g^{\mu\nu}}R^{\rho\sigma}+\int d^4x\sqrt{g}\,\frac{\delta g_{\rho\sigma}}{\delta g^{\mu\nu}}\frac{\delta R^{\rho\sigma}}{\delta g^{\gamma\varepsilon}}\,\,\,
\end{eqnarray}
and if we note that
\begin{equation}
\frac{\delta}{\delta g^{\gamma\varepsilon}}\left[-\frac{1}{2}\sqrt{g}\,g_{\mu\nu}\right]=
\frac{1}{4}\sqrt{g}\,g_{\gamma\varepsilon}g_{\mu\nu}-\frac{1}{4}\sqrt{g}(g_{\mu\gamma}g_{\nu\varepsilon}+
g_{\mu\varepsilon}g_{\nu\gamma}),
\end{equation}
one has the expression for the first integral, i.e.
\begin{equation}
\frac{\delta^2\sqrt{g}}{\delta g^{\gamma\varepsilon}\delta g^{\mu\nu}}R=
\frac{1}{4}\sqrt{g}(g_{\mu\gamma}g_{\nu\varepsilon}+g_{\mu\varepsilon}g_{\nu\gamma}-g_{\gamma\varepsilon}g_{\mu\nu})R,
\end{equation}
moreover, one has
\begin{equation}
\frac{\delta\sqrt{g}}{\delta g^{\mu\nu}}\frac{\delta g_{\rho\sigma}}{\delta g^{\gamma\varepsilon}}R^{\rho\sigma}=
-\frac{1}{4}\sqrt{g}\,g_{\mu\nu}(g_{\rho\gamma}g_{\sigma\varepsilon}+
g_{\rho\varepsilon}g_{\sigma\gamma})R^{\rho\sigma}=-\frac{1}{2}\sqrt{g}\,g_{\mu\nu}R_{\gamma\varepsilon}
\end{equation}
and similarly, the fourth integral is
\begin{equation}
\frac{\delta\sqrt{g}}{\delta g^{\gamma\varepsilon}}\frac{\delta g_{\rho\sigma}}{\delta g^{\mu\nu}}R^{\rho\sigma}=
-\frac{1}{4}\sqrt{g}\,g_{\gamma\varepsilon}(g_{\rho\mu}g_{\sigma\nu}+
g_{\rho\nu}g_{\sigma\mu})R^{\rho\sigma}=-\frac{1}{2}\sqrt{g}\,g_{\gamma\varepsilon}R_{\mu\nu}.
\end{equation}

At this stage, we note that the variation of the Ricci tensor respect to the metric can be written as
\begin{eqnarray}
\frac{\delta R^{\rho\sigma}}{\delta g^{\gamma\varepsilon}}&=&
\frac{1}{4}(\delta^\alpha_\gamma\delta^\sigma_\varepsilon
+\delta^\alpha_\varepsilon\delta^\sigma_\gamma)\nabla_\alpha\nabla^\rho
+\frac{1}{4}(\delta^\alpha_\gamma\delta^\rho_\varepsilon
+\delta^\alpha_\varepsilon\delta^\rho_\gamma)\nabla_\alpha\nabla^\sigma\nonumber \\
&-&\frac{1}{4}(\delta^\rho_\gamma\delta^\sigma_\varepsilon
+\delta^\rho_\varepsilon\delta^\sigma_\gamma)\nabla_\alpha\nabla^\alpha
-\frac{1}{2}g_{\gamma\varepsilon}\nabla^\rho\nabla^\sigma.
\end{eqnarray}

Thus, the third integral is
\begin{equation}
\frac{\delta\sqrt{g}}{\delta g^{\mu\nu}}g_{\rho\sigma}\frac{\delta R^{\rho\sigma}}{\delta g^{\gamma\varepsilon}}=
-\frac{1}{2}\sqrt{g}\left\{\frac{1}{2}g_{\mu\nu}(\nabla_\gamma\nabla_\varepsilon+\nabla_\varepsilon\nabla_\gamma)
-g_{\mu\nu}g_{\gamma\varepsilon}\nabla_\alpha\nabla^\alpha\right\}
\end{equation}
and for the fifth integral, if we note that
\begin{eqnarray}
\frac{\delta^2g_{\rho\sigma}}{\delta g^{\gamma\varepsilon}\delta g^{\mu\nu}}&=&
\frac{1}{2}g_{\sigma\nu}(g_{\rho\gamma}g_{\mu\varepsilon}+g_{\rho\varepsilon}g_{\mu\gamma})+
\frac{1}{2}g_{\rho\mu}(g_{\sigma\gamma}g_{\nu\varepsilon}+g_{\sigma\varepsilon}g_{\nu\gamma})\nonumber \\
&+&\frac{1}{2}g_{\sigma\mu}(g_{\rho\gamma}g_{\nu\varepsilon}+g_{\rho\varepsilon}g_{\nu\gamma})+
\frac{1}{2}g_{\rho\nu}(g_{\sigma\gamma}g_{\mu\varepsilon}+g_{\sigma\varepsilon}g_{\mu\gamma}),
\end{eqnarray}
one has the expression
\begin{equation}
\sqrt{g}\frac{\delta^2g_{\rho\sigma}}{\delta g^{\gamma\varepsilon}\delta g^{\mu\nu}}R^{\rho\sigma}=
\frac{1}{2}\sqrt{g}\,[g_{\mu\varepsilon} R_{\nu\gamma}+g_{\mu\gamma} R_{\varepsilon\nu}+g_{\nu\varepsilon} R_{\gamma\mu}+
g_{\nu\gamma} R_{\mu\varepsilon}].
\end{equation}

Moreover, for the last integral, one has
\begin{eqnarray}
\sqrt{g}\,\frac{\delta g_{\rho\sigma}}{\delta g^{\mu\nu}}\frac{\delta R^{\rho\sigma}}{\delta g^{\gamma\varepsilon}}&=&
\frac{1}{2}\sqrt{g}\,(g_{\mu\rho}g_{\sigma\nu}+g_{\rho\nu}g_{\sigma\mu})\frac{\delta R^{\rho\sigma}}{\delta g^{\gamma\varepsilon}}\nonumber \\
&=&\frac{1}{4}\sqrt{g}\,[g_{\varepsilon\mu}\nabla_\gamma\nabla_\nu+g_{\nu\gamma}\nabla_\varepsilon\nabla_\mu
+g_{\varepsilon\nu}\nabla_\gamma\nabla_\mu+g_{\gamma\mu}\nabla_\varepsilon\nabla_\nu\nonumber \\
&-&(g_{\gamma\mu}g_{\varepsilon\nu}+g_{\varepsilon\mu}g_{\gamma\nu})\nabla_\alpha\nabla^\alpha
-g_{\gamma\epsilon}(\nabla_\mu\nabla_\nu+\nabla_\nu\nabla_\mu)].
\end{eqnarray}

At this point, adding the terms obtained, one has the following operator:
\begin{eqnarray}
\frac{\delta^2S}{\delta g^{\gamma\varepsilon}\delta g^{\mu\nu}}=\mathcal{P}_{\mu\nu\gamma\varepsilon}&=&
-\frac{1}{4}\sqrt{g}\,[(g_{\gamma\mu}g_{\varepsilon\nu}+g_{\varepsilon\mu}g_{\gamma\nu}
-2g_{\mu\nu}g_{\gamma\varepsilon})\nabla_\alpha\nabla^\alpha\nonumber \\
&+&g_{\mu\nu}(\nabla_\gamma\nabla_\varepsilon+\nabla_\varepsilon\nabla_\gamma)
+g_{\gamma\varepsilon}(\nabla_\mu\nabla_\nu+\nabla_\nu\nabla_\mu)\nonumber \\
&-&g_{\varepsilon\mu}\nabla_\gamma\nabla_\nu
-g_{\nu\gamma}\nabla_\varepsilon\nabla_\mu-g_{\varepsilon\nu}\nabla_\gamma\nabla_\mu
-g_{\gamma\mu}\nabla_\varepsilon\nabla_\nu\nonumber \\
&+&2(g_{\mu\nu}R_{\gamma\varepsilon}+g_{\gamma\varepsilon}R_{\mu\nu}
-g_{\mu\varepsilon}R_{\nu\gamma}-g_{\mu\gamma}R_{\varepsilon\nu}
-g_{\nu\varepsilon}R_{\gamma\mu}-g_{\nu\gamma}R_{\mu\varepsilon})\nonumber\\
&-&(g_{\mu\gamma}g_{\nu\varepsilon}+g_{\mu\varepsilon}g_{\nu\gamma}-g_{\gamma\varepsilon}g_{\mu\nu})R].
\label{L7}
\end{eqnarray}

However, $\mathcal{P}_{\mu\nu\gamma\varepsilon}$ is a singular operator. One has a supplementary condition for the $\delta g_{\mu\nu}$, that is
\begin{equation}
P_\alpha^{\mu\nu}\delta g_{\mu\nu}=0.
\end{equation}

Thus, using the supermetric $G^{\mu\nu\rho\sigma}$, one has
\begin{equation}
P^{\mu\nu\alpha}=-\frac{1}{2}G^{\mu\nu\rho\sigma}(\lambda)\,Q_{\rho\sigma}^{\,\,\,\,\,\,\alpha},
\end{equation}
with
\begin{equation}
Q_{\rho\sigma}^{\,\,\,\,\,\,\alpha}=-(\delta_\rho^\alpha\nabla_\sigma+\delta_\sigma^\alpha\nabla_\rho)
\label{L1.5}
\end{equation}
and
\begin{equation}
G^{\mu\nu\rho\sigma}(\lambda)=\frac{1}{2}(g^{\mu\rho}g^{\nu\sigma}+g^{\mu\sigma}g^{\nu\rho}+\lambda g^{\mu\nu}g^{\rho\sigma}),
\label{L2}
\end{equation}
where, however, not all the values of $\lambda\in \textbf{R}$ are acceptable. In fact, the supermetric $G^{\mu\nu\rho\sigma}(\lambda)$ must be invertible, that is
\begin{equation}
G^{\mu\nu\rho\sigma}(\lambda)\,G_{\rho\sigma\varepsilon\tau}(f(\lambda))=
\frac{1}{2}(\delta_\tau^\mu\delta_\varepsilon^\nu+\delta_\varepsilon^\mu\delta_\tau^\nu),
\label{L3}
\end{equation}
with
\begin{equation}
G_{\rho\sigma\varepsilon\tau}(f(\lambda))=\frac{1}{2}[g_{\rho\varepsilon}g_{\sigma\tau}
+g_{\rho\tau}g_{\sigma\varepsilon}+f(\lambda)g_{\rho\sigma}g_{\tau\varepsilon}].
\label{L4}
\end{equation}

Thus, using (\ref{L2}) and (\ref{L4}) into (\ref{L3}), one has
\begin{equation}
G^{\mu\nu\rho\sigma}(\lambda)\,G_{\rho\sigma\varepsilon\tau}(f(\lambda))=
\frac{1}{2}(\delta_\tau^\mu\delta_\varepsilon^\nu+\delta_\varepsilon^\mu\delta_\tau^\nu)
+g^{\mu\nu}g_{\varepsilon\tau}\left[\frac{\lambda}{2}+\frac{f(\lambda)}{2}+\frac{n\lambda f(\lambda)}{4}\right],
\label{L5}
\end{equation}
where $n$ is the dimension of the space-time. Now, if we want that (\ref{L5}) is equal to (\ref{L3}), we must impose that
\begin{equation}
\left[\frac{\lambda}{2}+\frac{f(\lambda)}{2}+\frac{n\lambda f(\lambda)}{4}\right]=0.
\end{equation}

Thus, solving this equation for $f(\lambda)$, one has
\begin{equation}
f(\lambda)=-\frac{2\lambda}{2+n\lambda},
\end{equation}
which is the coefficient of the inverse supermetric. Eventually, we note that the supermetric is invertible if
\begin{equation}
\lambda\neq-\frac{2}{n}.
\end{equation}

At this stage, using (\ref{L1.5}) and (\ref{L2}), one has
\begin{eqnarray}
-\frac{1}{2}G^{\mu\nu\rho\sigma}(\lambda)Q_{\rho\sigma}^{\,\,\,\,\,\,\alpha}&=&
\frac{1}{4}[g^{\mu\rho}g^{\nu\sigma}\delta_\rho^\alpha\nabla_\sigma
+g^{\mu\sigma}g^{\nu\rho}\delta_\rho^\alpha\nabla_\sigma
+\lambda g^{\mu\nu}g^{\rho\sigma}\delta_\rho^\alpha\nabla_\sigma]\,\,\,\,\,\,\,\,\nonumber \\
&+&\frac{1}{4}[g^{\mu\rho}g^{\nu\sigma}\delta_\sigma^\alpha\nabla_\rho
+g^{\mu\sigma}g^{\nu\rho}\delta_\sigma^\alpha\nabla_\rho
+\lambda g^{\mu\nu}g^{\rho\sigma}\delta_\sigma^\alpha\nabla_\rho]
\end{eqnarray}
and, eventually, the supplementary condition becomes
\begin{equation}
P^{\mu\nu\alpha}(\lambda)=\frac{1}{2}[g^{\mu\alpha}\nabla^\nu+g^{\nu\alpha}\nabla^\mu+\lambda g^{\mu\nu}\nabla^\alpha].
\label{L6}
\end{equation}

Now, we compute the supplementary condition applied on the metric perturbations, i.e.
\begin{equation}
\Phi^\alpha(h_{\mu\nu},\lambda)=P^{\mu\nu\alpha}(\lambda)h_{\mu\nu},
\end{equation}
where $h_{\mu\nu}=\delta g_{\mu\nu}$. Thus, using (\ref{L6}),  $\Phi^\alpha$ becomes
\begin{eqnarray}
\Phi^\alpha(h_{\mu\nu},\lambda)&=&\frac{1}{2}[g^{\mu\alpha}\nabla^\nu h_{\mu\nu}+g^{\nu\alpha}\nabla^\mu h_{\mu\nu}+\lambda g^{\mu\nu}\nabla^\alpha h_{\mu\nu}]\nonumber \\
&=&\frac{1}{2}[\nabla^\nu h_\nu^\alpha+\nabla^\mu h_\mu^\alpha+\lambda \nabla^\alpha g^{\mu\nu} h_{\mu\nu}]=\frac{1}{2}\nabla^\mu[2h_\mu^\alpha+\lambda g_\mu^\alpha g^{\rho\sigma}h_{\rho\sigma}]\nonumber \\
&=&\nabla_\mu\left(h^{\mu\alpha}+\frac{\lambda}{2}g^{\mu\alpha}g^{\rho\sigma}h_{\rho\sigma}\right)
\end{eqnarray}
where if we put $\lambda=-1$, one has the \emph{de Donder's gauge}.

At this stage, we can to compute the invertible operator $\mathcal{F}_{\mu\nu\gamma\varepsilon}$, that is
\begin{equation}
\mathcal{F}_{\mu\nu\gamma\varepsilon}=\mathcal{P}_{\mu\nu\gamma\varepsilon}
-P_{\mu\nu}^{\,\,\,\,\,\,\alpha}g_{\alpha\beta}P^{\,\,\,\,\,\,\beta}_{\gamma\varepsilon}.
\end{equation}

Thus, if we note that
\begin{equation}
P_{\mu\nu}^{\,\,\,\,\,\,\alpha}=\frac{1}{2}(\delta_\mu^\alpha\nabla_\nu+\delta_\nu^\alpha\nabla_\mu+\lambda g_{\mu\nu}\nabla^\alpha),
\end{equation}
one has
\begin{eqnarray}
P_{\mu\nu}^{\,\,\,\,\,\,\alpha}g_{\alpha\beta}P^{\,\,\,\,\,\,\beta}_{\gamma\varepsilon}&=&
-\frac{1}{4}(g_{\mu\gamma}\nabla_\nu\nabla_\varepsilon
+g_{\mu\varepsilon}\nabla_\nu\nabla_\gamma+g_{\nu\gamma}\nabla_\mu\nabla_\varepsilon
+g_{\nu\varepsilon}\nabla_\mu\nabla_\gamma)\nonumber \\
&-&\frac{1}{4}\lambda g_{\gamma\varepsilon}(\nabla_\nu\nabla_\mu+\nabla_\mu\nabla_\nu)
-\frac{1}{4}\lambda g_{\mu\nu}(\nabla_\gamma\nabla_\varepsilon+\nabla_\varepsilon\nabla_\gamma)\nonumber \\
&-&\frac{1}{4}\lambda^2 g_{\mu\nu}g_{\gamma\varepsilon}\nabla_\alpha\nabla^\alpha.
\label{L8}
\end{eqnarray}

Now, using (\ref{L7}) and (\ref{L8}), the operator $\mathcal{F}_{\mu\nu\gamma\varepsilon}$ becomes
\begin{eqnarray}
\mathcal{F}_{\mu\nu\gamma\varepsilon}(\lambda)&=&
-\frac{1}{4}\sqrt{g}\,\{[g_{\gamma\mu}g_{\varepsilon\nu}+g_{\varepsilon\mu}g_{\gamma\nu}
-(2-\lambda^2)g_{\mu\nu}g_{\gamma\varepsilon}]\nabla_\alpha\nabla^\alpha\nonumber \\
&+&(1+\lambda)g_{\mu\nu}(\nabla_\gamma\nabla_\varepsilon+\nabla_\varepsilon\nabla_\gamma)
+(1+\lambda)g_{\gamma\varepsilon}(\nabla_\mu\nabla_\nu+\nabla_\nu\nabla_\mu)\nonumber \\
&-&g_{\mu\gamma}(\nabla_\nu\nabla_\varepsilon-\nabla_\varepsilon\nabla_\nu)
-g_{\mu\varepsilon}(\nabla_\nu\nabla_\gamma-\nabla_\gamma\nabla_\nu)\nonumber\\
&-&g_{\nu\gamma}(\nabla_\mu\nabla_\varepsilon-\nabla_\varepsilon\nabla_\mu)
-g_{\nu\varepsilon}(\nabla_\mu\nabla_\gamma-\nabla_\gamma\nabla_\mu)\nonumber\\
&+&2(g_{\mu\nu}R_{\gamma\varepsilon}+g_{\gamma\varepsilon}R_{\mu\nu}-g_{\mu\varepsilon}R_{\nu\gamma}
-g_{\mu\gamma}R_{\varepsilon\nu}-g_{\nu\varepsilon}R_{\gamma\mu}-g_{\nu\gamma}R_{\mu\varepsilon})\nonumber \\
&-&(g_{\mu\gamma}g_{\nu\varepsilon}+g_{\mu\varepsilon}g_{\nu\gamma}-g_{\gamma\varepsilon}g_{\mu\nu})R\}.
\end{eqnarray}

At this stage, we note the following relations:
\begin{eqnarray}
&&-g_{\mu\gamma}(\nabla_\nu\nabla_\varepsilon-\nabla_\varepsilon\nabla_\nu)
-g_{\mu\varepsilon}(\nabla_\nu\nabla_\gamma-\nabla_\gamma\nabla_\nu)
-g_{\nu\gamma}(\nabla_\mu\nabla_\varepsilon-\nabla_\varepsilon\nabla_\mu)\nonumber\\
&&-g_{\nu\varepsilon}(\nabla_\mu\nabla_\gamma-\nabla_\gamma\nabla_\mu)
=g_{\tau\gamma}(R^\tau_{\mu\nu\varepsilon}+R^\tau_{\nu\mu\varepsilon})
+g_{\mu\tau}(R^\tau_{\gamma\nu\varepsilon}+R^\tau_{\epsilon\nu\gamma})\nonumber\\
&&+g_{\tau\varepsilon}(R^\tau_{\mu\nu\gamma}+R^\tau_{\nu\mu\gamma})
+g_{\nu\tau}(R^\tau_{\gamma\mu\varepsilon}+R^\tau_{\varepsilon\mu\tau})
=2R_{\mu\gamma\nu\varepsilon}+2R_{\mu\varepsilon\nu\gamma}.
\label{L9}
\end{eqnarray}

Eventually, with $\lambda=-1$ and using (\ref{L9}), one has
\begin{eqnarray}
\mathcal{F}_{\mu\nu\gamma\varepsilon}&=&-\frac{1}{4}\sqrt{g}\,\{[g_{\gamma\mu}g_{\varepsilon\nu}
+g_{\varepsilon\mu}g_{\gamma\nu}-g_{\mu\nu}g_{\gamma\varepsilon}](\nabla_\alpha\nabla^\alpha-R)
+2R_{\mu\gamma\nu\varepsilon}+2R_{\mu\varepsilon\nu\gamma}\nonumber \\
&+&2(g_{\mu\nu}R_{\gamma\varepsilon}+g_{\gamma\varepsilon}R_{\mu\nu}-g_{\mu\varepsilon}R_{\nu\gamma}
-g_{\mu\gamma}R_{\varepsilon\nu}-g_{\nu\varepsilon}R_{\gamma\mu}
-g_{\nu\gamma}R_{\mu\varepsilon})\}.
\end{eqnarray}

\chapter{Heun functions}

The five multi-parameter Heun equations have been popping up with surprising frequency in applications during the last 15 years. Heun equations include as particular cases the Lame, Mathieu, spheroidal wave, hypergeometric, and with them most of the known equations of mathematical physics. Five Heun functions are defined as the solutions to each of these five Heun equations, computed as power series solutions around the origin satisfying prescribed initial conditions.

The General Heun equation, with four regular singular points in $\{0,1,a,\infty\}$, is
\b
\left[\frac{d^{\,2}}{dz^2}+\left(\frac{\gamma}{z}+\frac{\delta}{z-1}+\frac{\epsilon}{z-a}\right)\frac{d}{dz}
+\frac{\alpha\beta z-q}{z(z-1)(z-a)}\right]y(z)=0.
\e

The solution to this equation is implemented in this work as the \textit{HeunG} function. The sum of the exponents of the singularities of Heun's equation is equal to two and the parameter $\epsilon$ is expressed in terms of the other ones by
\b
\epsilon=\alpha+\beta+1-\gamma-\delta.
\e

The other four Heun equations are confluent cases, obtained from the general Heun equation above through confluence processes. The Heun Confluent equation, with regular singular points in $\{0,1\}$ and irregular in $\{\infty\}$, is
\b
\left[\frac{d^{\,2}}{dz^2}+\left(\frac{\gamma}{z}+\frac{\delta}{z-1}-\epsilon\right)\frac{d}{dz}
+\frac{q-\alpha\beta}{z-1}-\frac{q}{z}\right]y(z)=0,
\e
having for solution the \textit{HeunC} function. The Biconfluent equation, with regular singular point in $\{0\}$ and irregular in $\{\infty\}$, is
\b
\left[\frac{d^{\,2}}{dz^2}+\left(-2z-\beta+\frac{1+\alpha}{z}\right)\frac{d}{dz}
+\gamma-\alpha-2-\frac{1}{2}\frac{(1+\alpha)\beta+\delta}{z}\right]y(z)=0,
\e
having for solution the \textit{HeunB} function. The Doubleconfluent equation, with two irregular singular points in $\{-1,1\}$, is
\b
\left[\frac{d^{\,2}}{dz^2}-\left(\frac{\alpha+2z+\alpha z^2-2z^3}{(z^2-1)^2}\right)\frac{d}{dz}
+\frac{\delta+(2\alpha+\gamma)z+\beta z^2}{(z^2-1)^3}\right]y(z)=0,
\e
having for solution the \textit{HeunD} function. Eventually, the Triconfluent equation, with irregular singular point in $\{\infty\}$, is
\b
\left[\frac{d^{\,2}}{dz^2}-\left(\gamma+3z^2\right)\frac{d}{dz}
+\alpha+\beta z-3z\right]y(z)=0,
\e
having for solution the \textit{HeunT} function. The Heun functions, \textit{HeunG}, \textit{HeunC}, \textit{HeunB}, \textit{HeunD} and \textit{HeunT}, are defined as the solutions to the corresponding General, Confluent, Biconfluent, Doubleconfluent and Triconfluent Heun equations. These solutions are constructed as power series solutions around the origin, for certain initial conditions.

The power series solutions at the base of the functions' definitions have restricted radius of convergence in the \textit{HeunG}, \textit{HeunC} and \textit{HeunD} cases, where the numerical evaluation is done using analytic extensions, exploring closed form identities satisfied by these functions, as well as series expansions around different singularities. For arbitrary values of the parameters, however, closed form formulas for the connection constants relating series expansions around different singularities, are not known.

\section{Some important facts about Heun functions}

\begin{itemize}
\item The coefficients entering the series expansions represented by the Heun functions satisfy \textit{three term recurrence relations}. A solution to these recursion equations is not known in the general case, so a closed form for the series's coefficients is not available and the computation of - say - the \textit{nth} coefficient requires the explicit computation of all the previous ones.

\item They are more general than the rest of the functions of the mathematical language in that they contain most of them as particular cases. Consequently, the Heun equations cannot have their solution expressed (but as infinite sum power series) without using the corresponding Heun functions.
\item The Heun functions have a rich structure and so satisfy a rather large number of identities.
\item Because they have such a rich structure and include as particular so many functions, including the Mathieu, Lame, Spheroidal Wave and hypergeometric functions, the interrelations between them and the Heun ones are a source of many nontrivial identities between the former.
\item Due to the enlarged structure of singularities (if compared for instance with hypergeometric functions) the Heun functions are increasingly appearing in the modeling of different types of problems in applied mathematics.
\end{itemize}

\chapter{Wald's method on a de Sitter background}

Depending on the coordinate system used, there are many ways of viewing de Sitter space. It is has been studied largely because of the central role it plays in almost all inflationary scenarios of the early universe. Roughly speaking, the expansion is driven by a large cosmological constant which appears due to the energy density of a false vacuum. In this regard the description of de Sitter space by Robertson-Walker (RW) coordinates has tended to be the natural choice for most workers because of their obvious cosmological significance. In this coordinate system, constant time surfaces appear homogeneous and isotropic.

However, yet another picture of de Sitter space time is provided by static coordinates. Here homogenity of constant time surfaces is lost but space-time appears static within a horizon distance, a very different state of affairs from that prevailing in the RW description. Gal'tsov and N\'{u}\~{n}ez \cite{Nunez} have studied gravitational field perturbations in a de Sitter background described by static coordinates. To accomplish this its have employed the technique of Debye potentials introduced by Wald \cite{Wald}. This treatment is particularly attractive because it reduces the problem of solving the sourceless equations for fields of different spin \emph{s} to that of solving a single differential equation for the Debye potentials with free parameter \emph{s}.

\section{Wald's technique for the Debye potentials}

Wald has described \cite{Wald} a remarkable technique for solving field equations. This technique requires writing the differential equations in two ways, which are related to obtain a third equation for the so-called Debye potential. The solution of the original field equations are expressed as operators acting upon this potential.

Explicitly, Wald's method takes the field equations as its point of departure
\b
_{|s|}\mathcal{E}_{ab}\varphi^b=4\pi_{|s|}j_a,
\label{W1}
\e
where $s$ is the spin weight, $_{|s|}\mathcal{E}_{ab}$ are field operators, $\varphi^b$ is the ``middle'' potential and $_{|s|}j_a$ is the source for the field. Next Teukolsky's work \cite{Teuko}, which makes use of the Newman-Penrose formalism \cite{NP} to express the field equations, is then incorporated and the field variables $\psi$ are expressed as operators $_sM_a$ acting upon the middle potentials
\b
_s\psi={_sM_a}\varphi^a.
\label{W2}
\e

Wald went on to obtain a second way of writing the field equations
\b
\frac{1}{\rho\rho^\ast}{_s\cstok{\ }}{_sM_a}\varphi^a=4\pi{_s\tau^a}{_{|s|}j_a},
\label{W3}
\e
where $\rho$ is a spinor coefficient, $_s\cstok{\ }$ is a Teukolsky operator, $_s\tau^a$ is a source projection operator and the other quantities are defined as in Eqs. (\ref{W1}) and (\ref{W2}).

By operating on Eq. (\ref{W1}) with $_s\tau^a$, it is possible to equate the projected left-hand side of Eq. (\ref{W1}) with that of Eq. (\ref{W3})
\b
{_s\tau^a}{_{|s|}\mathcal{E}_{ab}\varphi^b}=\frac{1}{\rho\rho^\ast}{_s\cstok{\ }}{_sM_b}\varphi^b.
\label{W4}
\e

From here Wald proceeds to the operator identity,
\b
{_s\tau^a}{_{|s|}\mathcal{E}_{ab}}=\frac{1}{\rho\rho^\ast}{_s\cstok{\ }}{_sM_b}.
\label{W5}
\e

However, this step needs some justification. For example, in the case of the scalar function, $\nabla_\mu\varphi=\partial_\mu\varphi$ does not imply the operator identity $\nabla_\mu=\partial_\mu$. The validity of equation Eq. (\ref{W5}) has, however, been established for a wide class of space-time which includes the de Sitter space \cite{Nunez_thesis}.

The adjoint tilde of the operator identity (\ref{W5}) is now taken. By making use of the fact that the operator $_{|s|}\mathcal{E}_{ab}$ is self-adjoint for the fields of physical interest, we obtain
\b
{_{|s|}\mathcal{E}_{ab}}\,\,{_s\tilde{\tau}^a}={_s\tilde{M}_b}\,\,
{_s\widetilde{\cstok{\ }}}\widetilde{\left(\frac{1}{\rho\rho^\ast}\right)}.
\label{W6}
\e

If the function $_s\Xi$, satisfying
\b
{_s\widetilde{\cstok{\ }}}\widetilde{\left(\frac{1}{\rho\rho^\ast}\right)}{_{-s}\Xi}=0,
\label{W7}
\e
is introduced, then on account of Eq. (\ref{W6}), $_s\Xi$ also satisfy the equation
\b
{_{|s|}\mathcal{E}_{ab}}\,\,{_s\tilde{\tau}^a}\,{_{-s}\Xi}=0,
\label{W8}
\e
$_s\Xi$ is known as the Debye potential. For integer spin fields, the operator $_{|s|}\mathcal{E}_{ab}$ is real, in which case $({_s\tilde{\tau}^a}\,{_{-s}\Xi})^\ast$ also satisfies Eq. (\ref{W8}). For the spin $\frac{1}{2}$ field, however, this is not so.

In the integer spin case, we define
\b
_s\varphi^{a(\pm)}=\frac{1}{2\alpha}\{{_s\tilde{\tau}^a}\,{_{-s}\Xi}\pm ({_s\tilde{\tau}^a}\,{_{-s}\Xi})^\ast\},
\label{W9}
\e
where $\alpha=1$ for $(+)$ and $\alpha=i$ for $(-)$. So
\b
{_{|s|}\mathcal{E}_{ab}}\,\,{_s\varphi^{a(\pm)}}=0,
\label{W10}
\e
that is, the function $_s\varphi^{a(\pm)}$ satisfy the sourceless fields equations (\ref{W1}). For the spin $\frac{1}{2}$ field, the solutions are given by (\ref{W9}) without taking the complex-conjugate. Thus, the problem of solving the field equations reduces simply to solving Eq. (\ref{W7}), for the Debye potential.

\section{Wald's method in de Sitter space}

In a cosmological context it is natural to describe de Sitter space in Robertson-Walker (flat) coordinates
\b
ds^2=dt^2-e^{2\alpha t}(dr^2+r^2d\theta^2+r^2\sin^2\theta\, d\varphi^2).
\e

However,\, Wald's method is most directly applied in static coordinates ($\tau$, $\chi$, $\theta$, $\varphi$), where formally de Sitter can be treated as a member of the Kerr family. These are related by
\b
\tau=t-\frac{1}{2\alpha}\ln(1-\alpha^2r^2e^{2\alpha t});\quad \chi=re^{\alpha t};\quad \theta=\theta;\quad \varphi=\varphi.
\e
In this coordinates the line element takes the form
\b
ds^2=\frac{\Delta}{\chi^2}d\tau^2-\frac{\chi^2}{\Delta}d\chi^2-\chi^2(d\theta^2+\sin^2\theta\, d\varphi^2),
\e
where
\b
\Delta=\chi^2(1-\alpha^2\chi^2)
\e
and $\alpha$ is related to the cosmological constant $\Lambda$ by $\alpha^2=\frac{\Lambda}{3}$. The de Sitter horizon is given by $\chi_+^2=\alpha^{-2}$.

We now introduce the Kinnersley null tetrad given by
\begin{eqnarray}
l^\mu &=& \left(\frac{\chi^2}{\Delta},1,0,0\right),\nonumber\\ n^\mu &=& -\frac{\Delta}{2\chi^2}\left(-\frac{\chi^2}{\Delta},1,0,0\right),\nonumber\\
m^\mu &=& \frac{1}{\sqrt{2}\chi}\left(0,0,1,i\csc\theta\right).
\end{eqnarray}
With respect to this choice of tetrad the nonzero spin coefficients are
\b
\rho=-\frac{1}{\chi}; \quad \mu=\frac{\Delta}{2\chi^2}\rho; \quad \gamma=\mu+\frac{\Delta'}{4\chi^2}
\e
and
\b
\beta=\frac{\cot\theta}{2\sqrt{2}\chi}; \quad \alpha=-\beta,
\e
where the prime represents differentiation whit respect to $\chi$. We notice also that the Weyl tensor vanishes; thus in the Petrov classification, de Sitter space is a type N space (a particular case of type D space).

We work with the operators $D_p$, $D_p^+$, $L_q$ and $L_q^+$, which where introduced by Chandrasekhar \cite{Chandra} in the black hole context. In the case of de Sitter space, they are related to the Newman-Penrose operators $D$, $\Delta$, $\delta$ and $\delta^\ast$, in the following way
\b
D_p=D+p\frac{\Delta'}{\Delta}; \quad -\frac{\Delta\rho^2}{2}D_p^+=\Delta-2p\frac{\Delta'}{4\chi^2}
\e
and
\b
-\frac{\rho}{\sqrt{2}}L_q^+=\delta+2q\beta; \quad -\frac{\rho}{\sqrt{2}}L_q=\delta^\ast+2q\beta,
\e
or, more explicitly,
\begin{eqnarray}
D_p &=& \partial_\chi+p\frac{\Delta'}{\Delta}+\frac{\chi^2}{\Delta}\,\partial_r, \nonumber \\
D_p^+ &=& \partial_\chi+p\frac{\Delta'}{\Delta}-\frac{\chi^2}{\Delta}\,\partial_r, \nonumber \\
\mathcal{L}_q &=& \partial_\theta+q\cot\theta-i\csc\theta\,\partial_\varphi, \nonumber \\
\mathcal{L}_q^+ &=& \mathcal{L}_q^\ast.
\end{eqnarray}
With respect to the adjoint operation,
\b
\tilde{\nabla}_\mu=-\nabla_\mu,
\e
so that the adjoint Chandrasekhar operators are
\begin{eqnarray}
\tilde{D}_p &=& -\rho^2D_{-p}\rho^2, \nonumber \\
\tilde{D}_p^+ &=& -\rho^2D_{-p}^+\rho^2, \nonumber \\
\tilde{\mathcal{L}}_q &=& -\mathcal{L}_{1-q}^+, \nonumber \\
\tilde{\mathcal{L}}_q^+ &=& -\mathcal{L}_{1-q}.
\end{eqnarray}
The derivation of the Debye's potential has been already done for a large class of type D and type N space-times, which include the de Sitter space as a particular case of type N space \cite{Torres}, so we give just the results for the field perturbations in the absence of sources in terms of the corresponding Debye potential $_{\pm s}\Xi$.

For the gravitational case $s=\pm 2$, the field perturbations $_{\pm 2}h^{\mu\nu}$ are given by
\b
_{\pm 2}h^{\mu\nu(\pm)}=\frac{1}{2\alpha}\{{_{\pm 2}\tilde{\tau}^{\mu\nu}}_{\mp 2}\Xi\pm c.c.\},
\e
where
\begin{eqnarray}
_{+2}\tilde{\tau}^{\mu\nu} &=& l^\mu l^\nu\frac{\rho^2}{2}\mathcal{L}_1\mathcal{L}_2
+l^{(\mu}m^{\ast \nu)}\sqrt{2}\rho^{-1}\mathcal{L}_2\mathcal{D}_0 \rho^2 \nonumber \\
&+& m^{\ast\mu}m^{\ast\nu}\rho\mathcal{D}_0\rho^{-4}\mathcal{D}_0\rho^3, \nonumber \\
_{-2}\tilde{\tau}^{\mu\nu} &=& n^\mu n^\nu\frac{\rho^{-2}}{2}\mathcal{L}_1^+\mathcal{L}_2^+
-n^{(\mu}m^{\nu)}\frac{\Delta\rho^{-3}}{\sqrt{2}}\mathcal{L}_2^+\mathcal{D}_2^+\rho^2 \nonumber \\
&+& m^\mu m^\nu\frac{\rho\Delta^2}{4}\mathcal{D}_2^+\rho^{-4}\mathcal{D}_2^+\rho^3.
\end{eqnarray}

For the electromagnetic case $s=\pm 1$, the vector potential $_{\pm 1}A^\mu$ is given by
\b
_{\pm 1}A^{\mu(\pm)}=\frac{1}{2\alpha}\{{_{\pm 1}\tilde{\tau}^\mu}_{\pm 1}\Xi\pm c.c.\},
\e
where
\begin{eqnarray}
_{+1}\tilde{\tau}^\mu &=& -l^\mu \frac{\rho}{\sqrt{2}}\mathcal{L}_1
-m^{\ast \mu}\rho^{-1}\mathcal{D}_0 \rho, \nonumber \\
_{-1}\tilde{\tau}^\mu &=& n^\mu \frac{\rho^{-1}}{\sqrt{2}}\mathcal{L}_1^+
-m^{\mu}\frac{\Delta\rho^{-1}}{2}\mathcal{D}_1^+\rho.
\end{eqnarray}

The Debye potentials for each field satisfy a second-order differential equation, which leaving the spin projection $s$ as a parameter can be written as a single master equation, obtained from the Teukolsky equation
\b
_s\cstok{\ }^\ast{\,}_s\Xi=0,
\label{Teu}
\e
on account of the identity
\b
_s\widetilde{\cstok{\ }\rho^2}=\rho^2{\,}_{-s}\cstok{\ }^\ast,
\e
where the Teukolsky operator, $_s\cstok{\ }$, is given, for $s>0$, by
\b
_s\cstok{\ }=\Delta D_1D_s^++\mathcal{L}_{1-s}^+\mathcal{L}_s-2(2s-1)\chi\partial_r-2(2s-1)(s-1)\alpha^2\chi^2
\label{sp}
\e
and, for $s<0$, by
\b
_s\cstok{\ }=
\Delta D_{1+s}^+D_0+\mathcal{L}_{1+s}\mathcal{L}_{-s}^+-2(2s+1)\chi\partial_r-2(2s+1)\alpha^2\chi^2.
\label{sn}
\e
The explicit form of the operators $_s\tau^a$ and $_sM_a$, needed in these derivations, are given in \cite{Nunez}. Because we are working in static coordinates, we expect a time behavior $e^{-i\omega\tau}$; we observe, moreover, that the Debye potentials factorize into radial, angular, and temporal parts as follows:
\b
_s\Xi(x)=\int{_s}R(\chi)\,_{-s}Y(\theta,\varphi)e^{-i\omega\tau}d\omega.
\label{ra}
\e

Furthermore, it is clear that equation (\ref{Teu}) separate into radial and angular parts. Substituting (\ref{ra}) into (\ref{Teu}) and by using (\ref{sp}) and (\ref{sn}), we obtain for the radial functions $(s>0)$
\b
\left[\Delta D_1D_s^++2(2s-1)i\omega\chi-2(2s-1)(s-1)\alpha^2\chi^2\right]{_s}R(\chi)={_s}\lambda{_s}R(\chi)
\e
and, for $s<0$, we have
\b
\left[\Delta D_{1+s}^+D_0+2(2s+1)i\omega\chi-2(2s+1)(s+1)\alpha^2\chi^2\right]{_s}R(\chi)={_s}\lambda{_s}R(\chi).
\e

For the angular functions, we obtain
\begin{eqnarray}
\mathcal{L}_{1-s}\mathcal{L}_s^+\,_{-s}Y(\theta,\varphi)=-\,{_s}\lambda\,_{-s}Y(\theta,\varphi), \quad s>0\nonumber\\
\mathcal{L}_{1+s}^+\mathcal{L}_{-s}\,_{-s}Y(\theta,\varphi)=\,-{_s}\lambda\,_{-s}Y(\theta,\varphi), \quad s<0
\end{eqnarray}
where ${_s}\lambda$ is some separation constant. Eventually, it is possible to show that for the radial functions, one has \cite{Nunez}
\begin{eqnarray}
{_s}R(\chi)&=&{_s}C_1\chi^{-2s-1}
F\left(l+s+1,-l+s,1+s+\frac{i\omega}{2\alpha},\frac{1+\alpha\chi}{2\alpha\chi}\right)\nonumber\\
&\times&\left(\frac{1+\alpha\chi}{1-\alpha\chi}\right)^\frac{i\omega}{2\alpha}\nonumber\\
&+&{_s}C_2\Delta^{-s}\chi^{2s-1}
F\left(l-s+1,-l-s,1-s-\frac{i\omega}{2\alpha},\frac{1+\alpha\chi}{2\alpha\chi}\right)\nonumber\\
&\times&\left(\frac{1+\alpha\chi}{1-\alpha\chi}\right)^{-\frac{i\omega}{2\alpha}}.
\end{eqnarray}

\end{document}